\documentclass{revtex4-2}

\usepackage[utf8]{inputenc}
\usepackage[english]{babel}

\usepackage[hidelinks]{hyperref}

\usepackage{graphicx}
\usepackage{subcaption}
\usepackage{parskip}

\usepackage{float}
\usepackage{booktabs}
\usepackage{makecell}
\usepackage{multirow}

\usepackage[figurename=Fig.]{caption}
\usepackage[tablename=Tab.]{caption}
\usepackage[font=small]{caption}

\usepackage{amsmath}
\usepackage{amssymb}
\usepackage{amsfonts}
\usepackage{mathtools}

\usepackage{upgreek}
\usepackage{xcolor}
\usepackage[normalem]{ulem}

\begin{document}

\title{Matter Accumulations and Accretion Tori around Wormholes}
\date{\today}

\author{Kristian Gjorgjieski}
 \email{kristian.gjorgjieski@uol.de}
\author{Jutta Kunz}
 \email{jutta.kunz@uni-oldenburg.de}
\affiliation{Department of Physics, Carl von Ossietzky University of Oldenburg, 26111 Oldenburg, Germany}

\author{Petya Nedkova}
 \email{pnedkova@phys.uni-sofia.bg}
\affiliation{Department of Theoretical Physics, Sofia University, Sofia 1164, Bulgaria}

\begin{abstract}
    We study circular orbits and accretion structures around symmetric wormholes. As exemplary solutions we choose three different wormhole spacetimes, namely rotating traversable wormholes from the Teo class, the rotating Simpson-Visser metric with the parameter spectrum corresponding to wormholes and a static wormhole from beyond Horndeski theories. We show the existence of a spectrum of circular orbits at the wormhole throat for each of these wormhole solutions and analyze the boundaries of this spectrum across the respective wormhole parameter range. For each of the solutions we identified that vast regions of the parameter space correspond to stable orbits. The presence of this orbit spectrum can be linked through accretion disk models to matter accumulations which may form at the throat. We present here examples of such disk solutions by implementing the Polish Doughnut model. In some cases, these matter accumulations are encapsulating the throat and could imply central bright regions of the wormhole spacetime. Furthermore, the combined analysis of the equatorial Keplerian orbits and the throat orbits, leads to a set of different disk configurations, where matter accumulations at the throat may be present with outer tori around them, in some cases also as a connected structure. Our results may hint to possible instabilities of wormholes, especially if they have an ergoregion. Moreover, wormholes with such disks, may appear more star-like when it comes to their observational signature, due to the centralized and more spherical emission profile associated with the possible matter accumulations at the throat.
\end{abstract}

\maketitle


\section{Introduction}

Observational signatures in the form of emission spectra play a fundamental role when it comes to the search for astrophysical bodies. If we observe signatures from an object in the electro-magnetic spectrum and if they are not reflections, either the observed object is self-emitting, which is the case for stars, or it is surrounded by accumulations of emitting matter. These accumulations of matter are called accretion disks, which represent structures composed of hot gases and plasma that form around compact objects. Strong gravitational fields lead to the attraction of matter over long periods of time and the attracted matter settles on orbits around the central object. Usually these orbits are of circular shape, resulting in a circular flow of the disk particles within the disk. The morphology of the forming structures is representing therefore a disk or a torus. The properties of these disks or tori are mainly influenced by the properties of the spacetime, which in turn are determined by the properties of the central object. The theoretical analysis of possible accretion structures around hypothetical objects could therefore give hints on their observational signatures. For some objects this could perhaps be one of the few possibilities to actually differentiate and identify them, as it is the case for various black hole mimickers. 

The term Black hole mimicker generally describes a type of object, which might appear similar to a black hole for a distant observer. 
These objects are characterized by a missing radiation profile, as they do not emit or reflect any radiation. Thus, they have a central dark region similar to a black hole shadow. Their observable signatures originate from possible accretion structures, which may form around them. One class of such objects are wormholes. In general, wormholes are objects that connect two arbitrary distant regions of spacetime through a tunnel-like connection. Both ends of the wormhole can be located within the same connected spacetime or in two separate spacetimes. The latter case would correspond to a wormhole, that connects one universe to another. The first concepts of wormholes were introduced in the first half of the last century \cite{Flamm, Weyl, ERbridges}. However, they did not describe stable connections. In the second half of the last century, concepts were developed further \cite{Fuller, Ellis, Bronnikov} and a class of wormholes was conceptualized, which is referred to as traversable wormholes \cite{Morris-Throne, Visser1989, Teo}. These objects describe wormhole solutions, which are free of any horizons and singularities. Furthermore, it is possible for physical entities to cross from each of the two sides through the tunnel-like connection to the other side. This connection is also called the wormhole throat and it has to fulfill special constraints. For an open and lasting connection a negative energy density is required, which is a property of a proposed exotic type of matter that could stabilize the wormhole throat. However, there are various proposals how this constraint could be met without the introduction of exotic matter \cite{Cramer1995, Gravanis2007, Bronnikov_2015, Jose2021, Bakopoulos_2022}.

In this work we will investigate accretion structures and matter accumulations around different wormhole geometries. The analyzed wormhole solutions consist of wormholes from the Teo class \cite{Teo}, which is a rotating generalization of the Morris-Thorne wormhole \cite{Morris-Throne}. Furthermore, we analyze the rotating generalization of the Simpson-Visser metric within the wormhole regime \cite{Mazza_2021}, as well as a traversable wormhole from beyond Horndeski theories \cite{Bakopoulos_2022}. The phenomenology of these solutions in the electromagnetic spectrum has been studied in a series of works including investigations of the shadow \cite{Petya2013, Gyulchev:2018, Hyat2023}, the thin accretion disks \cite{Paul_2020}, the quasi-periodic oscillations \cite{Deligianni:2021, Deligianni:2021hwt}, and the polarization of the emission from an accretion disk \cite{Delijski:2022}.

As a model for accretion structures we will use the Thick Disk model, which models the accretion disk as a body composed of non-self gravitating fluid particles \cite{Fishbone1976,Abramowicz1978,Kozlowski1978,Abramowicz1980,Paczynsky1980,Paczynski1981,Paczysnki1982}. The solutions represent geometrically thick tori structures with a high opacity. Within the Thick Disk model, we will consider the Polish Doughnut solutions, which describe non-magnetized perfect fluid disks. These solutions are fully analytical and as such a useful model to capture qualitatively the main properties of possible accretion structures. Disk solutions are constructed on the basis of circular geodesics of the spacetime and their stability properties, which may lead to different kind of disk structures. Especially the presence of circular geodesics at the wormhole throat \cite{ThroatOrbits} may lead to distinctive accretion structures as we will show.

This work is structured as follows. In Section 2 we introduce the different wormhole solutions that we are analyzing and afterwards in Section 3 a brief introduction to the Polish Doughnut model will follow. Section 4 will present an analysis on the possibility of matter accumulations at the wormhole throat. This analysis will be expanded in Section 5 by the investigation of the Keplerian circular orbits of the spacetime and the possible accretion structures which may form. The crucial findings of our results will be summarized in a conclusion at the end of this paper. Throughout this work we will use geometrized units with $G = c = 1$, where $G$ is the gravitational constant and $c$ is the speed of light.


\section{Traversable wormholes}

Traversable wormhole solutions were popularized by Michael S. Morris and Kip S. Thorne in \cite{Morris-Throne}. By formulating a set of conditions for the metric functions they obtained a class of metrics, which describes the geometry of a wormhole and further on ensures its traversability for observers. This class of metric solutions is known as the Morris-Throne wormhole and it describes a static, spherically symmetric and asymptotically flat spacetime, whose squared line element can be written as

\begin{align}
    ds^2 = - e^{\Phi}dt^2 + \left(1 - \frac{b}{r} \right)^{-1} dr^2 + r^2(d\theta^2 + \sin^2\theta d\varphi^2),
\end{align}

where $\Phi$ and $b$ are functions of the radial coordinate $r$. This class of static solutions was generalized by Edward Teo in \cite{Teo} to axially symmetric solutions by including rotation of the wormhole spacetime. These rotating traversable wormholes describe a class of solutions, whose generalized squared line element can be written as

\begin{align}
    ds^2 = -N^2dt^2 + \left( 1 - \frac{b}{r}\right)^{-1}dr^2 + r^2 K^2 \left( d\theta^2 + \sin^2\theta(d\varphi - \omega dt)^2\right),
\end{align}

where the metric functions $N$, $b$, $K$ and $\omega$ are regular on the symmetry axis and only depend on $r$ and $\theta$. By considering the physical implications of these functions, it is possible to further define conditions, which they must obey in order to describe the geometrical characteristics of rotating wormholes and to ensure their traversability. The metric function $N$ is related to the gravitational redshift and should therefore be finite and nonzero throughout the definition range. The metric function $b$ is also called the shape function, since it determines the shape of the wormhole throat, which is located where $g_{rr}$ becomes singular. The shape function should fulfill the so-called flare-out condition at the throat, $\partial_r b < 1$. The metric function $K$ is a measure of the radial distance to the origin and should therefore be a non-negative and increasing function. The quantity $2 \pi r K \sin\theta$ can geometrically be interpreted as the proper circumference of a circle (around the coordinate origin) located at the coordinates $(r,\theta)$ with $\varphi$ going from $0$ to $2\pi$. For the circumference coordinate $R = rK$ it follows therefore, that $\partial_r R > 0$. The metric function $\omega$ describes the rotation of the wormhole and it is the angular velocity experienced by a free falling particle from infinity at the coordinates $(r,\theta)$. These constraints determine a class of metric functions, which describe two spacetime regions joined together at the wormhole throat. The radial coordinate takes thereby the range $r \in [r_0, \infty)$, with $r_0$ designating the location of the wormhole throat in the local coordinates. Furthermore, imposing also asymptotical flatness leads to further restrictions on the behavior of the metric functions in the limit for $r \rightarrow \infty $, namely

\begin{align}
    N \rightarrow 1 \ \ ; \ \ K \rightarrow 1 \ \ ; \ \  \frac{b}{r} \rightarrow 0 \ \ ; \ \ \omega \rightarrow 0.
\end{align}

In particular, if the expansions of $N$ and $\omega$ at spatial infinity have the form

\begin{align}
    N = 1 - \frac{M}{r} + \mathcal{O}\left(\frac{1}{r^2}\right) \ \ ; \ \ \omega = \frac{2J}{r^3} + \mathcal{O}\left(\frac{1}{r^4}\right), \label{eq:expansions}
\end{align}

then $M$ and $J$ can be identified as the ADM mass and the total angular momentum of the wormhole, respectively.
In this work we will investigate an exemplary wormhole solution of the Teo class, which fulfills all described constraints and whose metric functions are given by

\begin{align}
    N = \exp \left(- \frac{r_0}{r} \right) \ \ \ ; \ \ \ b = r_0 \ \ \ ; \ \ \ K = 1 \ \ \ ; \ \ \ \omega = \frac{2J}{r^3}, 
\end{align}

where the mass of the wormhole is given by $M = r_0$. Furthermore, it is possible to introduce a global radial coordinate $l$ by the relation,

\begin{align}
    \frac{d l}{dr} &= \pm \left( 1 - \frac{r_0}{r} \right)^{\frac{1}{2}} \\
    \Rightarrow \ l(r) &= \pm \left( r \sqrt{1 - \frac{r_0}{r}} + r_0 \ln \left( \sqrt{r - r_0} + \sqrt{r} \right)  \right)
\end{align}

which is well behaved across the wormhole throat. The global radial coordinate $l$ extends from $-\infty$ to $+\infty$ and gives the proper radial distance to the throat, with the wormhole throat being located at $l = 0$. We will define the region $l \in (-\infty,0]$ as the lower side of the wormhole and $l \in [0,\infty)$ as the upper side of the wormhole. For a normalized mass, $M = r_0 = 1$, the squared line element of the metric can then be written as,

\begin{align}
    \mathcal{TE}: \ \ ds^2 = -\exp \left( -\frac{2}{r(l)} \right) dt^2 + dl^2 + r(l)^2 \left(d \theta^2 + \sin^2 \theta \left(d \varphi - \frac{2a}{r(l)^3} dt \right)^2 \right),
\end{align}

where we introduced the dimensionless spin parameter $a = \frac{J}{M^2}$ and where we designate the metric as $\mathcal{TE}$ ($\mathcal{T}$ for Teo class and $\mathcal{E}$ for the exponential character of $N$). Since $r$ is a symmetric function of $l$, both sides are symmetric with respect to the throat and mirror images of each other. The wormhole spacetime described by this metric possesses an ergoregion, if the transcendental equation

\begin{align}
     4a^2 \sin^2 \theta - r(l)^4 \exp \left(-\frac{2}{r(l)} \right) = 0,
\end{align}

has a solution. Solving numerically leads to the threshold value $a_E = 0.184$ above which the spacetime possesses an ergoregion enclosing the wormhole throat. 

As a second wormhole solution we will analyze the generalization of the Simpson-Visser metric to rotating spacetimes, which was derived in \cite{Mazza_2021}. It describes a stationary and axially symmetric spacetime, which simplifies for vanishing rotation to the Simpson-Visser metric. For a normalized mass, $M = 1$, the squared line element of the metric is given by

\begin{align}
    \mathcal{RSV}: \ \  ds^2 = - \left( 1 - \frac{2 \sqrt{l^2 + \xi^2}}{\Sigma} \right) dt^2  + \frac{\Sigma}{\Delta} dl^2  + \Sigma d\theta^2 - \frac{4  a \sin^2 \theta \sqrt{l^2 + \xi^2}}{\Sigma} dt d \varphi + \frac{A \sin^2 \theta}{\Sigma} d\varphi^2,
\end{align}

where we designate the metric as $\mathcal{RSV}$ (standing for rotating Simpson-Visser) and where

\begin{align}
    \Sigma &= l^2 + \xi^2 + a^2 \cos^2 \theta \ \ \ ; \ \ \ \Delta = l^2 + \xi^2 + a^2 - 2  \sqrt{l^2 + \xi^2} \\
    A &= (l^2 + \xi^2 + a^2)^2 - \Delta a^2 \sin^2 \theta,
\end{align}

with $l$ as the global radial coordinate, $\xi$ as the regularization parameter of the spacetime and $a$ as the dimensionless spin parameter again. The spacetime structure described by this metric depends on the parameter range of $\xi$ and $a$. Here we will restrict ourselves to the parameter range, which corresponds to traversable wormholes. For wormholes, the spacetime is defined for $l \in (-\infty,\infty)$ with $l = 0$ designated to the wormhole throat. The upper and lower side are both asymptotically flat and also symmetric to the throat. The shape of the wormhole throat is described by the regularization parameter $\xi$, and the necessary condition for wormhole solutions is, that the outer horizon $l_H^+$ is exceeded by the wormhole throat, which corresponds to the inequality

\begin{align}
   l_H^+ =  1 + \sqrt{1 - a^2} < \xi.
\end{align}

Thus, for $\xi > 2$ the metric describes always a traversable wormhole, independent of $a$. This is also the case for $a > 1$, independent of $\xi$, as $l_H$ is not real for values of $a$ greater than one. Furthermore, $\xi = 1$ marks the threshold up to which wormhole solutions can exist for $a < 1$. The traversable wormhole solutions also can possess an ergoregion, if additionally the condition 

\begin{align}
    \xi < 1 + \sqrt{1 - a^2 \cos^2\theta} = l_E^+,
\end{align}

is satisfied. Thus, solutions with ergoregions only appear in the parameter range of $l_H^+ < \xi < 2$ and $a < 1$. In this work we will investigate both, solutions with an ergoregion and solutions without an ergoregion, for the rotating Simpson-Visser wormholes as well as for the Teo wormhole.

As a third class of wormhole solutions we will analyze traversable wormholes in beyond Horndeski theories, which were constructed in \cite{Bakopoulos_2022}. They are described by a static and spherically symmetric spacetime, whose squared line element can be written in general form as

\begin{align}
    ds^2 = -h dt^2  + \frac{dr^2}{hW^{-1}} + r^2 (d \theta^2 + \sin^2\theta d\varphi^2),
\end{align}

where the metric functions $h$ and $W$ only depend on $r$ and are constrained by the conditions the spacetime has to fulfill. In our case we demand metric functions, which could describe a traversable wormhole and are symmetric with respect to the wormhole throat, as well as asymptotically flat. We orient ourselves on \cite{Hyat2023} and specify the following metric functions,

\begin{align}
    h = 1 + \frac{r^2}{2 \alpha} \left( 1 - \sqrt{ \frac{8 \alpha M}{r^3}} \right) \ \ \ ; \ \ \ W^{-1} = 1 - \frac{r_0}{ \lambda r} \left(1 - \sqrt{h} \right),
\end{align}

where $M$ is the mass and the parameters $(r_0, \lambda,\alpha)$ determine the structure of the spacetime. In the case of a normalized mass, the parameter range for wormhole solutions is given by either $\lambda > 0$ and $\alpha > 1$, or $\lambda > 0$, $\alpha \leq 1$ and $r_H < r_0$, with $r_H$ being the horizon given by $r_H = 1 + \sqrt{1 - \alpha}$ and $r_0$ being the location of the throat. We restrict ourselves here to these conditions in order to analyze traversable wormholes. By the introduction of a global coordinate $l^2 = r^2 - r_0^2$, the wormhole spacetime is defined again from $-\infty$ to $+\infty$ with the throat located at $l = 0$. The squared line element of the metric can then be written as,

\begin{align}
    \mathcal{BH}: \ \ ds^2 = -H(l)dt^2 + F(l)^{-1} dl^2 + (l^2 + r_0^2) (d\theta^2 + \sin^2\theta d\varphi^2),
\end{align}

where we designate the metric as $\mathcal{BH}$ (for beyond Horndeski) and where $r_0$ represents the location of the wormhole throat. The metric functions $H$ and $F$ are for a normalized mass given by

\begin{align}
    H(l) = 1 + \frac{l^2 + r_0^2}{2 \alpha} \left( 1 - \sqrt{ \frac{8 \alpha }{(l^2 + r_0^2)^\frac{3}{2}}} \right) \ \ \ ; \ \ \ F(l) = \frac{ \left( H \left( 1 -  \frac{r_0}{\lambda \sqrt{l^2 + r_0^2}} \left(1 - \sqrt{h} \right) \right) \right)(l^2 + r_0^2)}{l^2}.
\end{align}

In our analysis we will vary the parameter $\alpha$ as well as the location of the throat $r_0$ and fix the parameter $\lambda = 1$, since it does not influence the properties of possible accretion structures around the wormhole.


\section{Circular orbits and Polish Doughnuts}
In this section we will present the basics of circular orbits and the Polish Doughnut model. Considering an axisymmetric and stationary spacetime, the line element for the equatorial plane ($d\theta = 0$) can be written in general form as

\begin{align}
    ds_{eq}^2 = g_{tt} dt^2 + g_{ll} dl^2 + 2 g_{t \varphi} dt d\varphi + g_{\varphi \varphi} d\varphi^2,
\end{align}

where the metric components $g_{tt}, \ g_{ll}, \ g_{t\varphi}$ and $g_{\varphi\varphi}$ only depend on the radial coordinate, since the polar coordinate is set to $\theta = \frac{\pi}{2}$. The motion of a force-free massive test particle is described by a timelike geodesic. If it is circular, the tangential four-velocity of the test particle can be written as $u^\mu = (u^t, 0, 0, u^\varphi) = u^t (\eta^\mu +  \Omega \xi^\mu)$, with $\eta^\mu$ and $\xi^\mu$ being the two Killing vector fields for the cyclic coordinates $t$ and $\varphi$, and $\Omega$ being the angular velocity of the test particle, $\Omega = \frac{u^\varphi}{u^t}$. The equations of motion can be derived by minimizing the variation of the action $\mathcal{S}$,

\begin{align}
    \mathcal{S} = \frac{1}{2} \int g_{\mu \nu} u^\mu u^\nu d\tau.
\end{align}

From the equation of motion one can derive with the cyclic coordinates $t$ and $\varphi$ the constants of motion, namely the energy and the angular momentum, given by

\begin{align}
    E = -g_{tt} \dot{t} + g_{t\varphi} \dot{\varphi} = - u_t \ \ \ ; \ \ \ L = - g_{t\varphi} \dot{t} + g_{\varphi \varphi} \dot{\varphi} = u_\varphi.
\end{align}

Using the constants of motion and the normalization condition, $u_\mu u^\mu = -1$, the equation of motion for the radial coordinate $l$ can be derived, which is then given in the form of

\begin{align}
    g_{ll} \dot{l}^2 - \frac{g_{\varphi\varphi} E^2  +  g_{tt}L^2 + 2 g_{t \varphi} E L}{g_{t\varphi}^2 - g_{\varphi\varphi} g_{tt}} + 1 = g_{ll} \dot{l}^2 + V = 0.
    \label{eq:veff}
\end{align}

Circular orbits are characterized by a fixed value of the radial coordinate, $\dot{r} = 0$. Therefore the effective potential $V$ (eq. (\ref{eq:veff})) and its derivative $\partial_l V$ must vanish for circular orbits. Stability requires furthermore $\partial^2_l V \geq 0$. In general, marginally stable orbits are thus given by $l_{ms} = \{l_0: V|_{l=l_0} = 0 \wedge \partial_l V|_{l=l_0} = 0 \wedge \partial_l^2 V|_{l=l_0} = 0 \}$. The innermost stable circular orbit is called the $l_{ISCO}$, and is defined as $l_{ISCO} = \min(l_{ms})$. For radial values with $l < l_{ISCO}$ no stable circular orbits exist. Unstable bound orbits exist up to the innermost marginally bound orbit. For normalized mass-energy marginally bound orbits are given by $l_{mb} = \{l_0: V|_{l=l_0} = 0 \wedge \partial_r V|_{l=l_0} = 0 \wedge E|_{l=l_0} = 1 \}$. For radial values with $l < l_{mb}$ a particle coming from infinity would fall into the central object or escape into infinity, depending on the boundary conditions. For a given spacetime the described systems of equations can be solved for the energy, the angular momentum and the radial coordinate $l$. However, starting with the assumption of circular motion the energy and angular momentum can be rewritten in terms of the angular velocity $\Omega$,

\begin{align}
    E &= - \frac{g_{t\varphi}\Omega + g_{tt}}{\sqrt{g_{\varphi\varphi}\Omega^2 - 2 g_{t\varphi}\Omega - g_{tt}}} \label{eq:E} \\
    L &=  \frac{g_{\varphi\varphi}\Omega + g_{t\varphi}}{\sqrt{g_{\varphi\varphi}\Omega^2 - 2 g_{t\varphi}\Omega - g_{tt}}}.
    \label{eq:L}
\end{align}

Considering geodesic motion, the angular velocity of a test particle on a circular geodesic can be derived by using the normalization condition and solving $a_\mu = 0$ for $\Omega$. The resulting Keplerian angular velocity $\Omega_K^\pm$ is then given by the metric components and the derivatives of the metric components by

\begin{align}
    \Omega_K^\pm = \frac{-\partial_l g_{t\varphi} \pm \sqrt{(\partial_l g_{t\varphi})^2 - \partial_l g_{tt} \partial_l g_{\varphi\varphi}}}{\partial_l g_{\varphi\varphi}},
\end{align}

where the $+$ sign is referring to prograde motion of the test particle and the $-$ sign to retrograde motion of the test particle. Inserting $\Omega_K^\pm$ into eqs. (\ref{eq:E}) and (\ref{eq:L}) gives the energy and the angular momentum of a Keplerian circular orbit. By using this pair of fixed values for the energy and angular momentum, the systems of equations governing the marginally stable and bound orbits can be reduced to one equation in each case, which can be solved for the radial coordinate. Using the energy and angular momentum the Keplerian specific angular momentum can be defined as

\begin{align}
    \ell_K^\pm = \frac{L_K^\pm}{E_K^\pm} = - \frac{g_{\varphi\varphi}\Omega_K^\pm + g_{t\varphi}}{g_{t\varphi}\Omega_K^\pm + g_{tt}},
\end{align}

which will play an important role in the classification of accretion disk solutions. The simplest form of thick disk solution is the Polish Doughnut model. In this model the accretion disk is modeled as a perfect fluid, governed by a barotropic equation of state, where the pressure is only a function of the rest mass density, and the stress-energy tensor of the disk is given solely by the fluid terms, leaving other inner-physical disk phenomena as viscosity and heat transfer out of account. Despite its simplicity, the Polish Doughnut model is a suitable tool to analyze accretion disks fully analytically and still capture their characterizing features. The construction of thick disks is derived from the relativistic Euler equations, which connect thermodynamic quantities to kinematic quantities of the disk. They can be written in integral form as

\begin{align}
    \ln|u_t| - \ln|(u_t)_{in}| - \int_{\ell_{in}}^\ell \frac{\Omega(\ell')}{1- \Omega(\ell') \ell'}d\ell' = - \int_{p_{in}}^p \frac{1}{\rho(p') h(p')} dp',
    \label{eq:euler_int}
\end{align}

where $u_t$ is the mass-normalized energy, $\ell$ is the specific angular momentum, $\Omega$ is the angular velocity, $p$ is the pressure, $\rho$ and $h$ are the rest mass density and specific enthalpy and where the subscript \textit{in} is referring to the inner edge of the disk. Assuming a constant specific angular momentum distribution of the disk, the integral term over the specific angular momentum vanishes and with the definition of an effective potential $\mathcal{W} = \ln|u_t|$, $\mathcal{W}_{in} = \ln|(u_t)_{in}|$ the eq. (\ref{eq:euler_int}) simplifies to

\begin{align}
    \mathcal{W} - \mathcal{W}_{in} =  - \int_{p_{in}}^p \frac{1}{\rho(p') h(p')} dp' = - \int_{h_{in}}^h \frac{1}{h'} dh' = -\ln h,
    \label{eq:euler}
\end{align}

since $p_{in} = h_{in} = 0$. The effective potential at the inner edge of the disk, $\mathcal{W}_{in}$, is a specifiable parameter, which is lower bounded by the effective potential at the disk center, and upper bounded by zero, since an effective potential of zero corresponds to marginally bound orbits. The effective potential $\mathcal{W}$ depends only on the metric components and on the specific angular momentum $\ell_0$ of the disk, which is a free parameter since we assume a uniform specific angular momentum distribution,

\begin{align}
    \mathcal{W} = \ln|u_t| = \frac{1}{2} \ln \left( \frac{g_{t\varphi}^2 - g_{tt}g_{\varphi\varphi}}{\ell^2 g_{tt} + 2 \ell g_{t\varphi} + g_{\varphi \varphi}}\right).
\end{align}

Since the specific enthalpy of the disk is fully determined by eq. (\ref{eq:euler}) through the effective potential, also the rest mass density and pressure are fully determined by the effective potential, since they can be written as a function of the specific enthalpy and therefore as a function of $\mathcal{W}$. As a consequence the equi-potential surfaces of the disk coincide with the equi-pressure and equi-density surfaces. The effective potential $\mathcal{W}$ visualizes therefore the disk morphology and is a suitable quantity for the analyses of disk solutions. In the Polish Doughnut model the disk properties are thus fully determined by the metric components, the specific angular momentum of the disk $\ell_0$ and the effective potential at the inner edge of the disk $\mathcal{W}_{in}$. Depending on the chosen values for $\ell_0$ and $\mathcal{W}_{in}$, disk solutions can vary highly for the same spacetime. The effective potential at the center of the disk, $W_c$, is determined by the specific angular momentum of the disk $\ell_0$. At the disk center the acting forces of the fluid particles cancel each other out and the motion becomes geodesic. The specific angular momentum at the disk center corresponds therefore to the specific angular momentum of a Keplerian circular orbit, $\ell_K^\pm$. The Keplerian specific angular momentum of the spacetime determines thus the possible types of disk solutions. The intersection of the disk momentum $\ell_0$ with $\ell_K$ marks the center location $r_c$ of the disk. If $\ell_0 \neq \ell_K(r) \forall r$ there is no disk solution. If $\ell_K$ is non-monotonic, there could be more intersections of $\ell_0$ with $\ell_K$. The slope of $\ell_K$ determines in this case the disk properties: If $\partial_r \ell_K^\pm > 0$, then the intersection with $\ell_0$ marks a center for prograde motion and a cusp for retrograde motion, if $\partial_r \ell_K^\pm < 0$ the intersection marks a center for retrograde motion and a cusp for prograde motion. At the cusp, the geodesic motion is not stable and the equi-potential surface of the disk intersects itself. Since the equi-potential surface is open towards the central object at the cusp, the unstable motion of the fluid particles could lead to an accretion process apart from any inner-physical effects of the accretion disk. The value of $\ell_0$ determines therefore the type of disk and the distribution of $\ell_K$ determines the possible types of disk solutions.


\section{Matter Accumulations at Wormhole Throats}

The existence of circular orbits at the throat could lead to matter accumulations, which form an equilibrium state at the throat. For symmetrical wormholes the pressure gradient vanishes exactly at the junction of the upper and lower side of the wormhole, as the acting forces of the inflowing matter cancel each other out. As a consequence a central bright region could be present for a wormhole spacetime, where the emission is originating from a accretion structure enveloping the wormhole throat. In \cite{ThroatOrbits} it was shown, that not only one, but a whole spectrum of circular orbits in the parameter space could exist at the wormhole throat, depending on the properties of the metric components at the throat. We will apply this analysis of circular orbits at the throat here to the selected wormhole solutions and expand it by including the Thick Disk model, which infers the existence of circular orbits to mass accumulations at the throat. For the analysis, we apply the parameter space boundaries given in \cite{ThroatOrbits} explicitly for the given wormhole solutions and evaluate the resulting parameter spectrum of solutions. The boundaries of existence depend on the sign of $g_{tt}$ at the throat, which corresponds to the absence or presence of an ergoregion around the wormhole throat. If there is no ergoregion present, the boundaries of existence for bound circular orbits are given by $\ell_0 \in [\ell_B^+,\ell_B^-]$, with $\ell_0$ being the specific angular momentum of a test particle at the throat, and where $\ell_B^\pm$ are given by

\begin{align}
    \ell_B^\pm = \frac{-g_{t \varphi} \pm \sqrt{(g_{t \varphi}^2 - g_{tt} g_{\varphi \varphi})(1 + g_{tt})}}{g_{tt}} \Bigg|_{l =0}.
\end{align}

If an ergoregion is present, the parameter space for bound circular orbits is given by $\ell_0 \in ((-\infty,\ell_B^-] \ \vee \ [\ell_B^+,\infty))$. The stability of the orbits is bounded by the limits $\ell_S^\pm$,

\begin{align}
    \ell_S^\pm = \frac{D \partial_l^2 g_{t \varphi} - g_{t \varphi}\partial_l^2 D \pm \sqrt{(g_{t \varphi} \partial_l^2 D - D \partial_l^2 g_{t \varphi})^2 - (g_{tt} \partial_l^2 D - D \partial_l^2 g_{tt})(g_{\varphi \varphi} \partial_l^2 D - D \partial_l^2 g_{\varphi \varphi})}}{g_{tt} \partial_l^2 D - D \partial_l^2 g_{tt}} \Bigg|_{l = 0},
\end{align}

where $D = g_{t\varphi}^2 - g_{tt} g_{\varphi \varphi}$. If the expression $g_{tt} \partial_l^2 D - D \partial_l^2 g_{tt}$ is positive at the throat, all possible orbits are stable within the interval $\ell_0 \in [\ell_S^+, \ell_S^-]$. If the expression is negative, all possible orbits are stable for $\ell_0 \in ((-\infty,\ell_S^-] \ \vee \ [\ell_S^+,\infty))$. 

\subsection{$\mathcal{TE}$ Wormhole}

\begin{figure}[H]
\centering
\begin{subfigure}{.325\textwidth}
  \centering
  \includegraphics[width=\linewidth,  height= 0.75\textwidth]{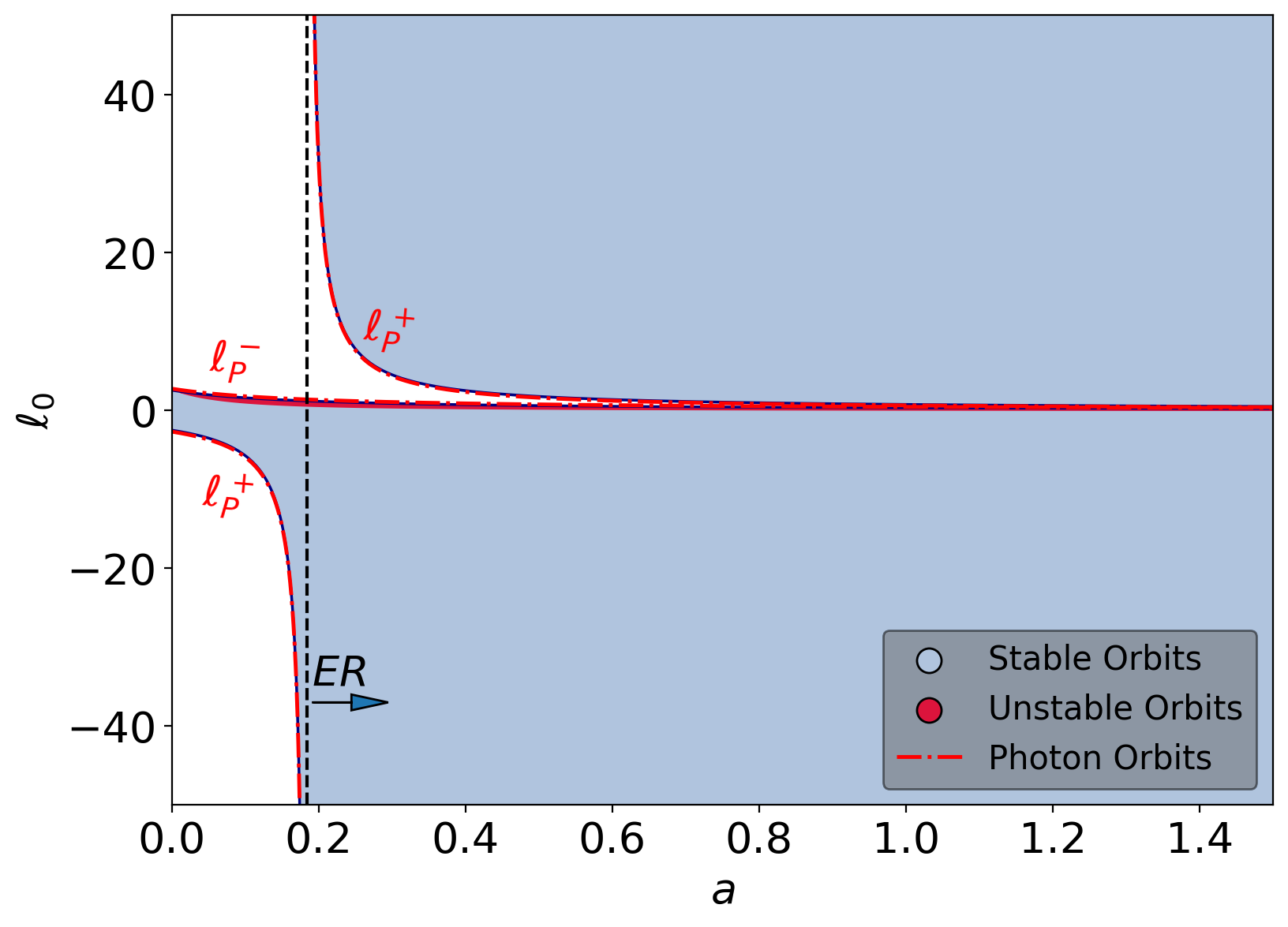}
  \caption{$\mathcal{TE}$: Parameter space for throat orbits}
\end{subfigure}
\begin{subfigure}{.325\textwidth}
  \centering
  \includegraphics[width=\linewidth,  height= 0.75\textwidth]{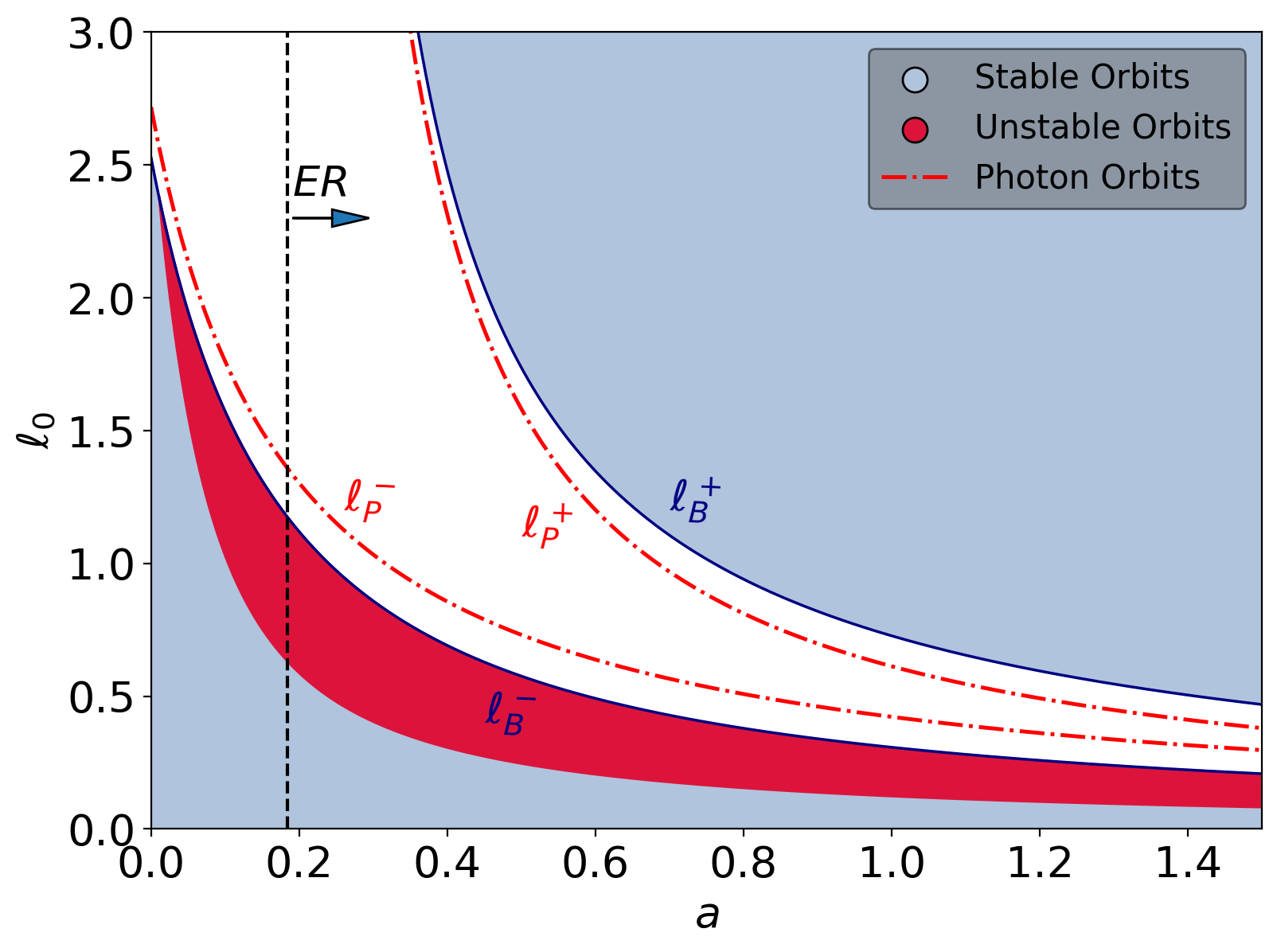}
  \caption{$\mathcal{TE}$: Close up on unstable region}
\end{subfigure}
\begin{subfigure}{.325\textwidth}
  \centering
  \includegraphics[width=\linewidth,  height= 0.75\textwidth]{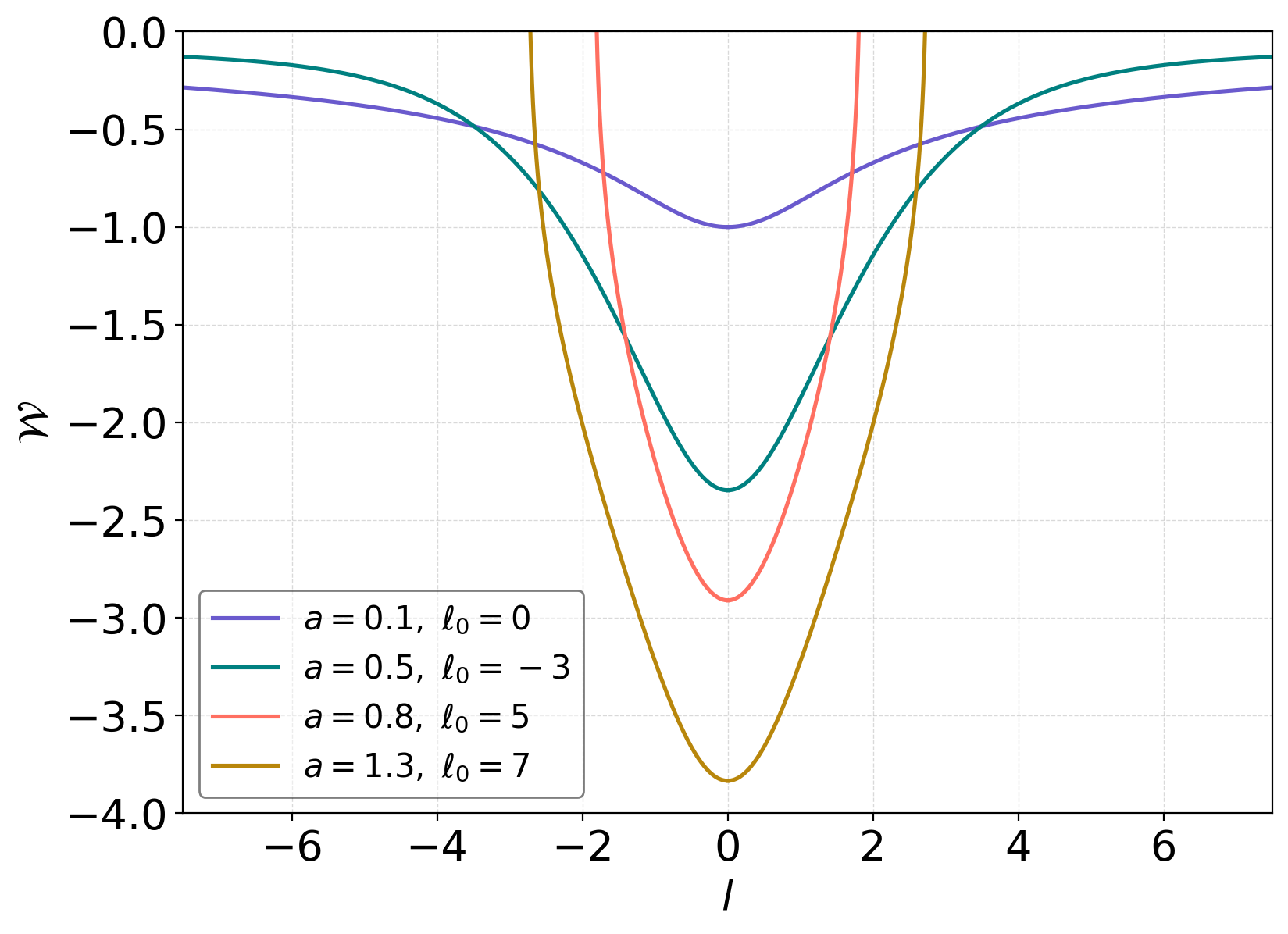}
  \caption{$\mathcal{TE}$: Exemplary disk solutions}
\end{subfigure}
\caption{(a) and (b) showcase the parameter space with regard to the spin parameter $a$ and the specific angular momentum $\ell_0$ of a test particle for circular orbits at the throat. The blue colored area showcases the region for which stable orbits exist, the red colored region showcases the region for which the orbits are unstable. The thin blue curves represent the boundaries of existence for bound orbits, $\ell_B^\pm$, and the red dotted dashed curves showcase the impact parameters for the photon orbits located at the throat, $\ell_P^\pm$. The dashed vertical line and the abbreviation \textit{ER} indicate the onset of an ergoregion surrounding the throat, for all values of $a$ right of the line. (c) presents the effective potential $\mathcal{W}$ in the equatorial plane for three exemplary disk solutions. The left half of the plot illustrates the lower side of the wormhole, the right half illustrates the upper side of the wormhole with the throat at $l = 0$.}
\label{fig:TE_phase}
\end{figure}

For the $\mathcal{TE}$ wormhole solutions the boundaries of existence for bound orbits at the throat are a function of the spin parameter $a$. They are approximately given by,

\begin{align}
    \ell_B^\pm \approx \frac{1.472a^2 + 2a \pm 0.318}{4a^2 - 0.135},
\end{align}

where the root of the denominator, $a = 0.184 \coloneqq a_E$, marks the onset of wormhole solutions with an ergoregion. The parameter space for orbits at the throat undergoes a phase transition at this critical value of $a$, which is depicted in Fig. \ref{fig:TE_phase}.

Up to the critical value $a_E$ the parameter space is confined between the boundaries $\ell_B^\pm$, from $a_E$ onwards the parameter space is located outside the interval between $\ell_B^\pm$. For increasing $a$ this interval shrinks and the interval size converges asymptotically to zero. As a consequence the parameter space for bound orbits at the throat approaches asymptotically $\ell_0 \in (-\infty,\infty)$. A small region of unstable orbits exists, which only extends to positive values of $\ell_0$ and also shrinks with increasing $a$. The subfigure \ref{fig:TE_phase} (c) showcases exemplary disk solutions for selected points of the parameter space. These disk solutions represent equilibrium states of matter accumulations around the throat of the wormhole. Matter flows in from both sides of the throat, leading to a vanishing pressure gradient and thus to a balance of forces at the wormhole throat. The following Fig. \ref{fig:TE_rho} showcases exemplary cross section plots of the density distribution of such throat disks.

\begin{figure}[H]
\centering
\begin{subfigure}{.325\textwidth}
  \centering
  \includegraphics[width=\linewidth,  height= 0.85\textwidth]{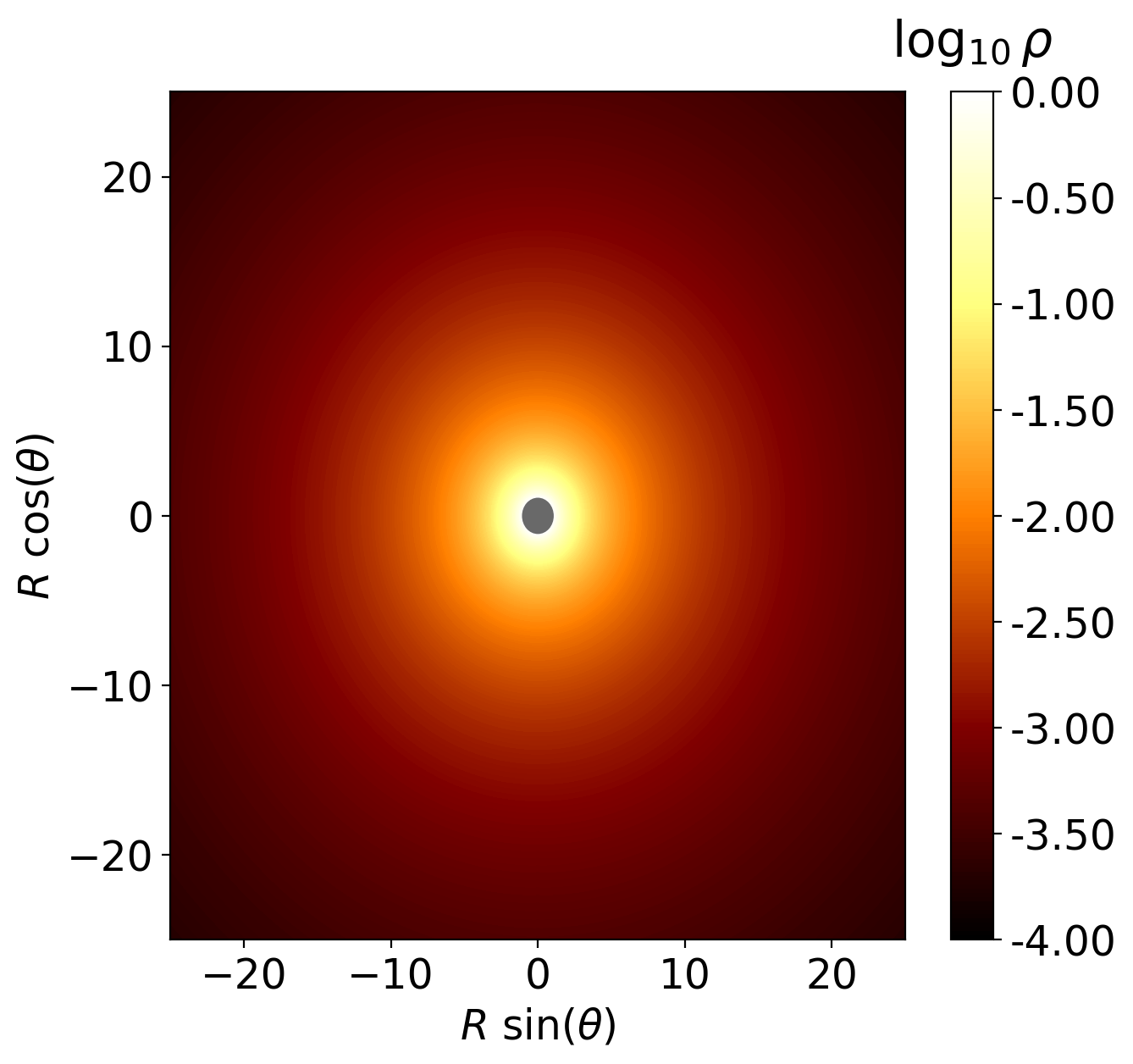}
  \caption{$\mathcal{TE}$: Spherical disk solution with $\ell_0 = 0$}
\end{subfigure}
\begin{subfigure}{.325\textwidth}
  \centering
  \includegraphics[width=\linewidth,  height= 0.85\textwidth]{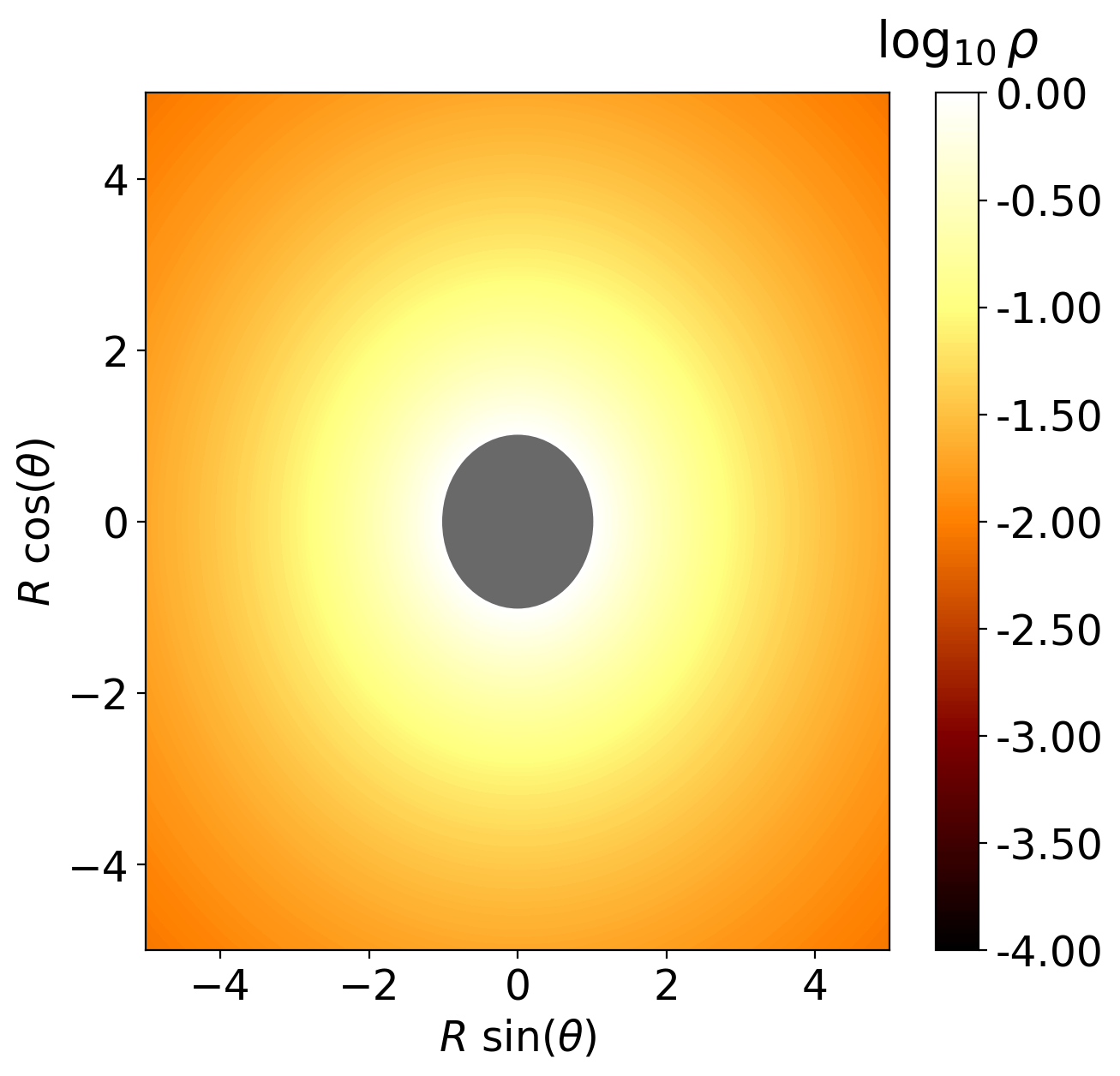}
  \caption{$\mathcal{TE}$: Close up on spherical disk solution}
\end{subfigure}
\begin{subfigure}{.325\textwidth}
  \centering
  \includegraphics[width=\linewidth,  height= 0.85\textwidth]{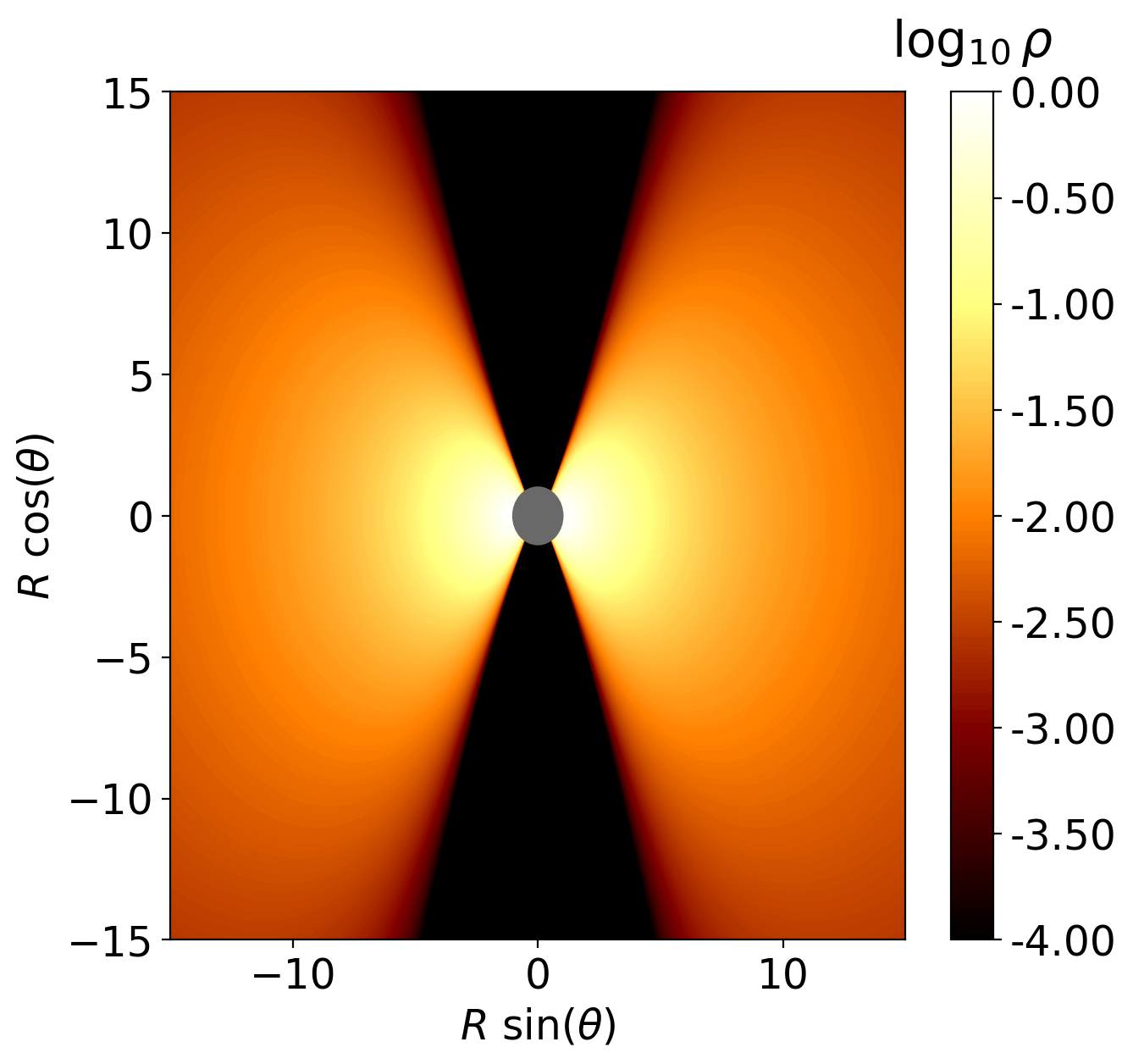}
  \caption{$\mathcal{TE}$: Disk solution for $a = 0.1$ and $\ell_0 =1$}
\end{subfigure}
\caption{Meridional cross section of the density distribution for selected disk solutions. $(a)$ and $(b)$ showcase a cross section of the density distribution for $\ell_0 = 0$ and arbitrary $a$. $(c)$ showcases a disk solutions for the selected point $a = 0.1$ and $\ell_0 = 1$ of the parameter space. The radial coordinate here is given by $R = \sqrt{l^2 + r_0^2}$, where $l$ is the global radial coordinate and $r_0$ designates the location of the throat, with $r_0 = 1$ for this case. The grey circle represents thus the wormhole throat.} 
\label{fig:TE_rho}
\end{figure}

In case of the spherical disk solutions the specific angular momentum of the disk vanishes. As a consequence all centrifugal terms in the effective potential vanish and only terms remain which solely depend on the radial coordinate. The resulting disk solutions correspond to spherical structures in which a hydrostatic equilibrium is established at the throat between gravitational attraction and pressure. Thus, they do not represent a typical accretion disk, but more a star-like structure when it comes to their observational signature and possible radiation profile. For a non-vanishing specific angular momentum, the disk structures enclose the wormhole throat with matter-less areas near the poles. It represents an open torus, which encapsulates the throat, with a high degree of spherical symmetry. Thus, we conclude that these matter accumulations around the throat in general may appear more star-like due to their central bright region.

\subsection{$\mathcal{RSV}$ Wormholes}

In case of the $\mathcal{RSV}$ wormholes the boundaries of existence for bound orbits are functions of the two parameters of the spacetime, namely the spin parameter $a$ and the regularization parameter $\xi$, 

\begin{align}
     \ell_B^\pm(a,\xi) = \frac{2 \left(a \pm \sqrt{a^{2} + \xi^{2} - 2 \xi}\right)}{2- \xi }.
\end{align}

We analyze the behavior of the parameter space for fixed values of $\xi$ and varying $a$, as well as in a three-dimensional presentation where both $a$ and $\xi$ are varied. For fixed values of $\xi$ we picked $\xi \in \{1.8,2.1,2.5\}$. The $\xi = 1.8$ solutions represent wormholes with an ergoregion, however wormhole solutions only exist for $a > 0.6$, as solutions with a smaller spin parameter represent black holes. The following Fig. \ref{fig:RSV_xi1.8_phase} showcases phase diagrams and exemplary disk solutions for $\xi = 1.8$.

\begin{figure}[H]
\centering
\begin{subfigure}{.325\textwidth}
  \centering
  \includegraphics[width=\linewidth,  height= 0.75\textwidth]{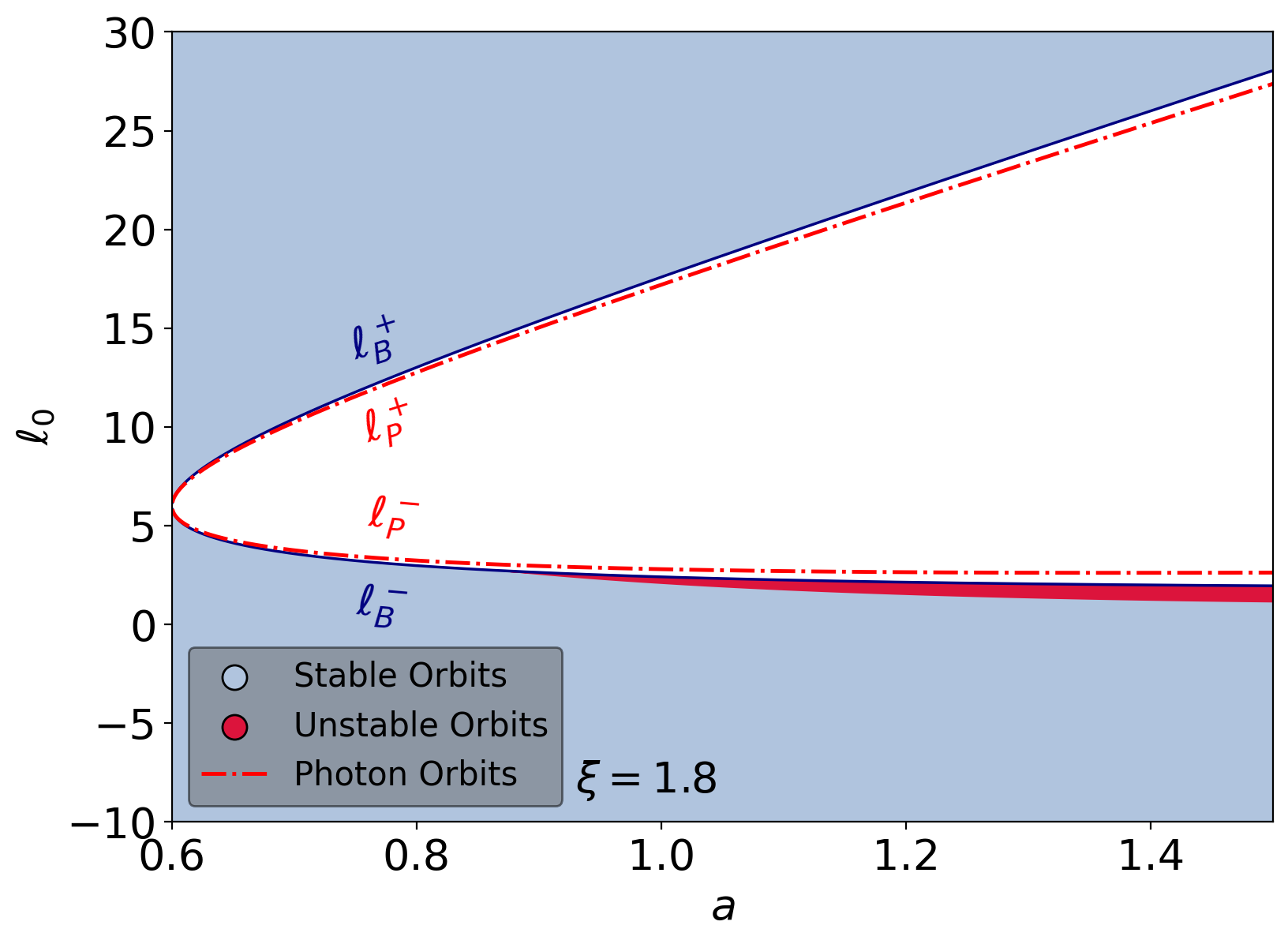}
  \caption{$\mathcal{RSV}$: Parameter space for throat orbits}
\end{subfigure}
\begin{subfigure}{.325\textwidth}
  \centering
  \includegraphics[width=\linewidth,  height= 0.75\textwidth]{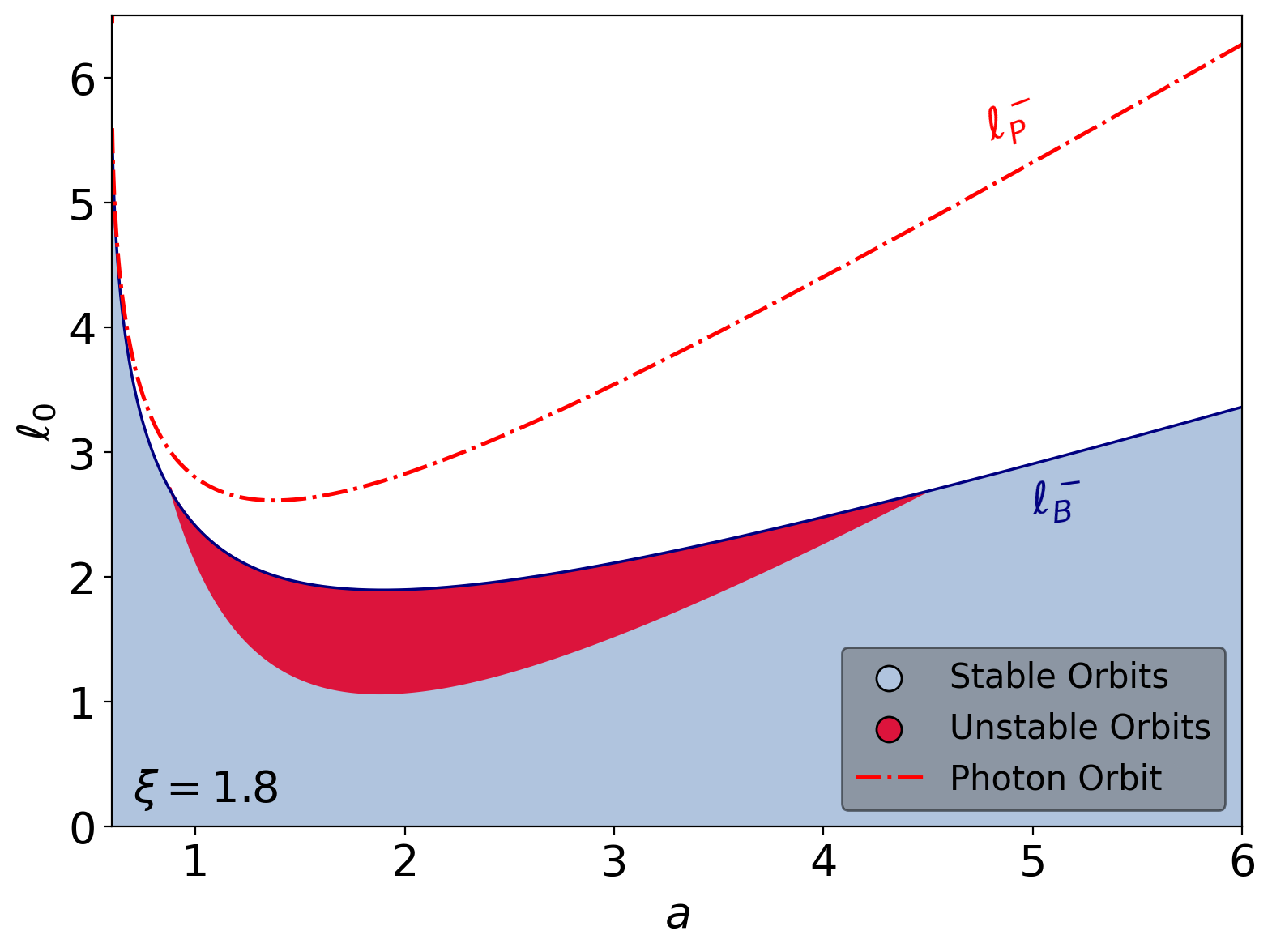}
  \caption{$\mathcal{RSV}$: Close up on unstable region}
\end{subfigure}
\begin{subfigure}{.325\textwidth}
  \centering
  \includegraphics[width=\linewidth,  height= 0.75\textwidth]{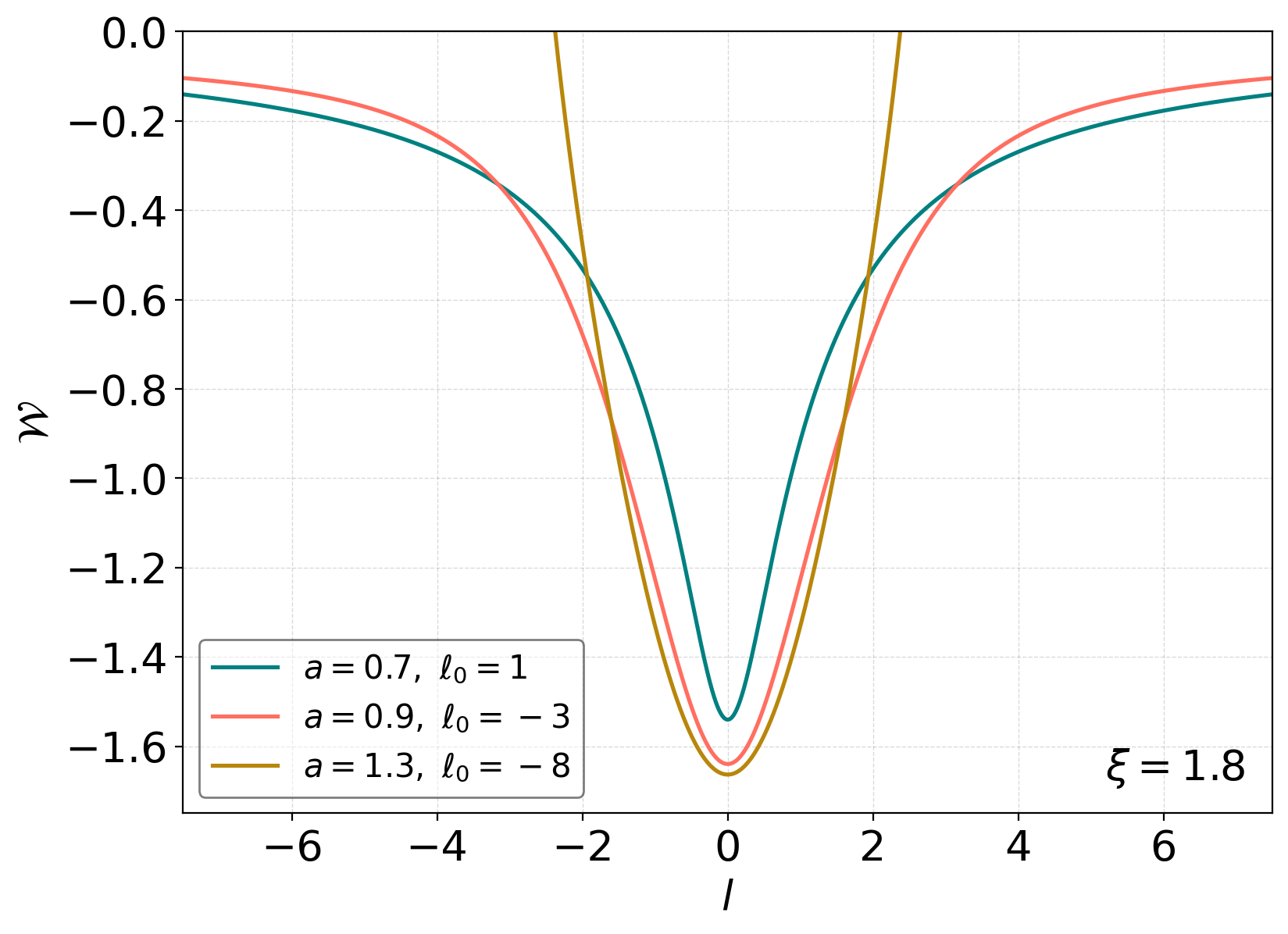}
  \caption{$\mathcal{RSV}$: Exemplary disk solutions}
\end{subfigure}
\caption{$\xi = 1.8$: (a) and (b) showcase the parameter space with regard to the spin parameter $a$ and the specific angular momentum $\ell_0$ of a test particle for circular orbits at the throat. The blue colored area showcases the region for which stable orbits exist, the red colored region showcases the region for which the orbits are unstable. The thin blue curves represent the boundaries of existence for bound orbits, $\ell_B^\pm$, and the red dotted dashed curves showcase the impact parameters for the photon orbits located at the throat, $\ell_P^\pm$. (c) showcases the effective potential $\mathcal{W}$ in the equatorial plane for three exemplary solutions. The left half of the plot illustrates the lower side of the wormhole, the right half illustrates the upper side of the wormhole with the throat at $l = 0$.}
\label{fig:RSV_xi1.8_phase}
\end{figure}

In case of $\xi = 1.8$ the wormhole has an ergoregion, which surrounds the wormhole throat. As a consequence the parameter space for bound circular orbits at the throat extends from $(-\infty,\ell_B^-]$ and $[\ell_B^+,\infty)$. With increasing rotation the parameter space shrinks, since the interval between the boundaries $\ell_B^\pm$ grows. A narrow region of unstable orbits exists within the interval $a \in (0.877, 4.483)$. However, this region only extends to positive values of $\ell_0$, thus all retrograde orbits $(\ell_0 < 0)$ are stable orbits. The existence of such a spectrum for bound orbits, which extends to infinity, may hint to instabilities of the spacetime, as the possibilities for the formation of matter structures around the wormhole throat scale with it. The following Fig. \ref{fig:RSV_xi2.1_phase} showcases the parameter space for $\xi = 2.1$.

\begin{figure}[H]
\centering
\begin{subfigure}{.325\textwidth}
  \centering
  \includegraphics[width=\linewidth,  height= 0.75\textwidth]{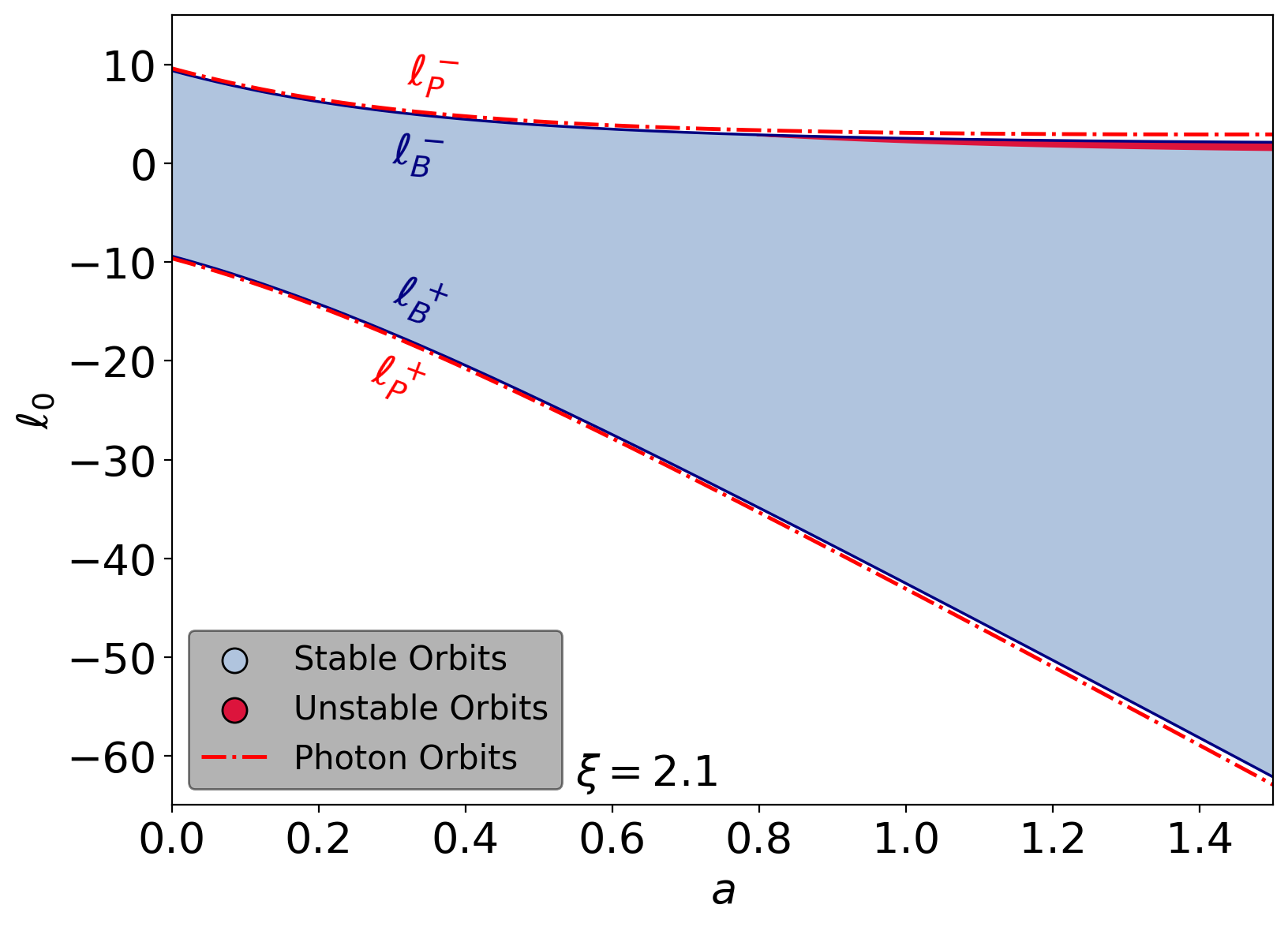}
  \caption{$\mathcal{RSV}$: Parameter space for throat orbits}
\end{subfigure}
\begin{subfigure}{.325\textwidth}
  \centering
  \includegraphics[width=\linewidth,  height= 0.75\textwidth]{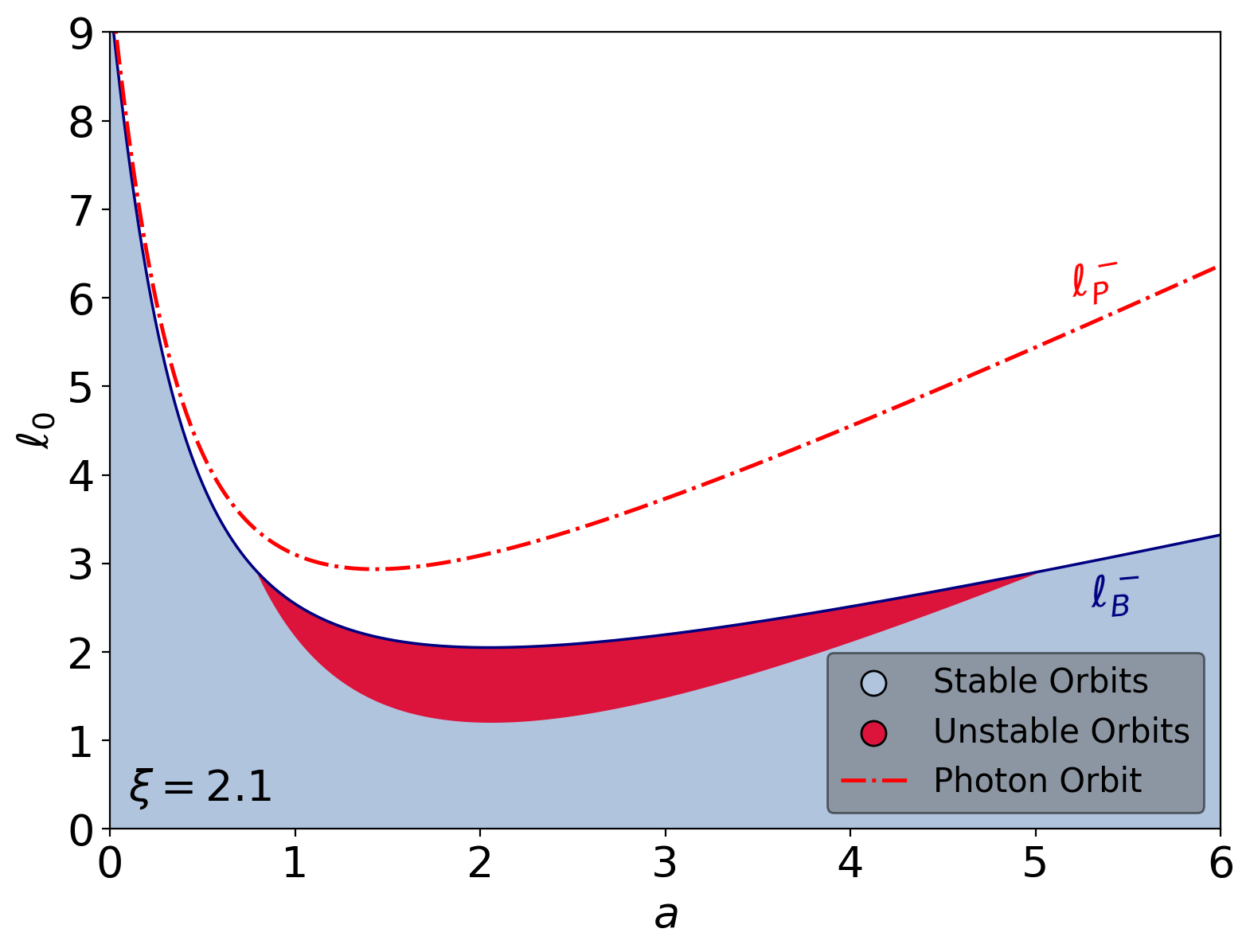}
  \caption{$\mathcal{RSV}$: Close up on unstable region}
\end{subfigure}
\begin{subfigure}{.325\textwidth}
  \centering
  \includegraphics[width=\linewidth,  height= 0.75\textwidth]{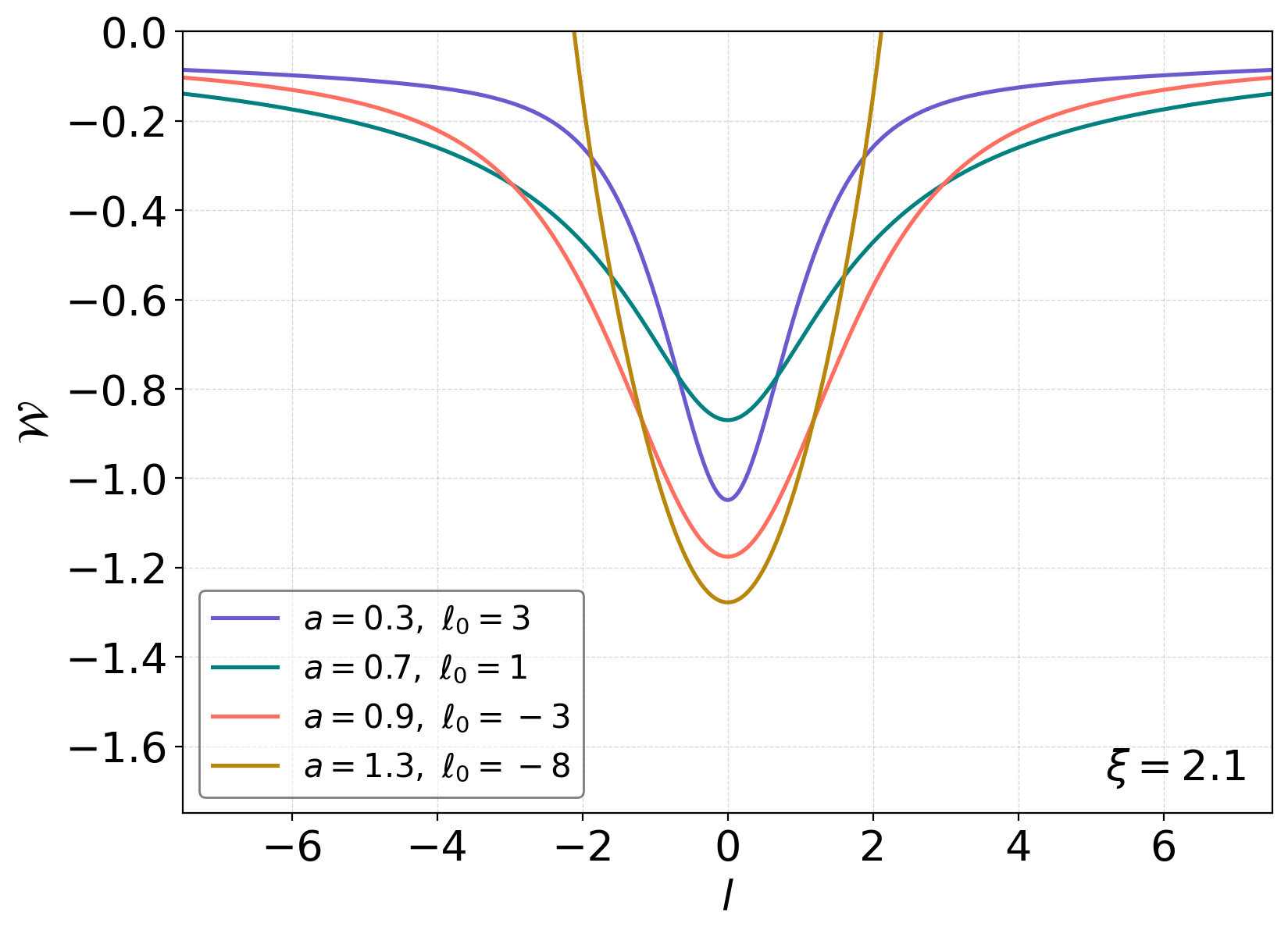}
  \caption{$\mathcal{RSV}$: Exemplary disk solutions}
\end{subfigure}
\caption{$\xi = 2.1$: (a) and (b) showcase the parameter space with regard to the spin parameter $a$ and the specific angular momentum $\ell_0$ of a test particle for circular orbits at the throat. The blue colored area showcases the region for which stable orbits exist, the red colored region showcases the region for which the orbits are unstable. The thin blue curves represent the boundaries of existence for bound, $\ell_B^\pm$, and the red dotted dashed curves showcase the impact parameters for the photon orbits located at the throat, $\ell_P^\pm$. (c) showcases the effective potential $\mathcal{W}$ in the equatorial plane for three exemplary solutions. The left half of the plot illustrates the lower side of the wormhole, the right half illustrates the upper side of the wormhole.}
\label{fig:RSV_xi2.1_phase}
\end{figure}

For the $\xi = 2.1$ wormhole the parameter space for bound orbits is confined between the boundaries $\ell_B^+$ and $\ell_B^-$. Almost the whole parameter space for bound orbits also consists of stable orbits, only a small region for positive specific angular momenta corresponds to unstable orbits. Unstable orbits exist within the interval $a \in (0.796,5)$. An increasing rotation corresponds to a growing of the parameter space, which in turn increases the possible configurations for matter accumulations at the wormhole throat. Since the course of the lower limit $\ell_B^+$ is steeper with respect to $a$, the parameter space shifts more and more to negative values for the specific angular momentum. This could increase the likelihood of counter-rotating disk structures around the wormhole throat. The next figure, Fig. \ref{fig:RSV_xi2.5_phase} showcases the phase space for $\xi = 2.5$.

\begin{figure}[H]
\centering
\begin{subfigure}{.325\textwidth}
  \centering
  \includegraphics[width=\linewidth,  height= 0.75\textwidth]{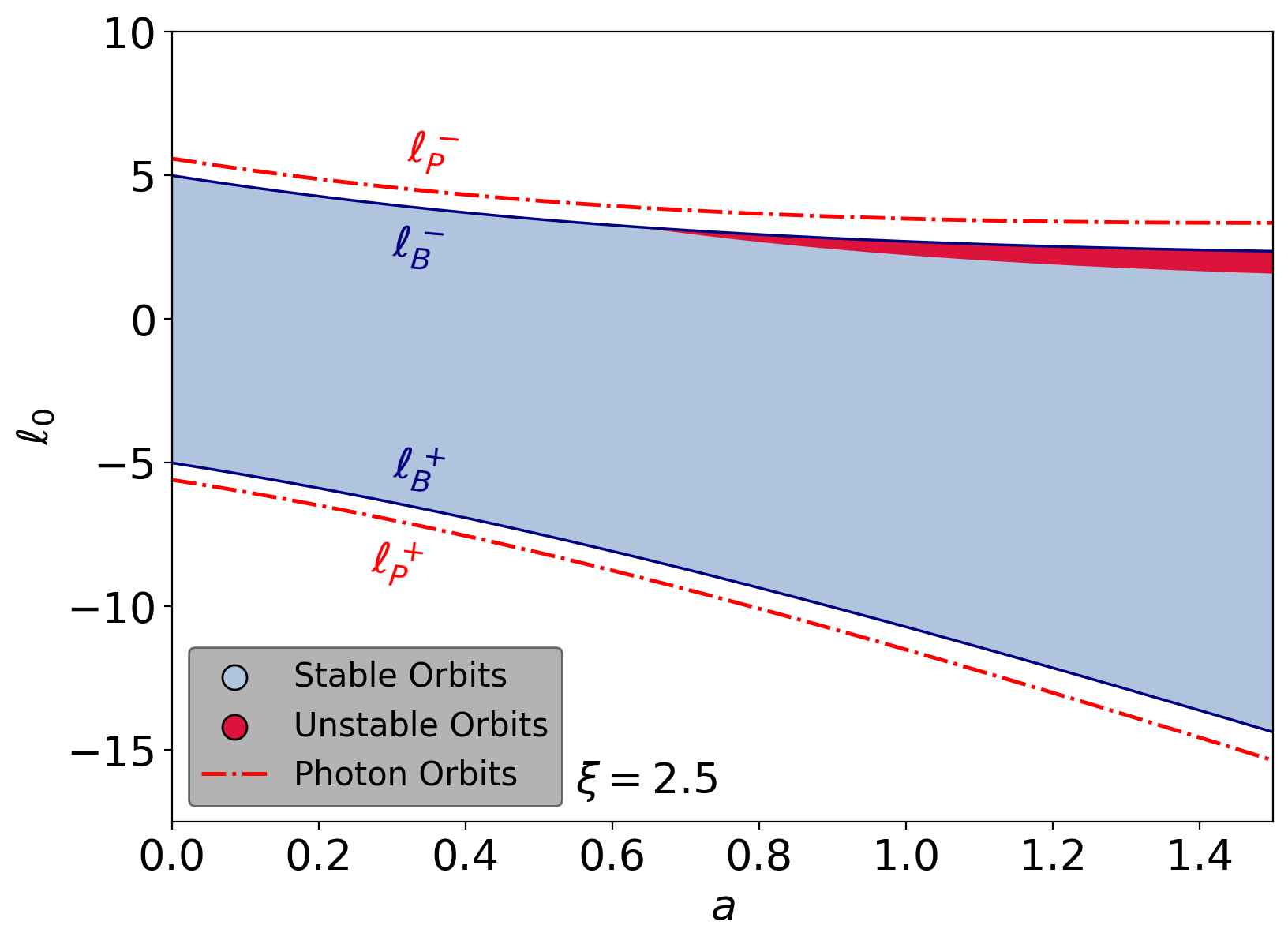}
  \caption{$\mathcal{RSV}$: Parameter space for throat orbits}
\end{subfigure}
\begin{subfigure}{.325\textwidth}
  \centering
  \includegraphics[width=\linewidth,  height= 0.75\textwidth]{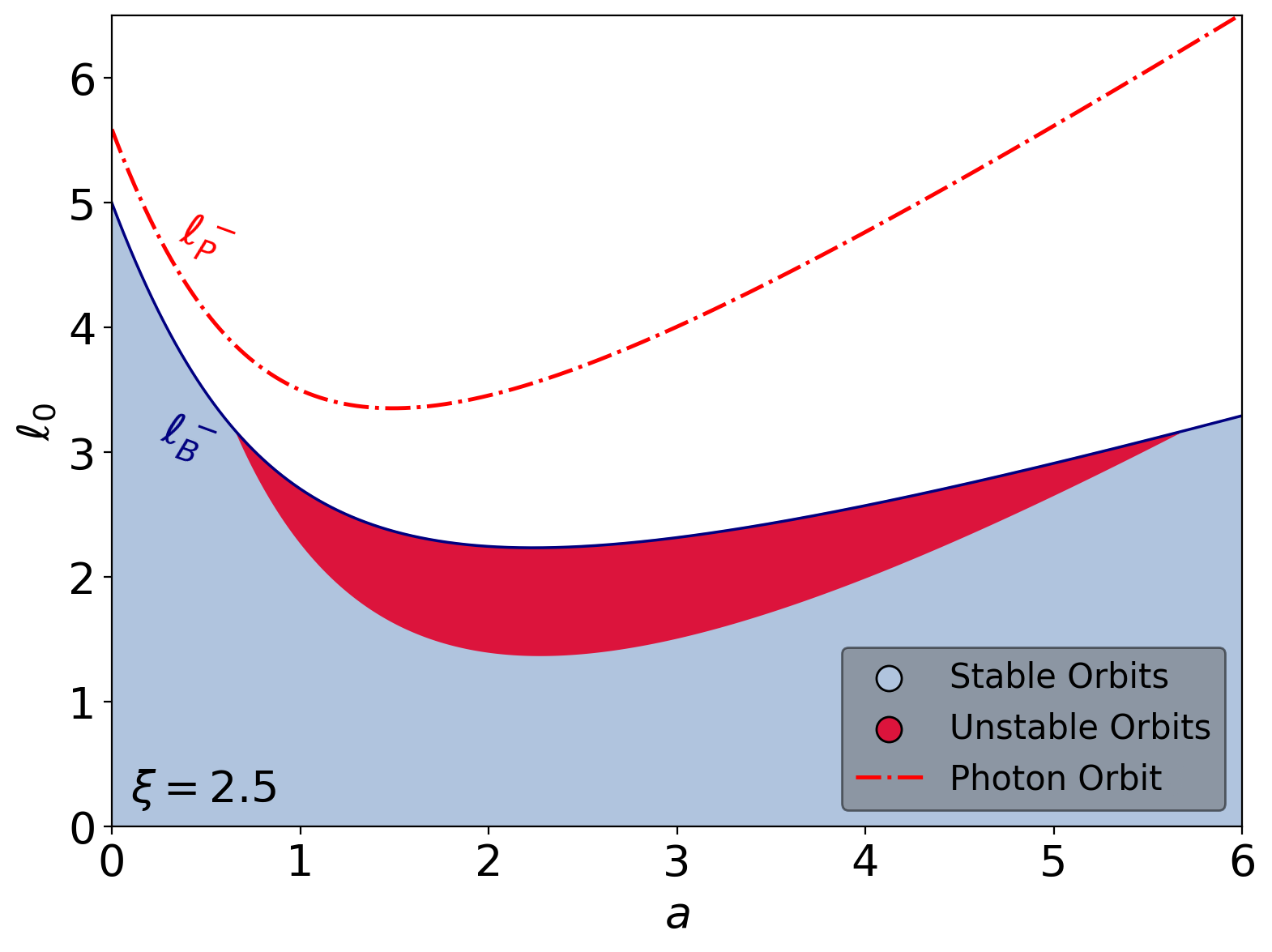}
  \caption{$\mathcal{RSV}$: Close up on unstable region}
\end{subfigure}
\begin{subfigure}{.325\textwidth}
  \centering
  \includegraphics[width=\linewidth,  height= 0.75\textwidth]{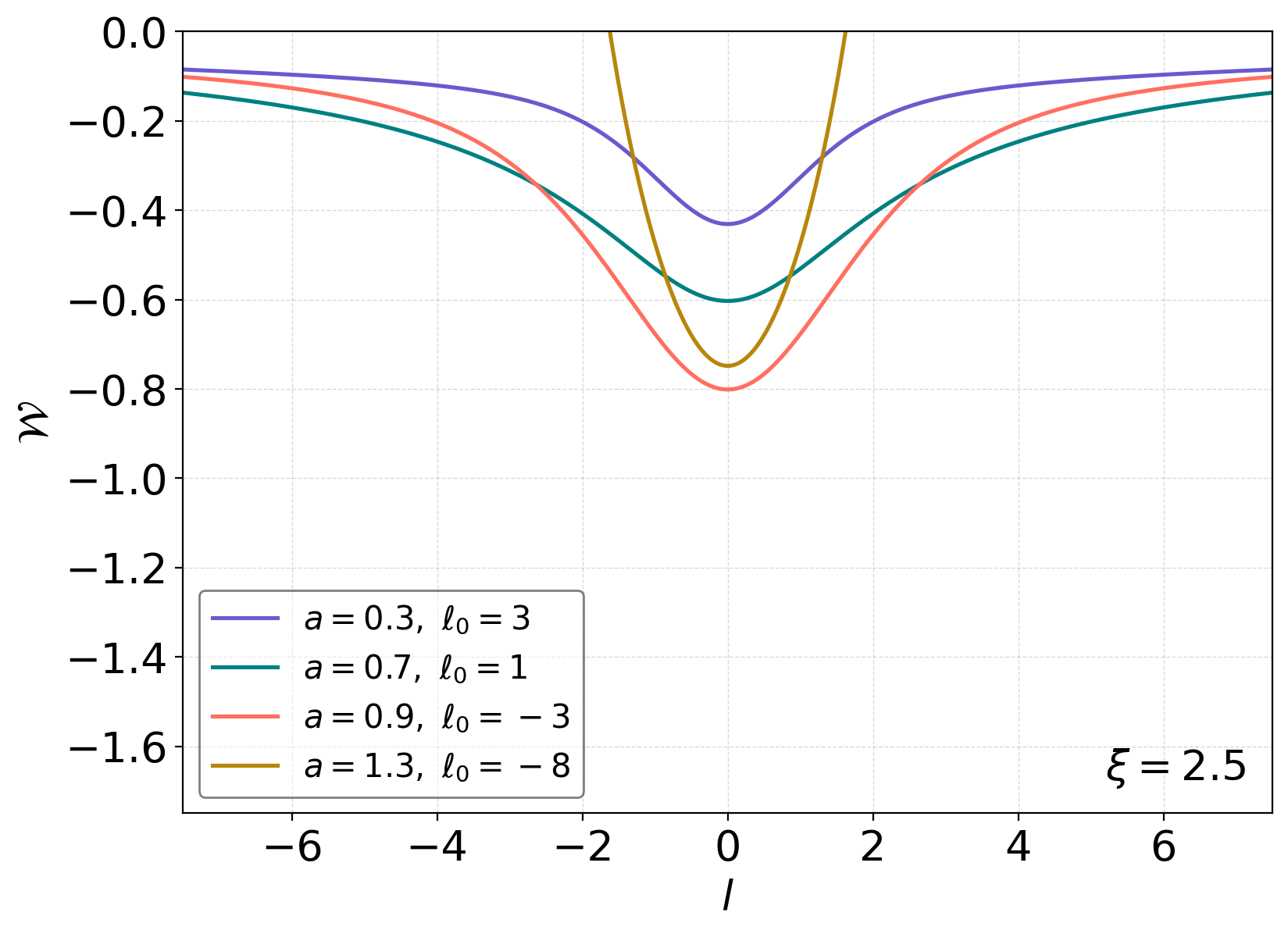}
  \caption{$\mathcal{RSV}$: Exemplary disk solutions}
\end{subfigure}
\caption{$\xi = 2.5$: (a) and (b) showcase the parameter space with regard to the spin parameter $a$ and the specific angular momentum $\ell_0$ of a test particle for circular orbits at the throat. The blue colored area showcases the region for which stable orbits exist, the red colored region showcases the region for which the orbits are unstable. The thin blue curves represent the boundaries of existence for bound orbits, $\ell_B^\pm$, and the red dotted dashed curves showcase the impact parameters for the photon orbits located at the throat, $\ell_P^\pm$. (c) showcases the effective potential $\mathcal{W}$ in the equatorial plane of three exemplary solutions. The left half of the plot illustrates the lower side of the wormhole, the right half illustrates the upper side of the wormhole.}
\label{fig:RSV_xi2.5_phase}
\end{figure}

For the $\xi = 2.5$ wormhole the parameter space is less affected by the spin parameter $a$ and furthermore it shrinks down compared to the $\xi = 2.1$ wormhole. The region of unstable orbits continues to be limited only to positive values of the specific angular momentum, however it extends now from $a \in (0.664,5.663)$. Compared to the $\xi = 1.8$ and $\xi = 2.1$ wormholes, the exemplary disk solutions for the same parameters are less energetically bound as the minima are less deep. We note, that an increase in $\xi$ decreases the influence of $a$ as well as the parameter spectrum for circular orbits at the throat. Moreover an increase in $\xi$ corresponds to a larger parameter region for unstable orbits as well as to less energetically bound disk configurations around the wormhole throat. From this we conclude, that the likelihood of stable matter accumulations around the wormhole throat is greater for smaller values of $\xi$, as the parameter space gives more possibilities for disk configurations, which are also stronger bound.

\begin{figure}[H]
\centering
\begin{subfigure}{.325\textwidth}
  \centering
  \includegraphics[width=\linewidth,  height= \textwidth]{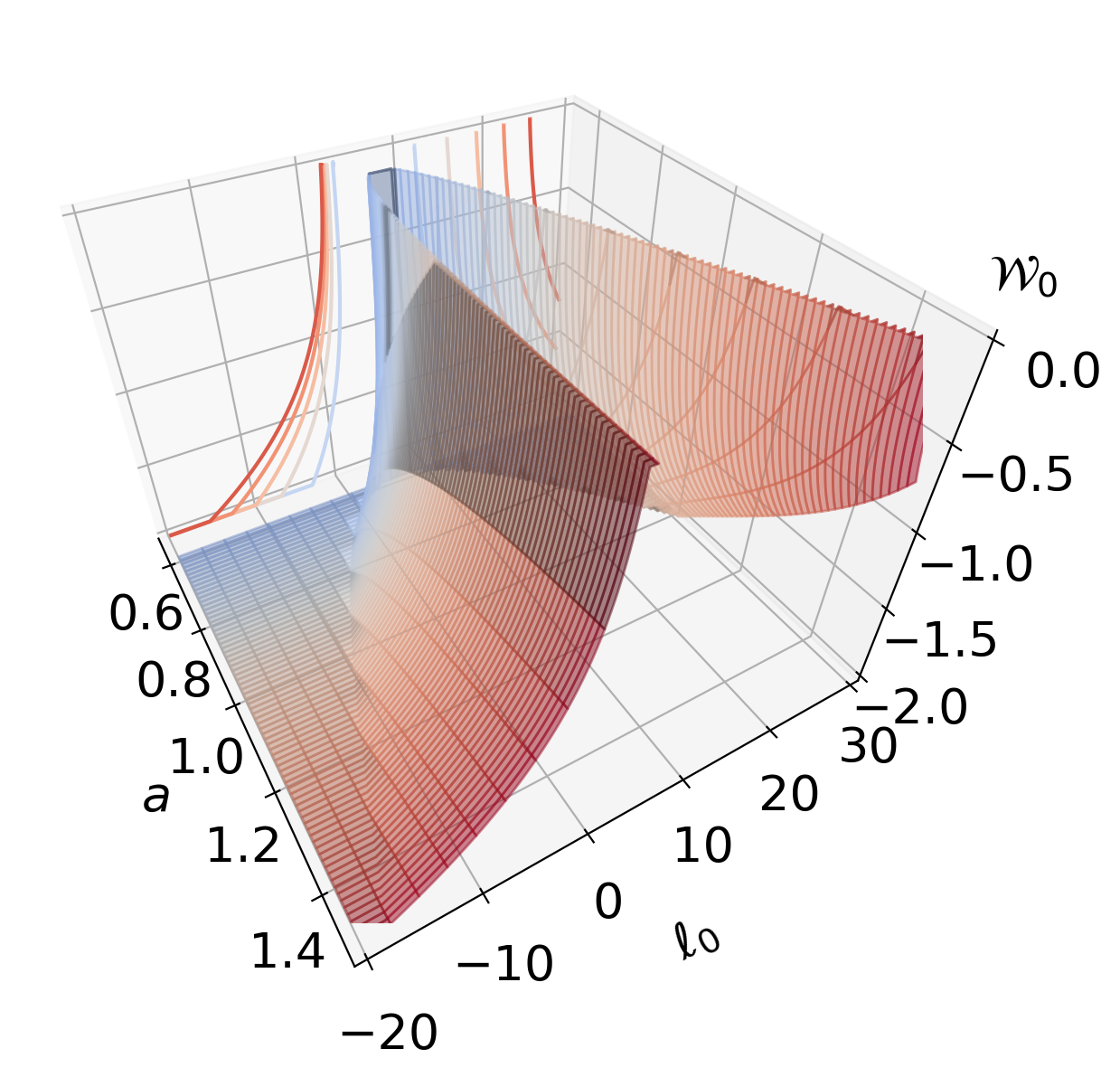}
  \caption{$\mathcal{RSV}$: $\xi = 1.8$}
\end{subfigure}
\begin{subfigure}{.325\textwidth}
  \centering
  \includegraphics[width=\linewidth,  height= \textwidth]{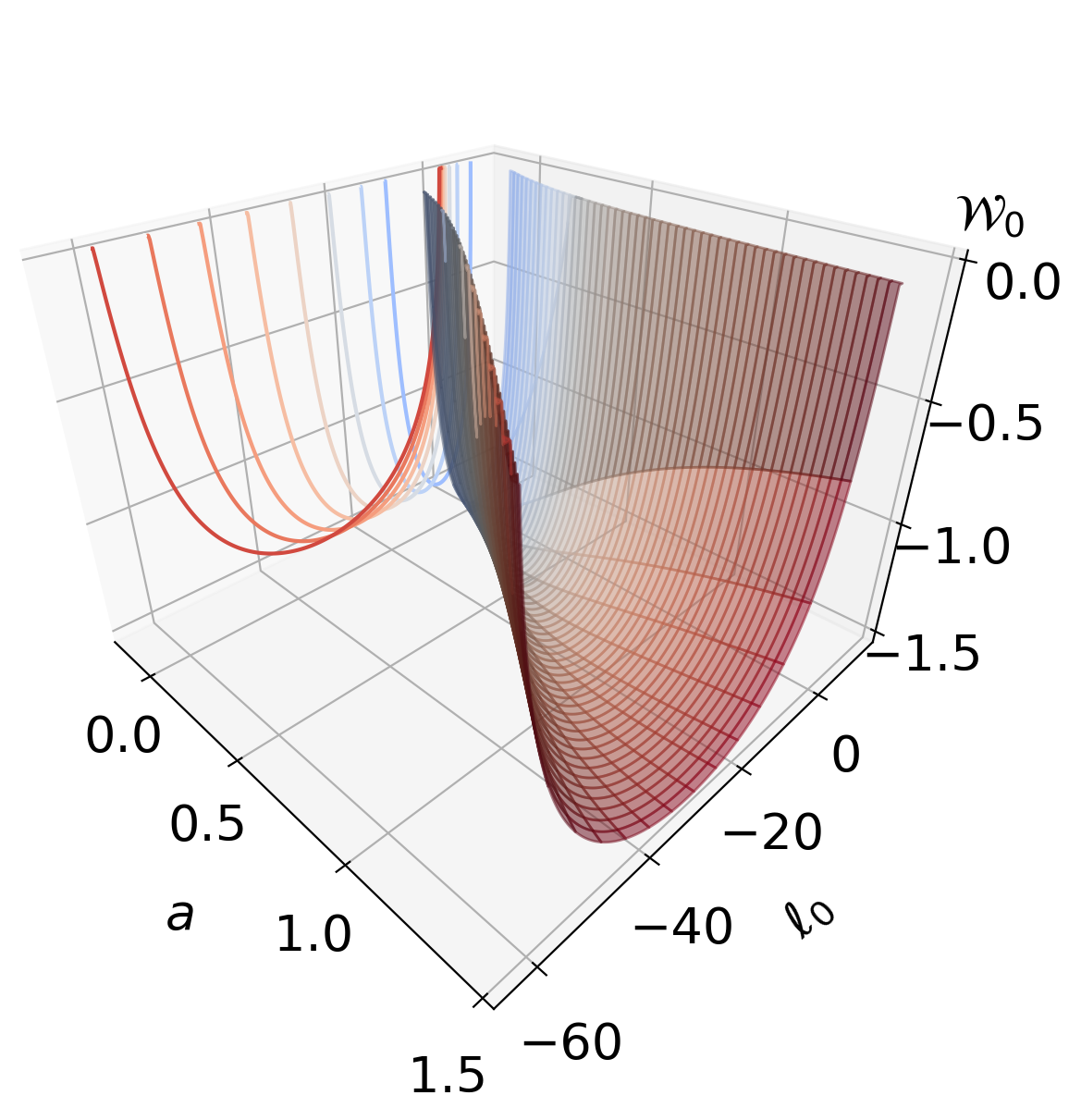}
  \caption{$\mathcal{RSV}$: $\xi = 2.1$}
\end{subfigure}
\begin{subfigure}{.325\textwidth}
  \centering
  \includegraphics[width=\linewidth,  height= \textwidth]{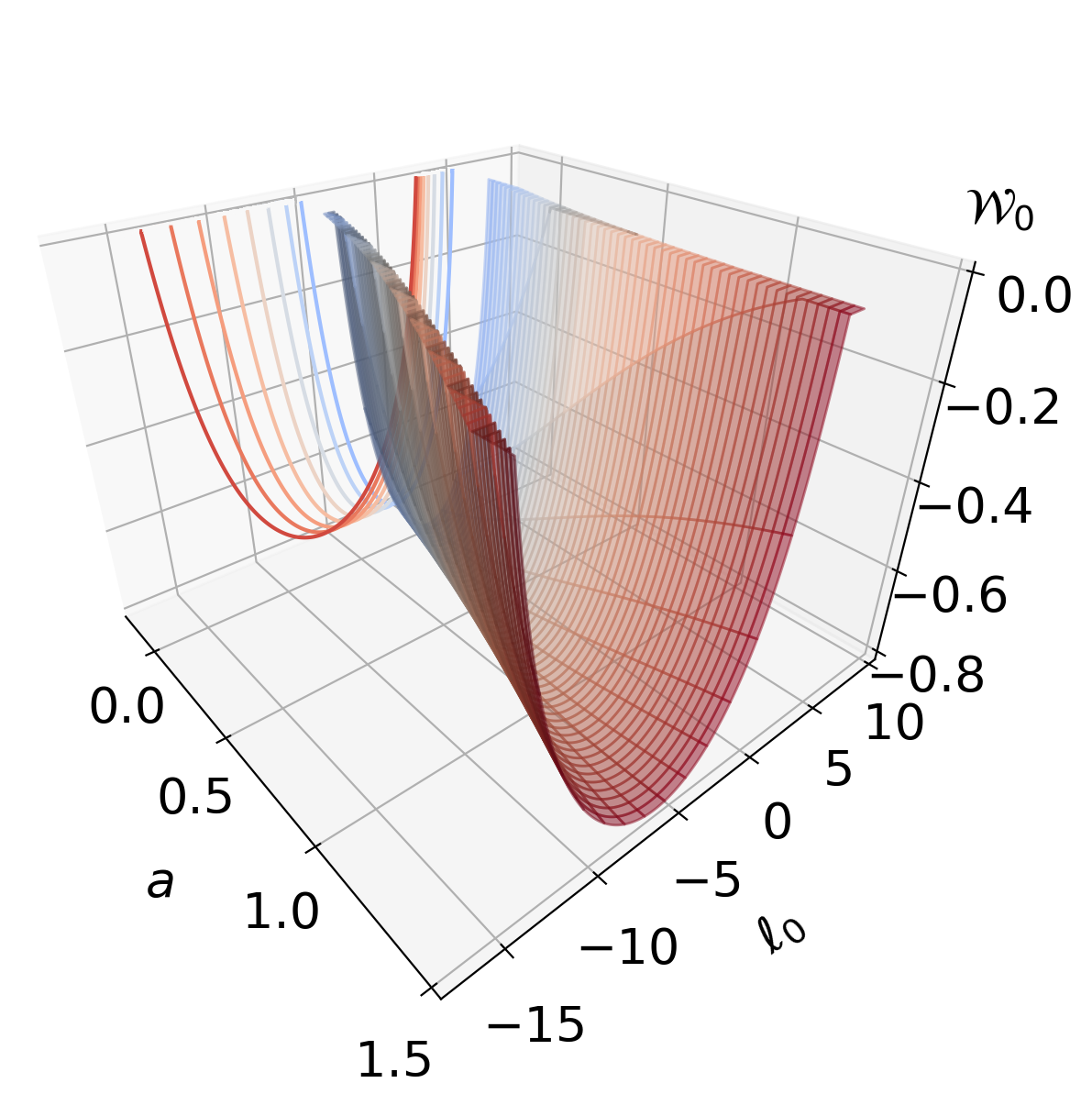}
  \caption{$\mathcal{RSV}$: $\xi = 2.5$}
\end{subfigure}
\caption{Contour map of the effective potential at the throat $\mathcal{W_0}$ for $\xi \in \{1.8, 2.1,2.5\}$ across the parameter space with regard to $a$ and $\ell_0$. The colors of the color map represent the value of the $a$, with blue tones corresponding to small values of $a$ and red tones to higher values of $a$. The curves in the background in the $\ell_0$-$\mathcal{W}_0$-plane are projections of the three-dimensional plot for fixed $a$, colored in the same way by their corresponding value of $a$. It should be noted, that in subfigure $(a)$ some regions reach deeper than the lower limit of the $\mathcal{W}_0$ axis, but in view of clearer depiction of the structure the lower limit of the axis was set to $\mathcal{W}_0 = -2$.}
\label{fig:RSV_3D_W_0}
\end{figure}

To illustrate this, the Fig. \ref{fig:RSV_3D_W_0} presents three-dimensional plots, showcasing the behavior of the minima of the effective potential, e. g. the values of the effective potential at the throat $\mathcal{W}_0$, for all possible disk configurations across the parameter space for the analyzed $\xi$ values.

If an ergoregion is present, disk configurations which are closer located to the boundaries for bound circular orbits in the parameter space have a smaller absolute value of the effective potential at the throat, representing therefore less energetically bound accumulations of matter at the throat. Far away from the boundaries disk configurations are more deeply bound. This drop-off is steeper for slower rotating wormholes. In case of the absence of an ergoregion, the effective potential is a quadratic function in the parameter space with regard to $\ell_0$, illustrated by the projected curves in the $\ell_0 - \mathcal{W}_0-$plane, which are parabolas. The effective potential at the wormhole throat thus is minimized exactly for the midpoint of the interval between $\ell_B^+$ and $\ell_B^-$, given by $\ell_0^{min} = \frac{\ell_B^+ + \ell_B^-}{2}$. Disk solutions closer to the center of the parameter space correspond therefore to more deeply bound configurations.

For a more comprehensive picture of the parameter space behavior in relation to the wormhole parameters, the following Fig. \ref{fig:RSV_parameter_space} showcases a three-dimensional representation of the boundaries $\ell_B^\pm$ across the parameter space with respect to $a$ and $\xi$.

\begin{figure}[H]
\centering
\begin{subfigure}{.48\textwidth}
  \centering
  \includegraphics[width=\linewidth,  height= 0.8\textwidth]{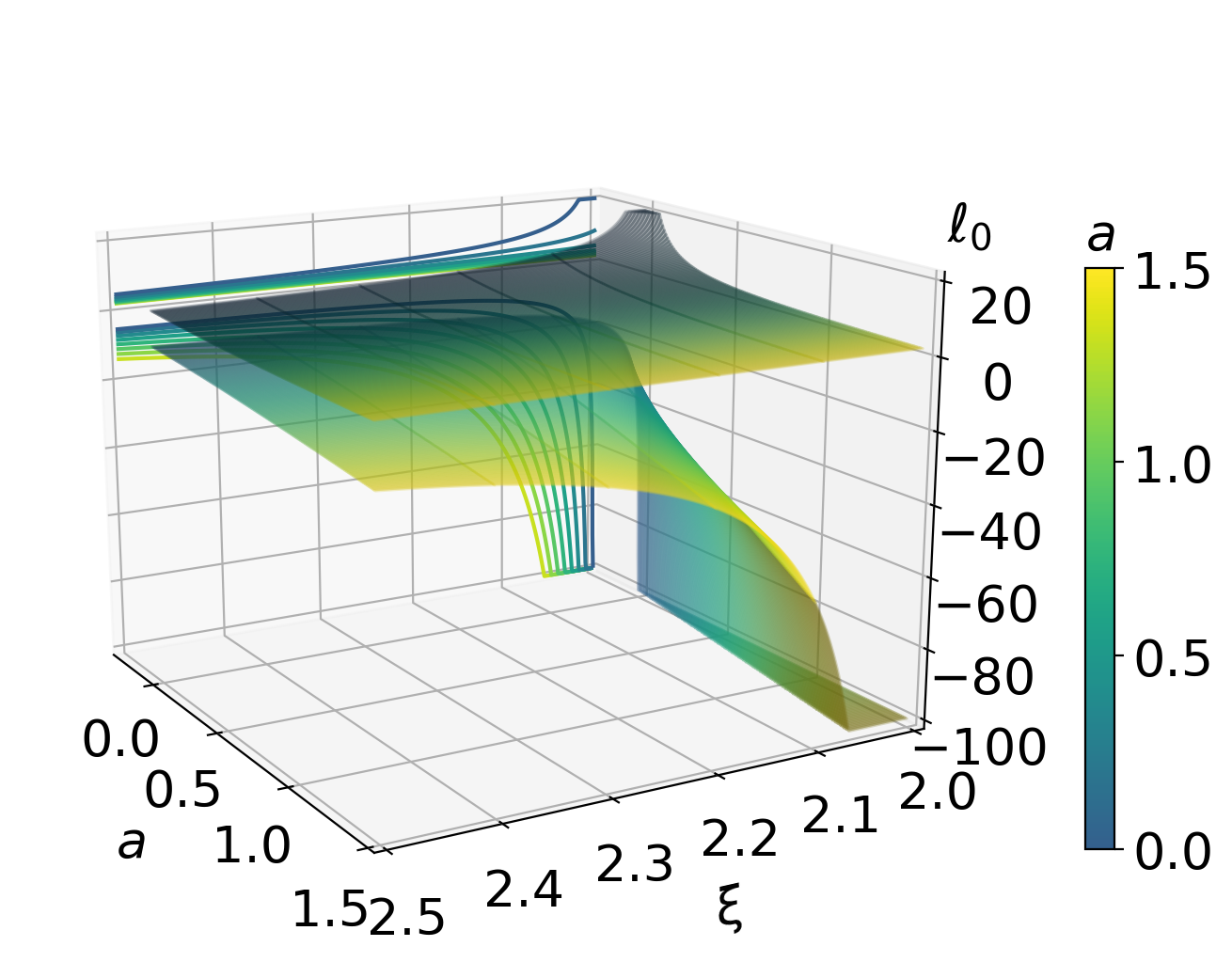}
\end{subfigure}
\caption{$\mathcal{RSV}$: Boundaries for bound circular orbits inside parameter space for varying $a$ and $\xi$. The surfaces represent the boundaries $\ell_B^\pm$, whereby the upper surface is $\ell_B^-$ and the lower surface is $\ell_B^+$. The surfaces are colored with respect to the value of $a$ within the parameter space, blueish tones represent lower values of $a$, whereas yellowish tones represent higher values of $a$. The curves in the $\xi$-$\ell_0$-plane represent projections of the surfaces for fixed values of $a$ and they are colored in their corresponding values of $a$. Bound circular orbits only exist between the surfaces.}
\label{fig:RSV_parameter_space}
\end{figure}

\subsection{$\mathcal{BH}$ Wormholes}

\begin{figure}[H]
\centering
\begin{subfigure}{.325\textwidth}
  \centering
  \includegraphics[width=\linewidth,  height= 0.75\textwidth]{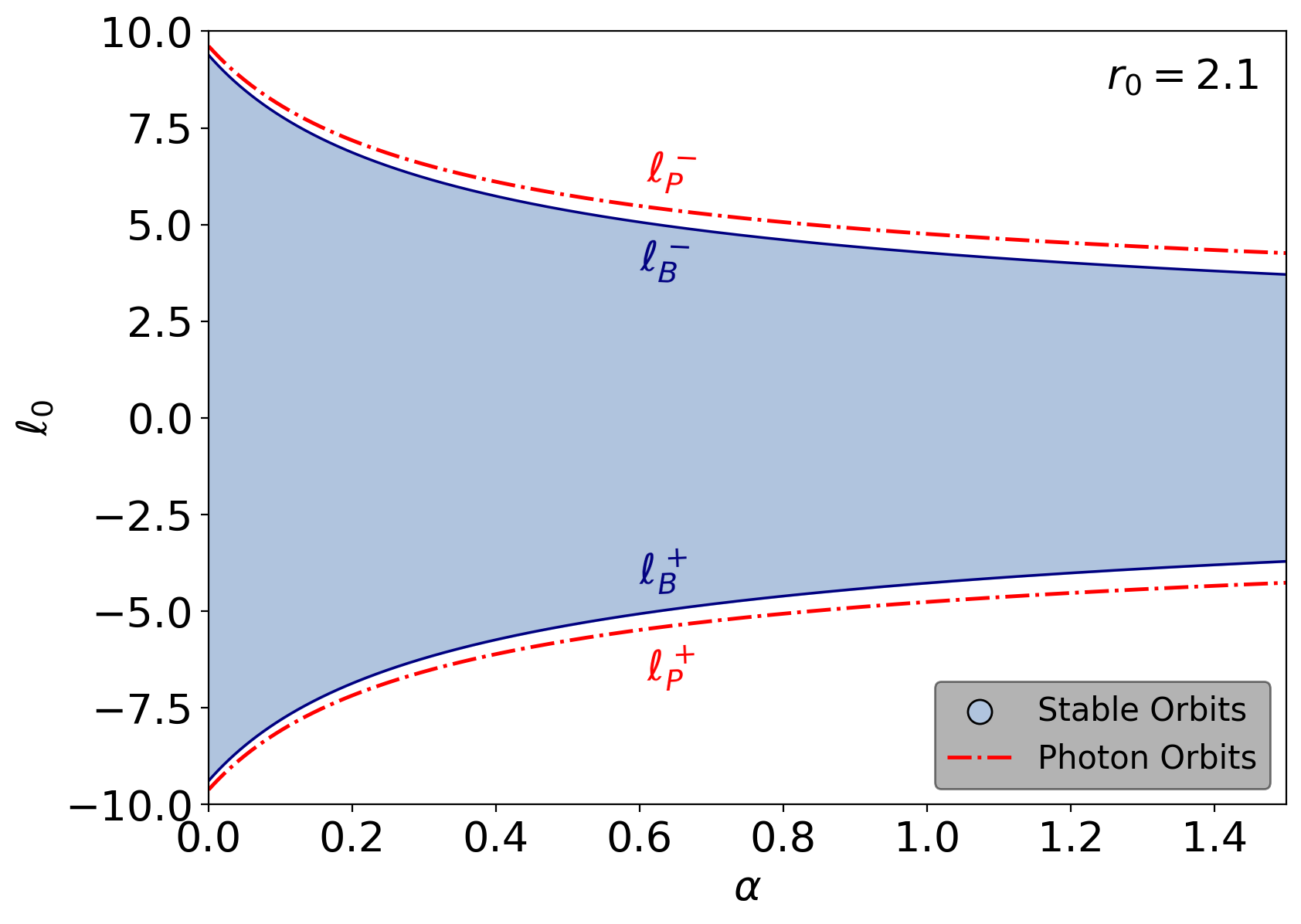}
  \caption{$\mathcal{BH}$: Throat orbits for $r_0 = 2.1$}
\end{subfigure}
\begin{subfigure}{.325\textwidth}
  \centering
  \includegraphics[width=\linewidth,  height= 0.75\textwidth]{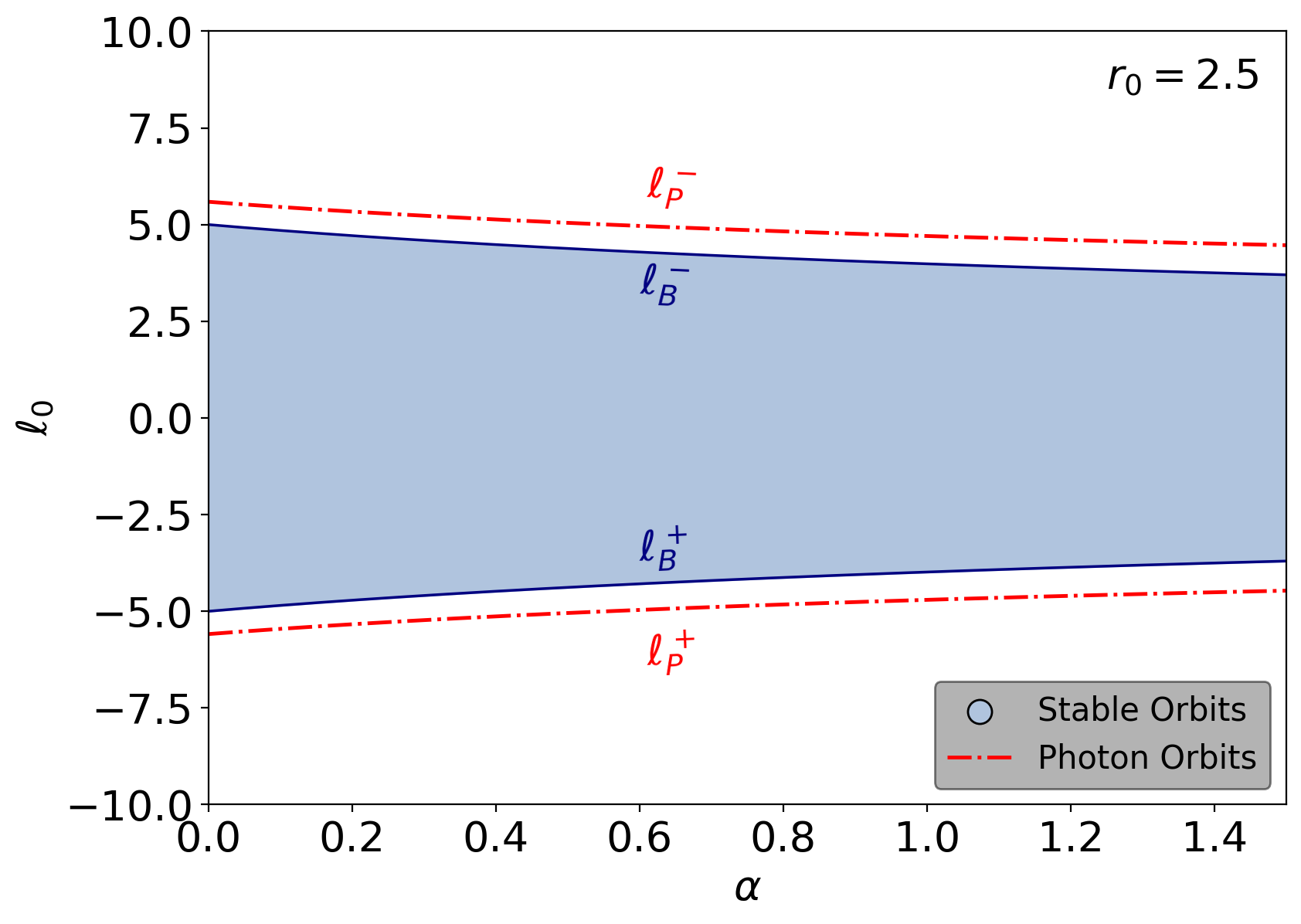}
  \caption{$\mathcal{BH}$: Throat orbits for $r_0 = 2.5$}
\end{subfigure}
\begin{subfigure}{.325\textwidth}
  \centering
  \includegraphics[width=\linewidth,  height= 0.75\textwidth]{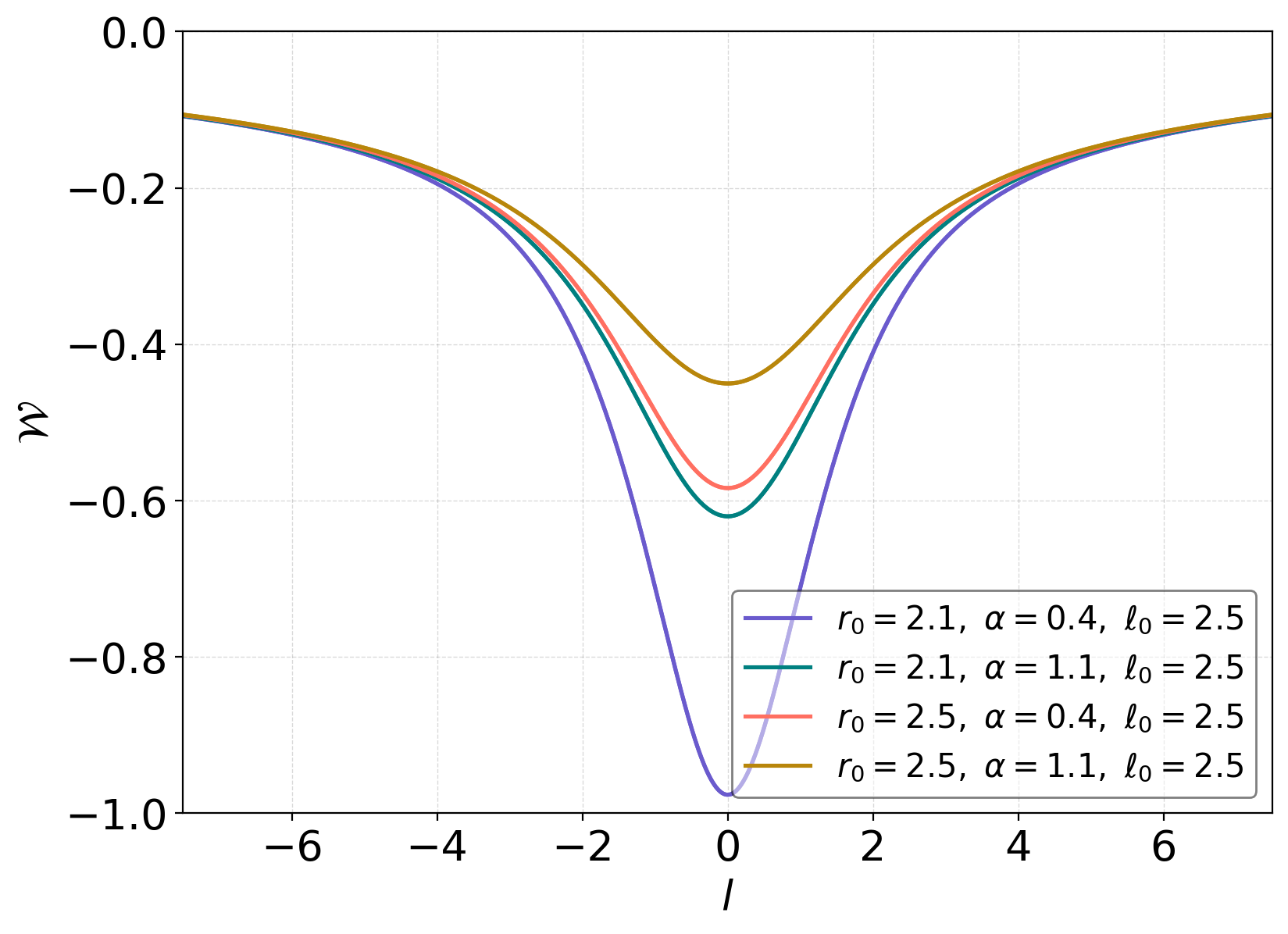}
  \caption{$\mathcal{BH}$: Exemplary disk solutions}
\end{subfigure}
\caption{(a) and (b) showcase the parameter space with regard to $\alpha$ for wormhole solutions where the throat is located at $r_0 = 2.1$ and $r_0 = 2.5$, respectively. The colored region represents the region for which bound orbits exist at the throat. The blue curves are the boundaries $\ell_B^\pm$ and the red dashed dotted curves indicate the impact parameters of the photon orbits, $\ell_P^\pm$. (c) showcases the effective potential $\mathcal{W}$ of disk solutions for different parameters.}
\label{fig:BH_phase}
\end{figure}

The $\mathcal{BH}$ wormholes are static wormholes and the parameter space for bound circular orbits at the throat depends here on the wormhole parameter $\alpha$ and the location of the throat $r_0^2$,

\begin{align}
    \ell_B^\pm = \pm r_0^2\sqrt{\frac{1 - \sqrt{1 + \frac{8 \alpha}{r_0^3}}}{2 \alpha + r_0^2 \left(1 - \sqrt{1 + \frac{8 \alpha}{r_0^3}} \right)}},
\end{align}

where the influence of $\alpha$ on the boundaries decreases for an increasing throat radius $r_0$. The parameter space for orbits at the throat becomes thus more constant for greater values of $r_0$ with regard to $\alpha$, which is depicted in Fig. \ref{fig:BH_phase}. Since greater values for $\alpha$ are considered unphysical, we restrict our analyses to the interval $\alpha \in [0,1.5]$.

\begin{figure}[H]
\centering
\begin{subfigure}{.48\textwidth}
  \centering
  \includegraphics[width=\linewidth,  height= 0.8\textwidth]{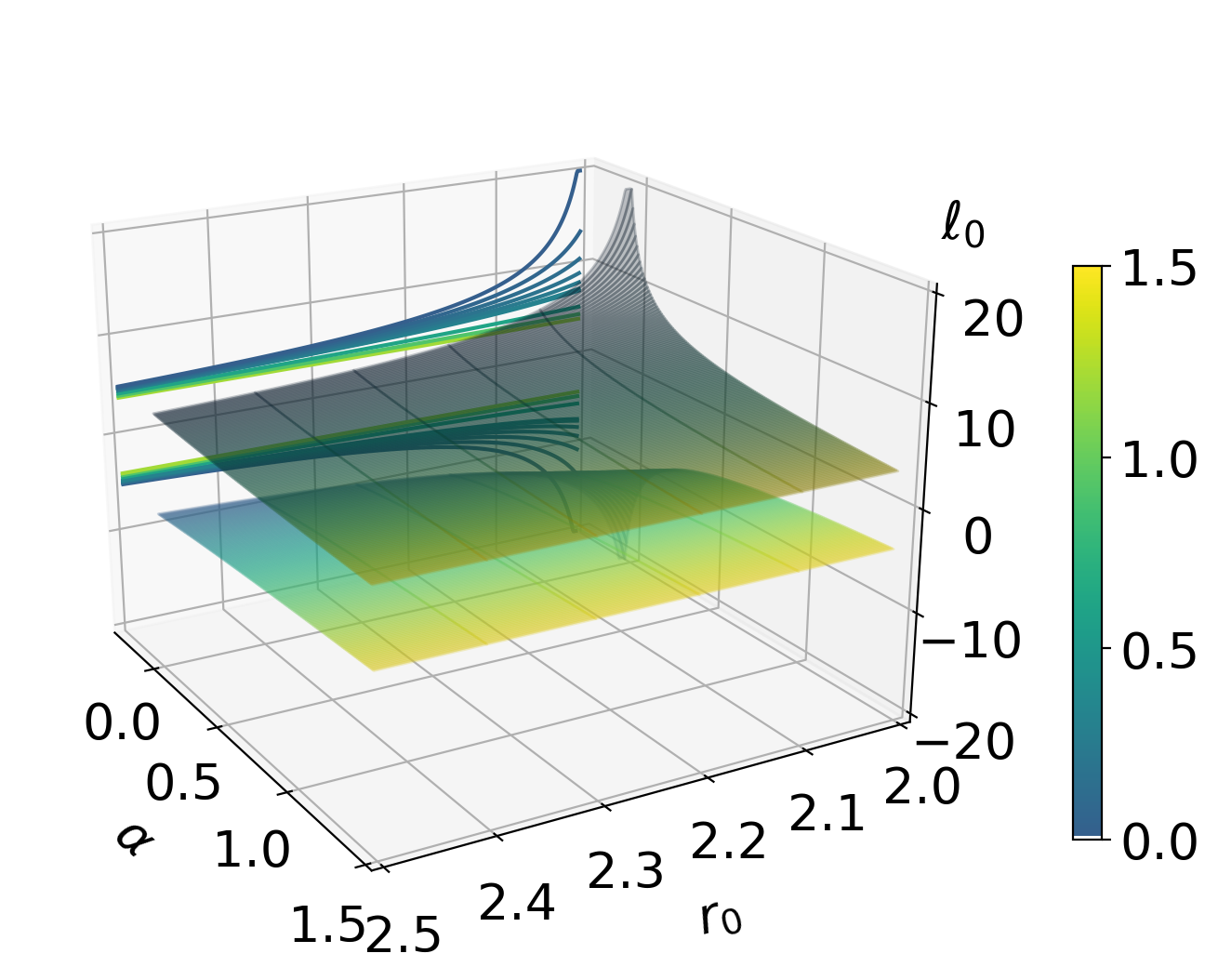}
\end{subfigure}
\caption{$\mathcal{BH}$: Boundaries for bound circular orbits inside the parameter space for varying $\alpha$ and throat radius $r_0$. The surfaces represent the boundaries $\ell_B^\pm$, whereby the upper surface is $\ell_B^-$ and the lower surface is $\ell_B^+$. The surfaces are colored with respect to the value of $\alpha$ within the parameter space, blueish tones represent lower values of $\alpha$, whereas yellowish tones represent higher values of $\alpha$. The curves in the $\xi$-$\ell_0$-plane represent projections of the surfaces for fixed values of $a$ and they are colored in their corresponding values of $\alpha$. Bound circular orbits only exist between the surfaces.}
\label{fig:BH_parameter_space}
\end{figure}

Since the $\mathcal{BH}$ wormholes are static wormholes, the parameter space for bound orbits at the throat is symmetric with respect to $\ell_0 = 0$, as there is no difference between the rotation orientation due to the spherical symmetry of the spacetime. Throughout the interval $r_0 \in (2,2.5]$ all orbits are stable for the analyzed range of $\alpha$ values. The parameter space is a decreasing function of both, the wormhole parameter $\alpha$ and the throat radius $r_0$. As a consequence, the maximal parameter space is given for $\alpha = 0$, with $\ell_0 \in [-9.38,9.38]$ for $r_0 = 2.1$ and $\ell_0 \in [-5,5]$ for $r_0 =  2.5$. The shape of the effective potential for different disk configurations across the parameter space appears similar, as illustrated exemplarily in subfigure \ref{fig:BH_phase} (c), only close to the throat their shapes deviate from each other. A three-dimensional representation of the parameter space boundaries can be found in Fig.~\ref{fig:BH_parameter_space}. In the limit $r_0 \rightarrow 2$ we have $|\ell_B^\pm| \rightarrow \infty$.

\section{Accretion Tori around Wormholes}
In this section we will continue our analysis by investigating thick accretion disks around the wormholes. We analyze hereby the Keplerian circular orbit distribution within the equatorial plane to determine the types of possible accretion structures and furthermore combine this analysis with the results from the previous section. As we will show, matter accumulations at the throat may form greater structures with accretion tori around the throat.

\subsection{$\mathcal{TE}$ wormhole}
\begin{figure}[H]
\centering
\begin{subfigure}{.325\textwidth}
  \centering
  \includegraphics[width=\linewidth,  height= 0.75\textwidth]{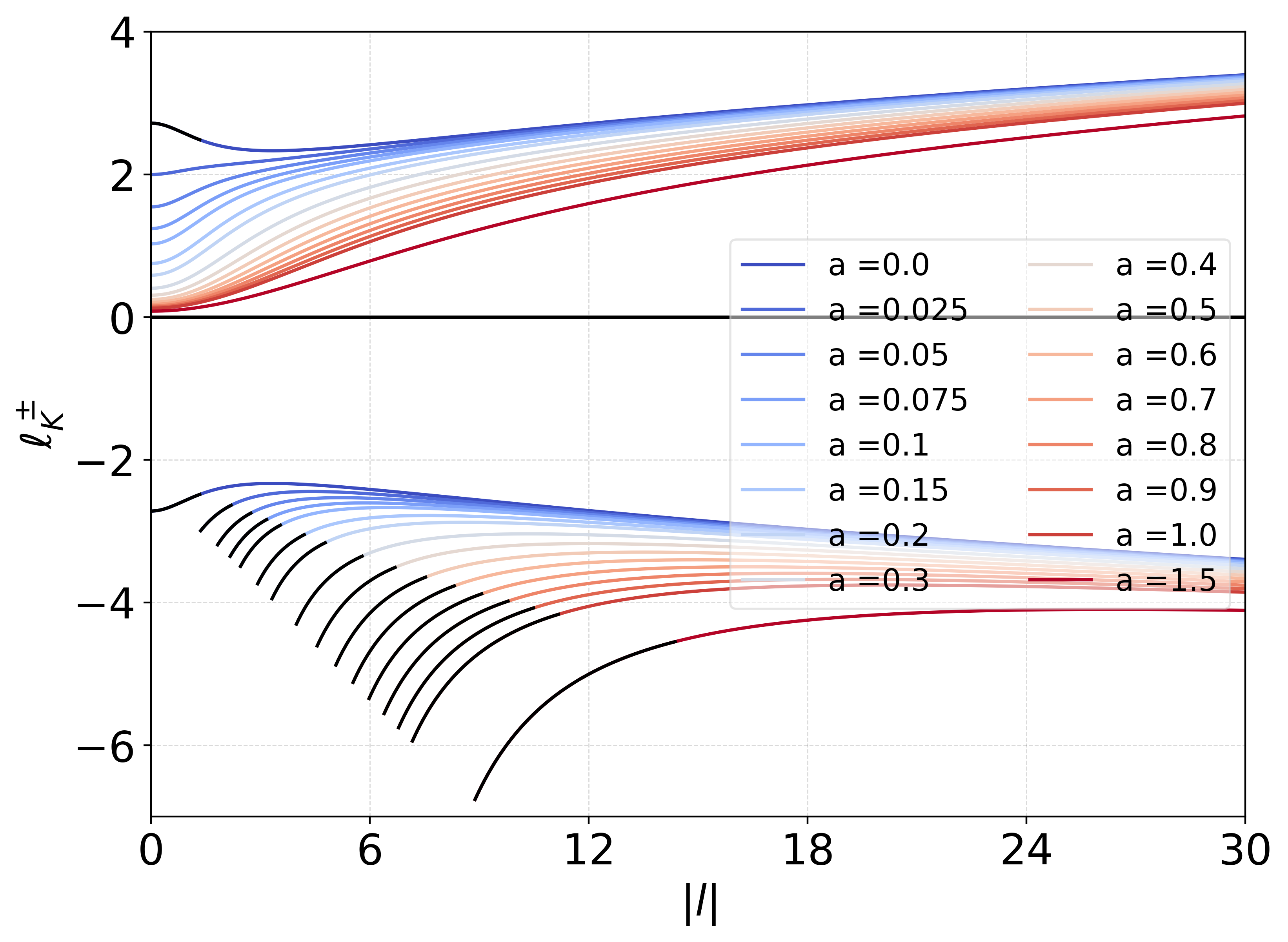}
  \caption{$\mathcal{TE}$: Equatorial $\ell_K^\pm$}
\end{subfigure}
\begin{subfigure}{.325\textwidth}
  \centering
  \includegraphics[width=\linewidth,  height= 0.75\textwidth]{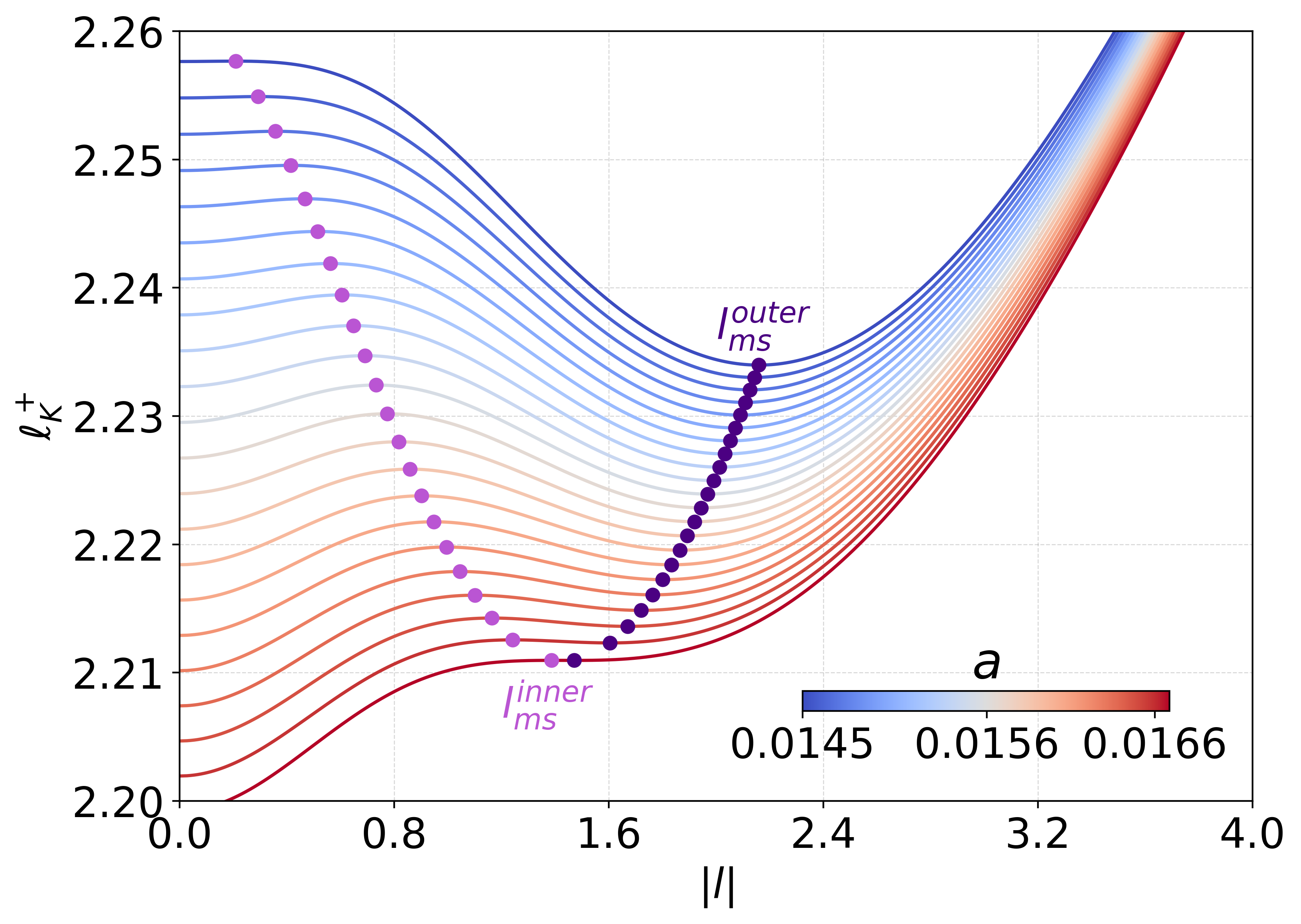}
  \caption{$\mathcal{TE}$: Solutions with two $R_{ms}$}
\end{subfigure}
\begin{subfigure}{.325\textwidth}
  \centering
  \includegraphics[width=\linewidth,  height= 0.75\textwidth]{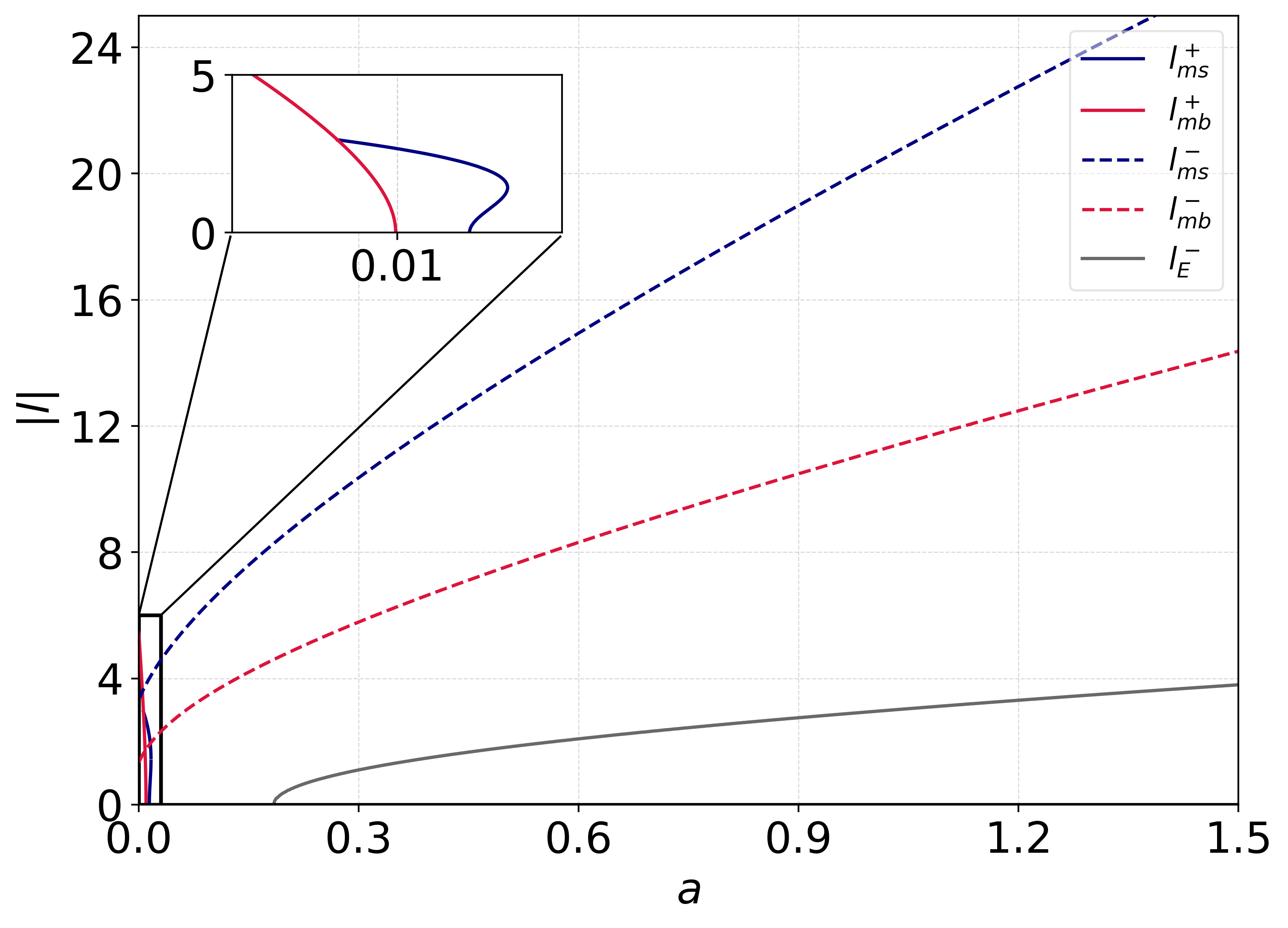}
  \caption{$\mathcal{TE}$: Marginally bound and stable orbits}
\end{subfigure}
\caption{(a) Keplerian specific angular momentum $\ell_K$ in the equatorial plane for different spin parameter $a$. The black curve sections mark regions where no bound orbits exist. The extrema indicate marginally stable orbits. (b) $\ell_K^+$ for solutions with an inner and outer marginally stable orbit. The pink and purple circles indicate the inner and outer marginally stable orbit, $l_{ms}^{inner}$ and $l_{ms}^{outer}$. (c) Marginally stable and marginally bound orbits versus the spin parameter $a$. The solid curves correspond to prograde motion, the dashed curves to retrograde motion. Blue colored curves mark marginally stable orbits $l^\pm_{ms}$, while red colored curves mark the marginally bound orbits $l^\pm_{mb}$. The grey solid curve marks the radial location of the ergo-sphere in the equatorial plane, $l_E$. The zoom in the plot showcases a close up of the marginally stable and bound orbits for prograde motion, which only exist for small values of $a$.}
\label{fig:TE_ellK}
\end{figure}

In case of the $\mathcal{TE}$ wormhole solutions, the Keplerian specific angular momentum, $\ell_K$, extends to the throat for prograde motion across the spin parameter range. For retrograde motion, the $\ell_K$ distribution follows a similar shape as for rotating black holes for higher spin parameter. This behavior can be linked to the presence of an ergoregion, which pushes the innermost stable circular orbits for retrograde motion outwards for higher spin parameter. Furthermore, in case of prograde motion, a small spin parameter space exists, for which an inner and outer marginally stable orbit is present, as depicted in Fig. \ref{fig:TE_ellK}.

\begin{figure}[H]
\centering
\begin{subfigure}{.48\textwidth}
  \centering
  \includegraphics[width=0.8\linewidth,  height= 0.6\textwidth]{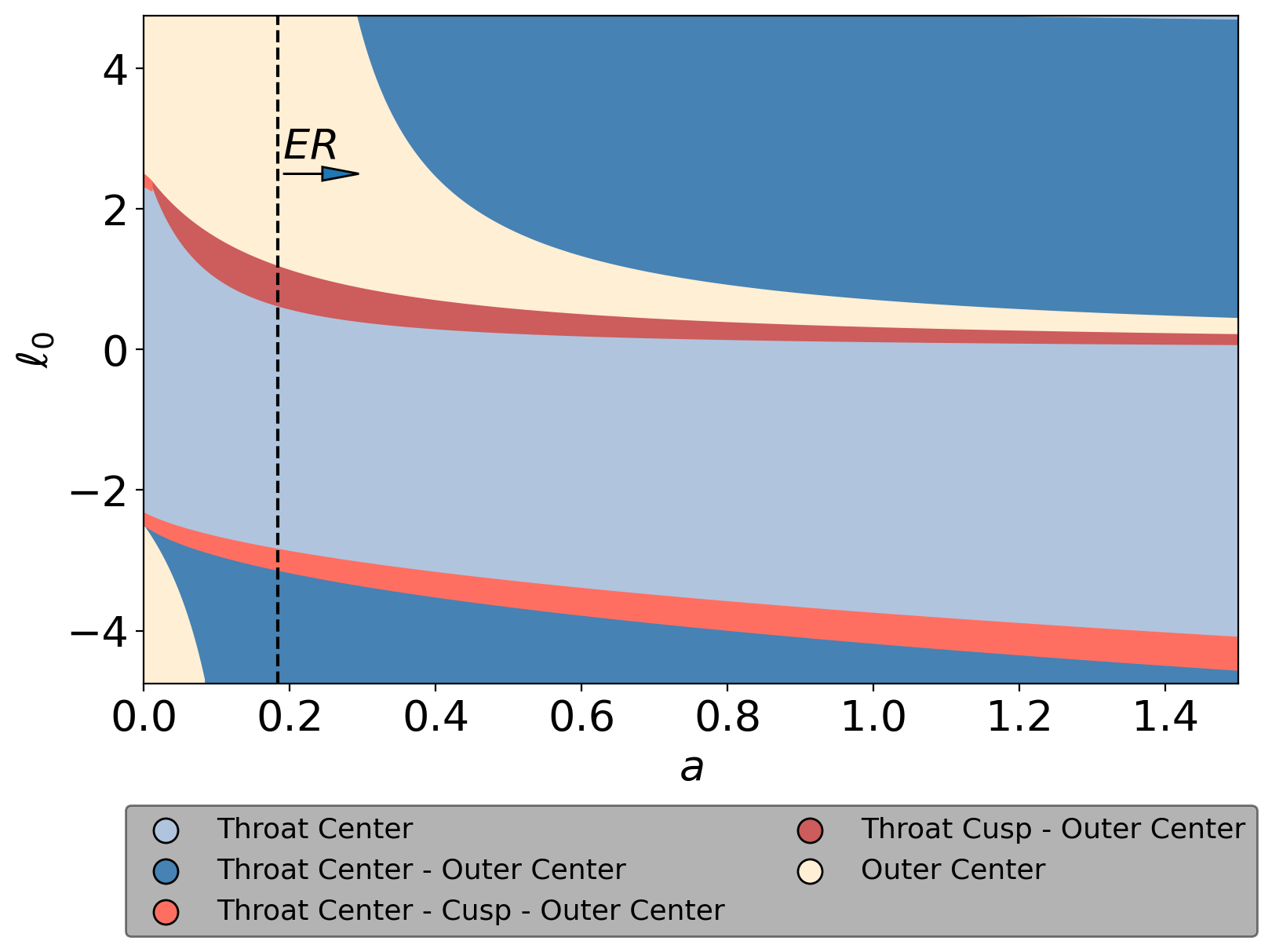}
  \caption{$\mathcal{TE}$: Parameter space of different disk configurations}
\end{subfigure}
\begin{subfigure}{.48\textwidth}
  \centering
  \includegraphics[width=0.8\linewidth,  height= 0.6\textwidth]{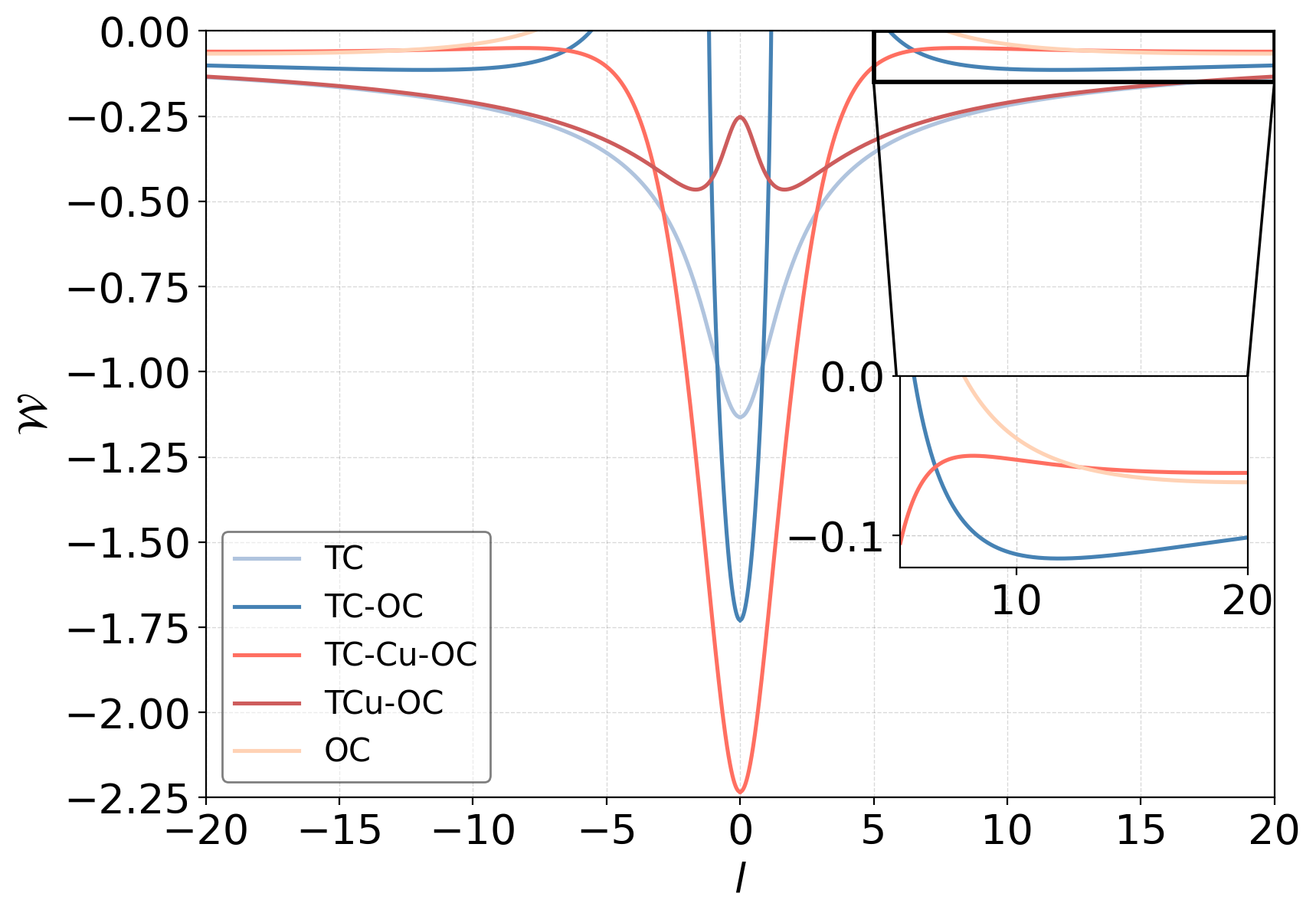}
  \caption{$\mathcal{TE}$: Types of disk configurations}
\end{subfigure}
\caption{(a) Phase space for the different possible disk configurations composed of matter accumulations at the throat and accretion tori around the wormhole. The `Throat Center' region is a region of the parameter space, for which only matter accumulations at the throat exist. The `Throat Center - Outer Center' is a region of parameter space, where matter accumulations at the throat and a accretion disk around the wormhole exist. The `Throat Center - Cusp - Outer Center' region, is a region where matter accumulations at the throat are connected to an outer accretion torus through a cusp. The `Throat Cusp - Outer Center' region, is a region where the effective potential has a maximum at the throat. Thus, the disk configuration is a connected structure, with an outer disk reaching up to the throat, where it has a cusp. The `Outer Center' region, is a region where no matter accumulations at the throat are present and only the usual disk solutions, associated with the $\ell_K^\pm$ distribution, exist. It should be noted, that since the retrograde marginally stable orbits are far extending in the equatorial plane for higher values of $a$, the maximal and minimal values of $\ell_K^\pm$ considered for this plot, are chosen so that all marginally stable orbits are captured for this diagram. (b) Effective potential in the equatorial plane for examples of the different possible types of disk configurations. The colors of the curves and the abbreviations in the legend match the colors and the description of the parameter space plot in (a). On the right side a zoom in the plot is inserted, which highlights the region close to zero and showcases the the minima and maxima of the `TC-Cu-OC' and `TC-OC' configuration.}
\label{fig:TE_phase_space_combined}
\end{figure}

For $a \in [0.0145,0.0167]$ $\ell_K^+$ has an inner and outer extremum corresponding to an inner and outer marginally stable orbit. The circular orbits in between the inner and outer marginally stable orbit are unstable. In general, only a small range of the parameter range harbors unbound orbits for prograde motion ($a < 0.01$). Furthermore, for $a \geq 0.017$ all orbits are stable and the $\ell_K^+$ distribution extends up to the throat. As a consequence, the disk solutions associated with the $\ell_K^+$ distribution are mostly composed of equilibrium tori, and since $\ell_K^+$ extends to the throat, accretion structures could be arbitrary close to the throat. Moreover, combining the results from the last section with the $\ell_K^\pm$ distribution leads to to a combination of disk configurations, composed of the matter accumulations at the throat and the accretion tori around the throat. An extended parameter space diagram for the different scenarios is presented in Fig. \ref{fig:TE_phase_space_combined}.

The extended parameter space is divided into five different regions, which correspond each to different disk configurations. Table \ref{tab:disk_configurations} showcases the conditions by which the different disk configurations are governed. Besides the basic configurations of an usual accretion torus around the wormhole, and the matter accumulations at the throat from the last section, also combined structures exist. Three different scenarios are here possible. In two of them, the matter present at the throat is much more compactified than in the outer torus. As a consequence the density values are several orders of magnitude higher at the wormhole throat. The structure at the throat is either closed towards the outside and disconnected from the outer torus, or connected by a cusp with the outer torus. The latter scenario corresponds to a continuous accretion structure, composed of two density maxima. In the third scenario the effective potential possesses a maximum at the throat and one minimum close to the throat. The accretion disk is thus connected through a cusp to the other side of the wormhole.

\begin{table}[H]
\centering
\vline
\begin{tabular}{c|c|c}
     \toprule
      \text{Disk Configuration} & \text{Motion at Throat} & \text{$\ell_0$ Range} \\
     \midrule
     \text{Throat Center} & \text{Stable} & $\ell_0\in [\ell_B^+,\ell_B^-] \ \wedge \ \ell_0 \notin \ell_K^\pm$ \\
     \text{Throat Center - Outer Center} & \text{Stable} & $\ell_0\in [\ell_B^+,\ell_B^-] \ \wedge \ \ell_0 \in \ell_K^\pm$ \\
     \text{Throat Center - Cusp - Outer Center} & \text{Stable} & $\ell_0\in [\ell_B^+,\ell_B^-] \ \wedge \ |\ell_0| \in (|\ell_{ms}|,|\ell_{mb}|]$ \\
     \text{Throat Cusp - Outer Center} & \text{Unstable} & $\ell_0 \in [\ell_B^+,\ell_B^-] \ \wedge \ \ell_0 \in \ell_K^\pm$ \\
     \text{Outer Center} & $-$ & $\ell_0 \notin [\ell_B^+,\ell_B^-] \ \wedge \ \ell_0 \in \ell_K^\pm$\\
     \bottomrule
\end{tabular}
\vline
\caption{Properties at the throat and parameter range for the different possible disk configurations. It should be noted that this only applies for wormholes without an ergoregion. If an ergoregion is present, then the condition $\ell_0 \in [\ell_B^+,\ell_B^-]$ must be replaced with the condition $\ell_0 \notin (\ell_B^-,\ell_B^+)$, and the condition $\ell_0 \notin [\ell_B^+, \ell_B^-]$ must be replaced with $\ell_0 \in (\ell_B^-,\ell_B^+)$.}
\label{tab:disk_configurations}
\end{table}

\subsection{$\mathcal{RSV}$ Wormholes}

\begin{figure}[H]
\centering
\begin{subfigure}{.325\textwidth}
  \centering
  \includegraphics[width=\linewidth,  height= 0.75\textwidth]{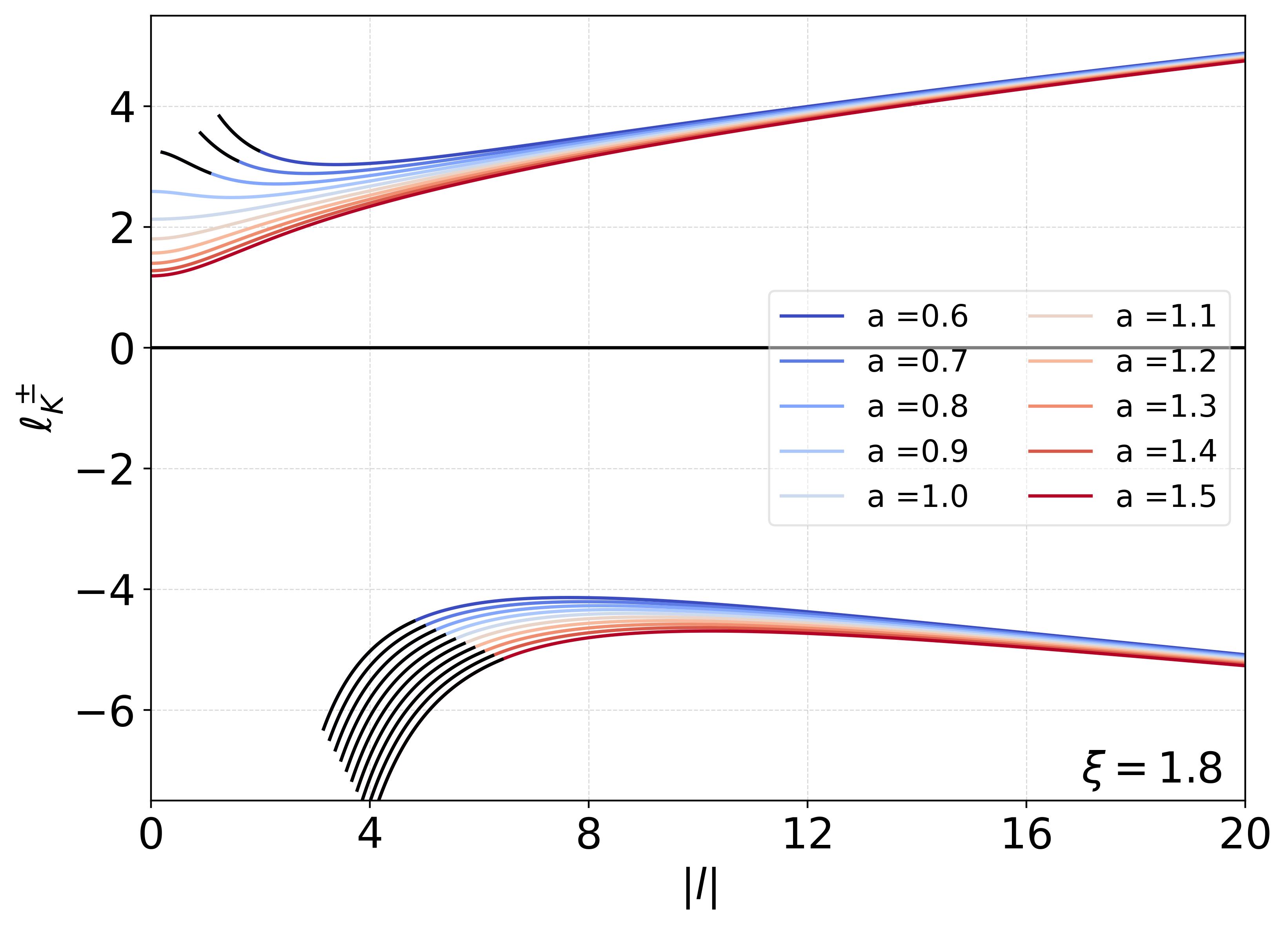}
  \caption{$\mathcal{RSV}$: Equatorial $\ell_K^\pm$ for $\xi = 1.8$}
\end{subfigure}
\begin{subfigure}{.325\textwidth}
  \centering
  \includegraphics[width=\linewidth,  height= 0.75\textwidth]{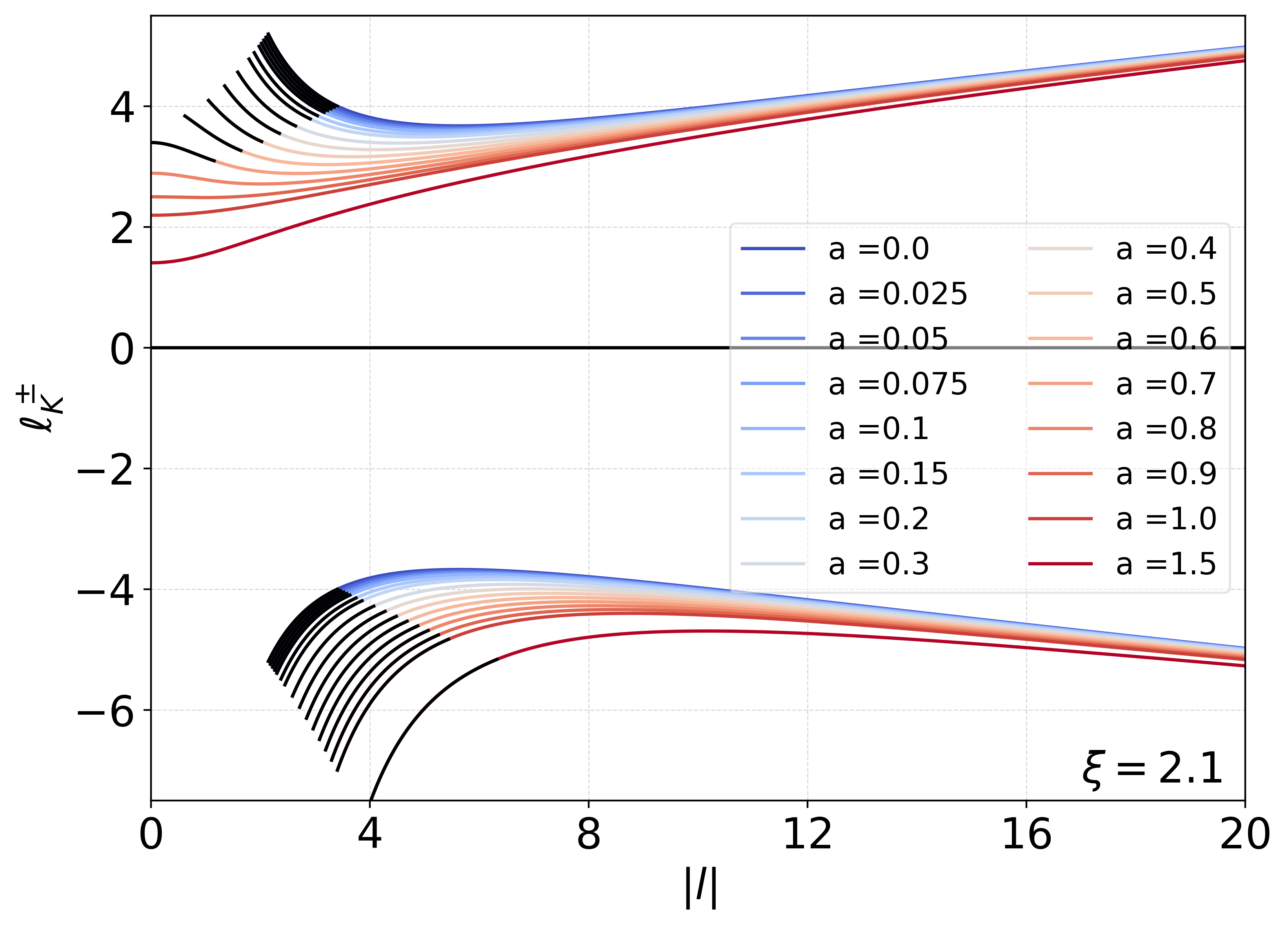}
  \caption{$\mathcal{RSV}$: Equatorial $\ell_K^\pm$ for $\xi = 2.1$}
\end{subfigure}
\begin{subfigure}{.325\textwidth}
  \centering
  \includegraphics[width=\linewidth,  height= 0.75\textwidth]{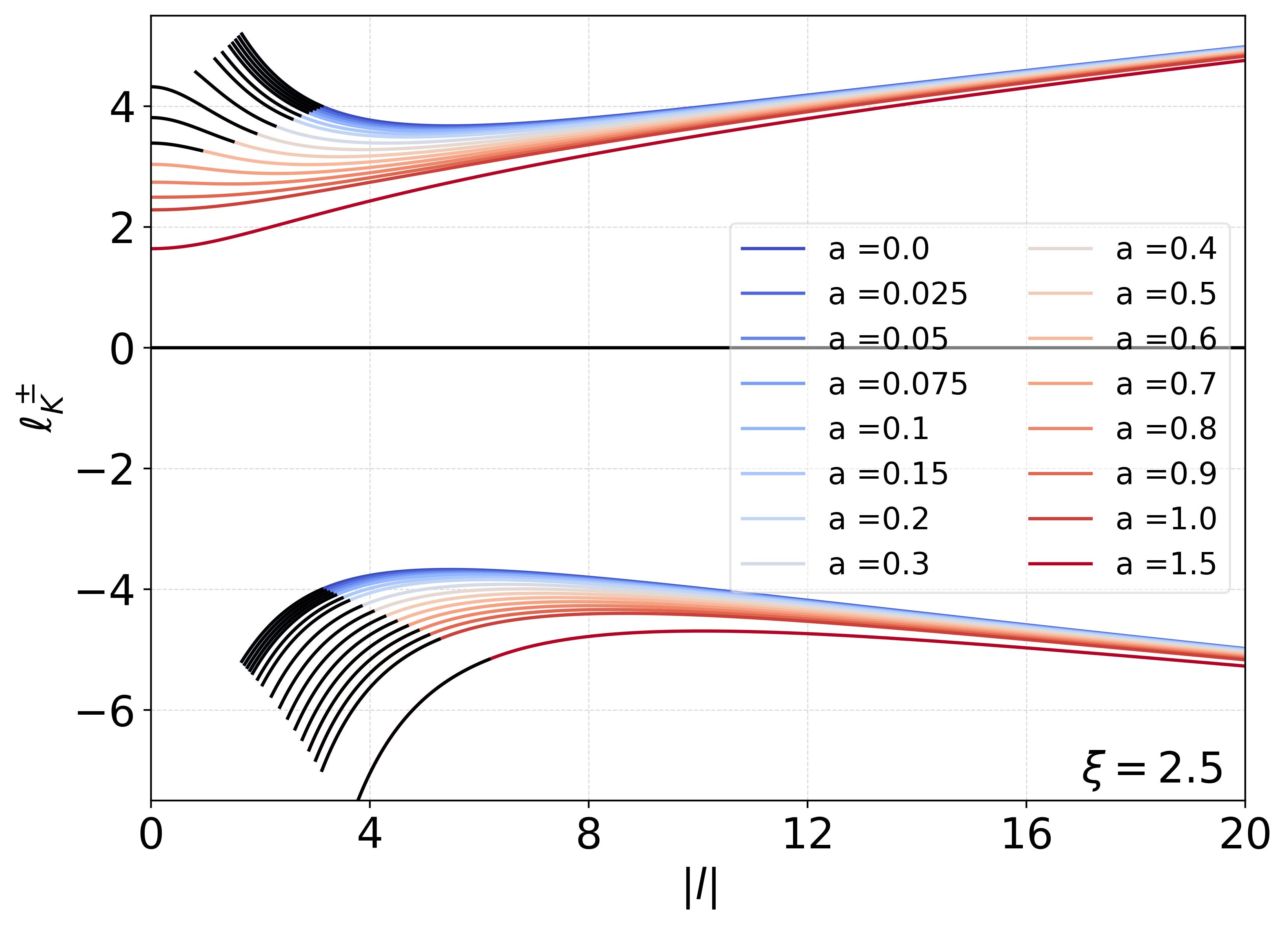}
  \caption{$\mathcal{RSV}$: Equatorial $\ell_K^\pm$ for $\xi = 2.5$}
\end{subfigure}
\caption{Keplerian specific angular momentum $\ell_K^\pm$ for the $\mathcal{RSV}$ wormholes and different values of $\xi$. Black curve sections mark regions where no bound orbits exist. The extrema indicate marginally stable orbits.}
\label{fig:RSV_ellK}
\end{figure}

In Fig. \ref{fig:RSV_ellK} the $\ell_K^\pm$ distribution for the $\mathcal{RSV}$ wormholes is shown. As for the $\mathcal{TE}$ wormholes, the distribution for the retrograde $\ell_K$ appears similar to the one for rotating black holes, with a marginally stable and marginally bound orbit for all spin parameters. There exist therefore retrograde disk solutions with a cusp for all spin parameter values. For higher values of $\xi$ the differences for different values of $a$ grow and also orbits closer to the throat exist.

\begin{figure}[H]
\centering
\begin{subfigure}{.48\textwidth}
  \centering
  \includegraphics[width=0.8\linewidth,  height= 0.6\textwidth]{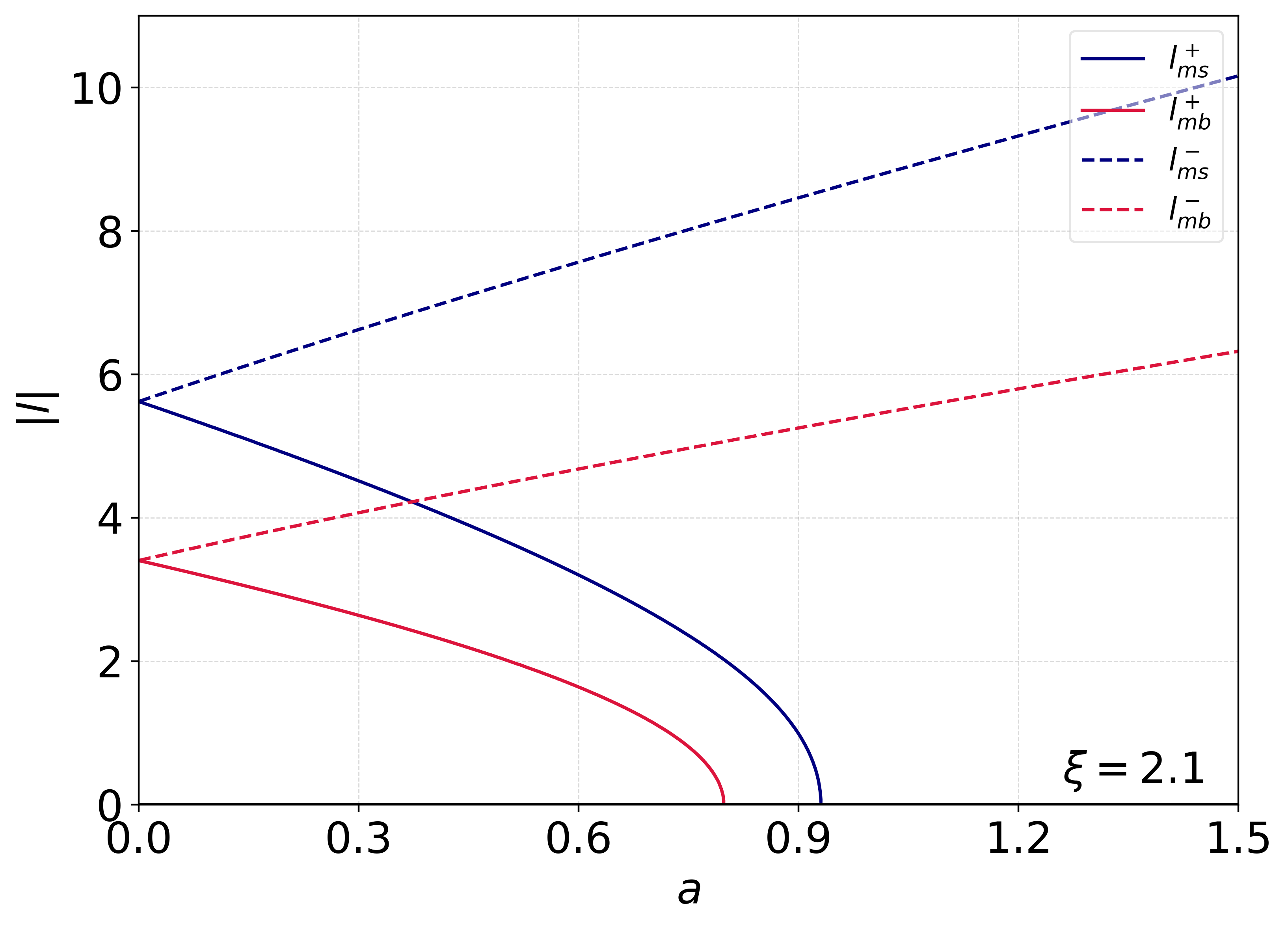}
  \caption{$\mathcal{RSV}$: Marginally bound and stable orbits}
\end{subfigure}
\begin{subfigure}{.48\textwidth}
  \centering
  \includegraphics[width=0.8\linewidth,  height= 0.6\textwidth]{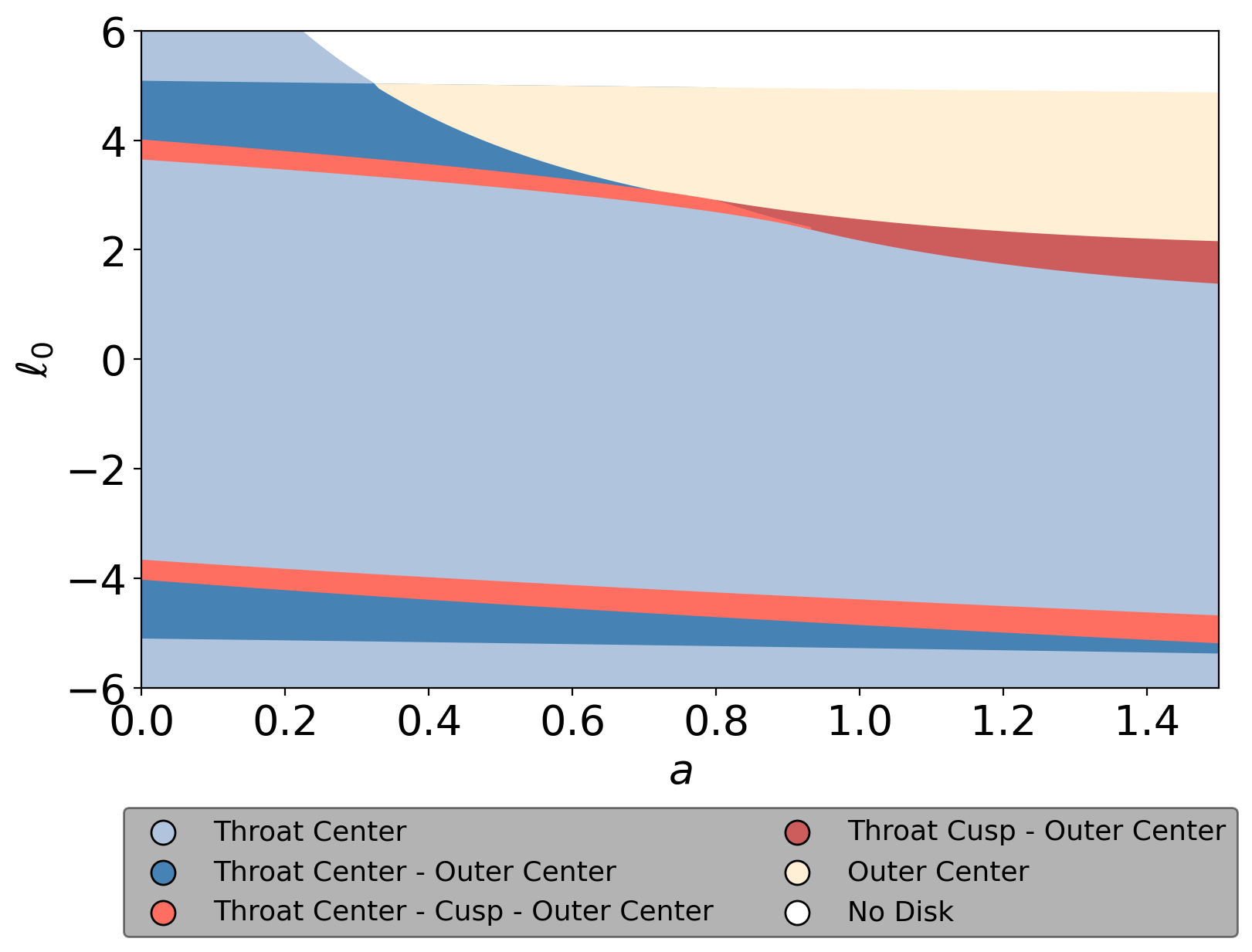}
  \caption{$\mathcal{RSV}$: Parameter space of different disk configurations for $\xi = 2.1$}
\end{subfigure}
\caption{(a) Marginally bound and stable orbits for retrograde and prograde motion with regard to the spin parameter $a$ for $\xi = 2.1$. The solid curves mark the marginally orbits for prograde motion, the dashed curves mark the marginally orbits for retrograde motion. (b) Extended parameter space for the different possible combined disk configurations. The different types are the same as for the $\mathcal{TE}$ wormholes depicted in Fig. \ref{fig:TE_phase_space_combined}. Since retrograde marginally stable orbits are far extending in the equatorial plane for higher values of $a$, the maximal and minimal values of $\ell_K^\pm$ considered for this plot, are chosen so that all marginally stable orbits are captured for this diagram.}
\label{fig:RSV_combined_phase}
\end{figure}

In case of prograde motion marginally bound and marginally stable orbits exist only up to a certain value of the spin parameter. Thus, circular orbits extend up to the throat for faster rotating wormholes, with all orbits also being stable. The critical values of $a$ for which the unbound and unstable regions disappear depend on the parameter $\xi$. For greater values of $\xi$ marginally bound and stable orbits exist up to higher values of $a$, thus the parameter spectrum for unstable motion grows. Since the high spin parameter solutions are strictly monotonic, only equilibrium tori may form around the wormhole. However, combining again the matter accumulations at the throat with the disks associated with the equatorial $\ell_K^\pm$ distribution, leads again to a set of different disk configurations, which are depicted with a plot of the marginally orbits in Fig. \ref{fig:RSV_combined_phase}. Since the distribution of the marginally orbits, as well as the $\ell_K^\pm$ distribution is similar for the selected values of $\xi$, we choose here $\xi = 2.1$ as an exemplary solution. However, it should be noted that for the $\xi = 1.8$ solution, the parameter space diagram has a different form due to the presence of the ergoregion. Nevertheless, the possible disk configurations remain the same, also with a similar volume inside the parameter space. The possible disk configurations are the same as for the $\mathcal{TE}$ wormhole solutions. As a major difference, the `Throat Center - Cusp - Outer Center' configuration has a larger parameter space volume for disk configurations with prograde flow.

\subsection{$\mathcal{BH}$ Wormholes}

\begin{figure}[H]
\centering
\begin{subfigure}{.325\textwidth}
  \centering
  \includegraphics[width=\linewidth,  height= 0.75\textwidth]{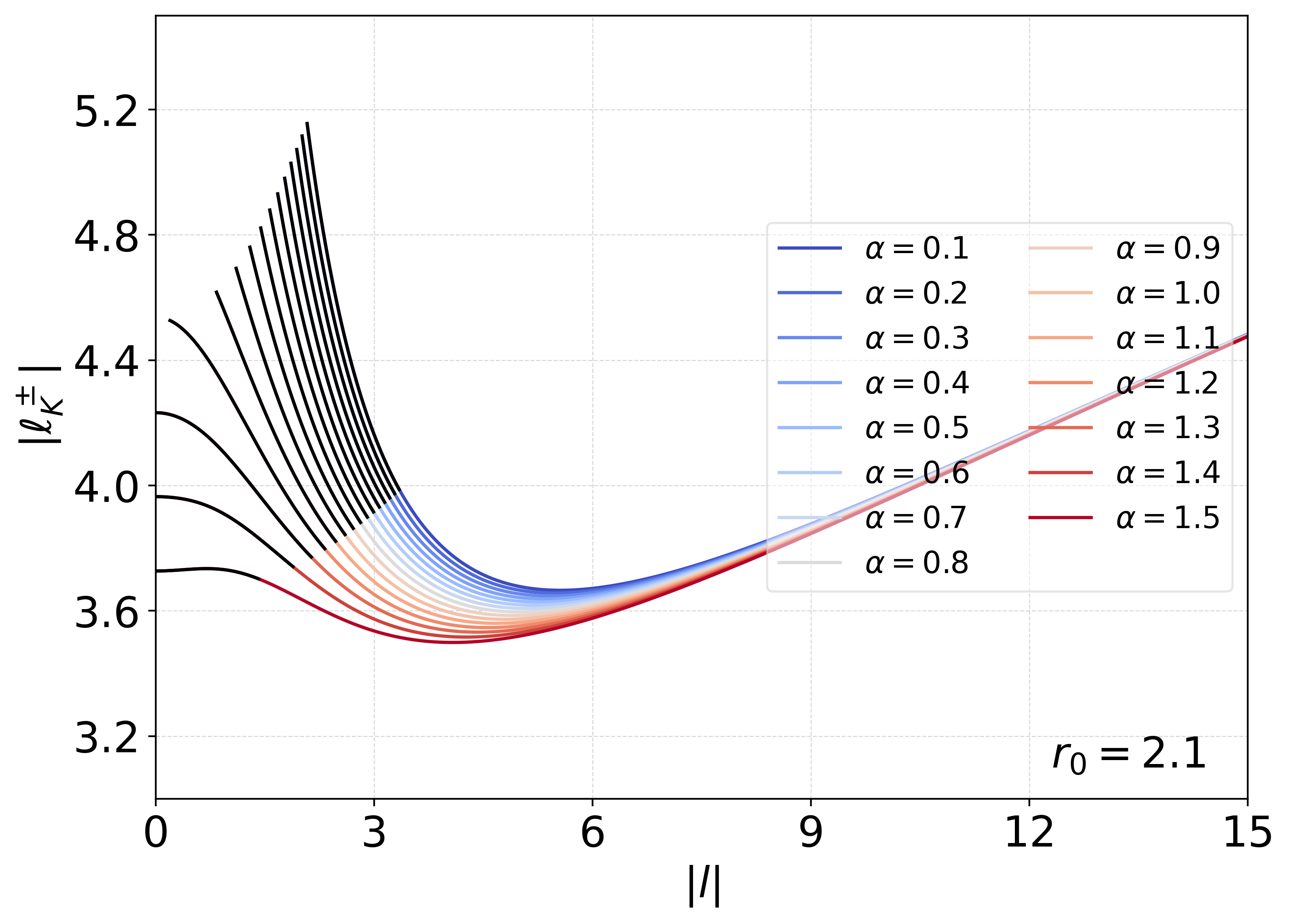}
  \caption{$\mathcal{BH}$: Equatorial $\ell_K^\pm$ for $r_0 = 2.1$}
\end{subfigure}
\begin{subfigure}{.325\textwidth}
  \centering
  \includegraphics[width=\linewidth,  height= 0.75\textwidth]{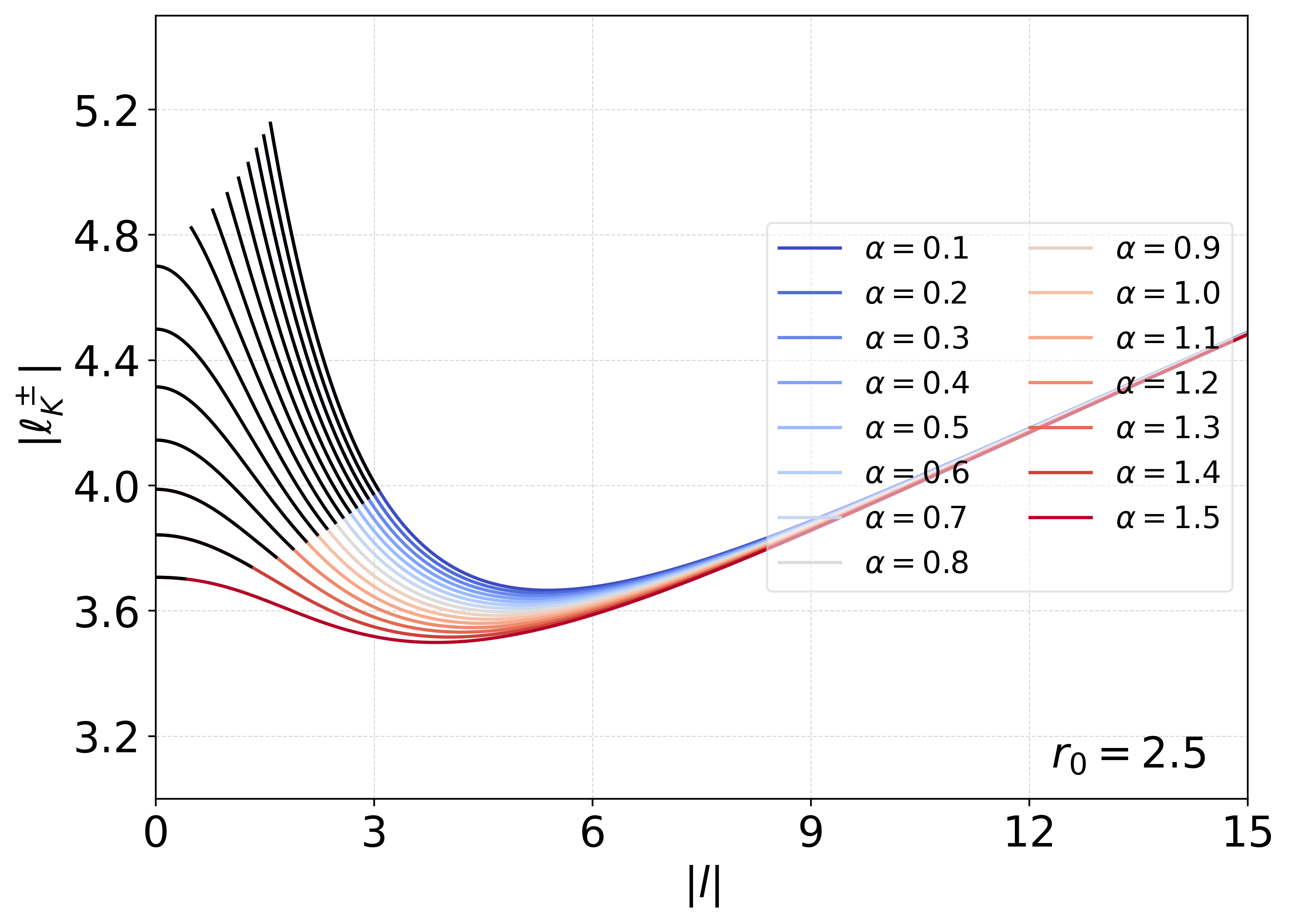}
  \caption{$\mathcal{BH}$: Equatorial $\ell_K^\pm$ for $r_0 = 2.5$}
\end{subfigure}
\begin{subfigure}{.325\textwidth}
  \centering
  \includegraphics[width=\linewidth,  height= 0.75\textwidth]{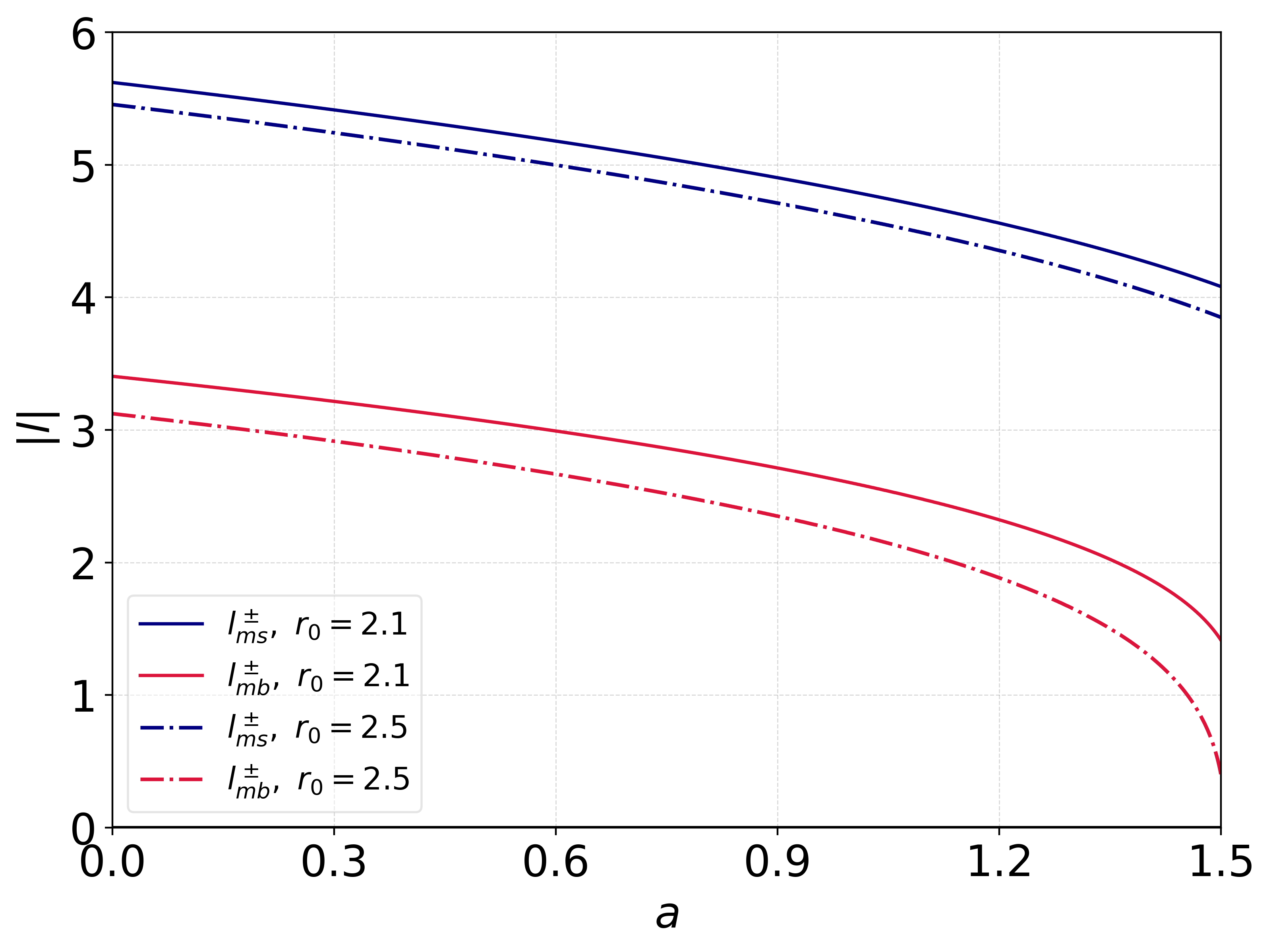}
  \caption{$\mathcal{BH}$: Marginally bound and stable orbits}
\end{subfigure}
\caption{(a) and (b) showcases the Keplerian specific angular momentum $\ell_K^\pm$ for the $\mathcal{BH}$ wormholes and for the throat radii $r_0 \in \{2.1, 2.5\}$. Black curve sections mark regions where no bound orbits exist. The extrema indicate marginally stable orbits. (c) Marginally bound and stable orbits for the selected throat radii. Red colored curves illustrate the marginally bound orbits and blue colored curves the marginally stable orbits.}
\label{fig:BH_ellK}
\end{figure}

In the case of the $\mathcal{BH}$ wormhole solutions, the $\ell_K^\pm$ distribution (depicted in Fig. \ref{fig:BH_ellK}) is again similar across the wormhole parameter range. For small values of $\alpha$ the distribution closely resembles that of a Schwarzschild black hole and with increasing $\alpha$ the distribution moves closer to the throat. The difference between the different throat radii $r_0$ is only noticeable for higher values of $\alpha$. For both analyzed values of $r_0$ the distributions have a region with no bound orbits, which moves closer to the throat for higher values of $\alpha$. Thus, in both cases marginally bound and stable orbits exist across the whole analyzed parameter range for $\alpha$ (depicted in Fig. \ref{fig:BH_ellK} (c)). As a consequence also accretion tori with a cusp can exist across the whole analyzed parameter range.

\begin{figure}[H]
\centering
\begin{subfigure}{.48\textwidth}
  \centering
  \includegraphics[width=0.8\linewidth,  height= 0.6\textwidth]{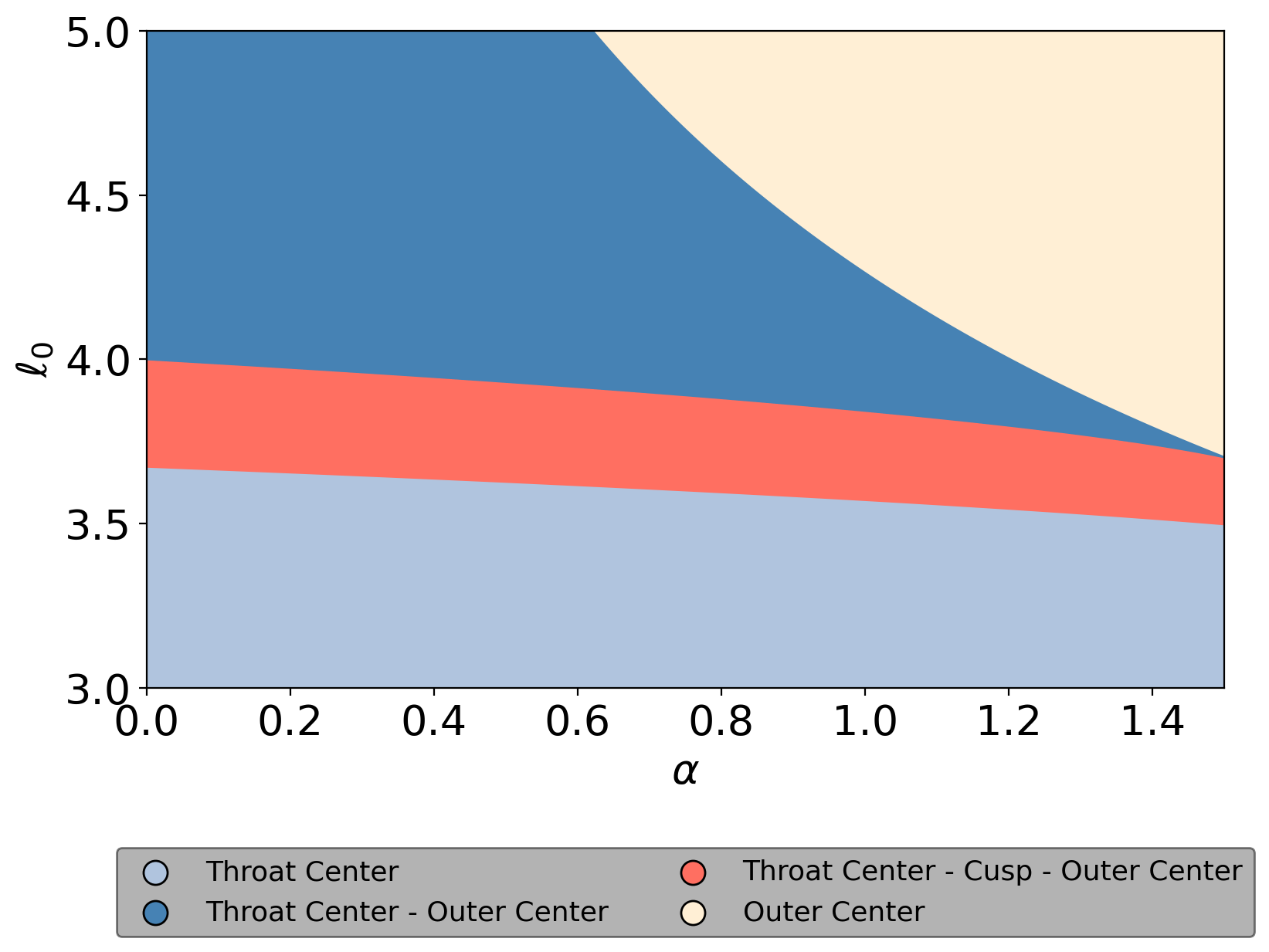}
\end{subfigure}
\caption{Parameter space of different disk configurations for $r_0 = 2.1$. The possible disk configurations and the colors representing them are the same as for the $\mathcal{TE}$ and $\mathcal{RSV}$ wormholes. The upper limit of $\ell_K^\pm$ considered for this parameter space diagram is selected such, that it corresponds to a circular orbit at a distance to the throat which is given by $r = 10 r_0$.}
\label{fig:BH_combined_phase}
\end{figure}

Since the distribution of $\ell_K^\pm$ is similar for the different throat radii, we select the $r_0 = 2.1$ solution as an example for the parameter space diagram of different disk configurations, which is shown in Fig. \ref{fig:BH_combined_phase}. Most of the parameter space corresponds to matter accumulations around the throat, only for a small region combined accretion structures exist. Since all orbits at the throat are stable for the analyzed parameter range of $\alpha$, only disk configurations with a density maximum at the throat exist. Thus, the configuration with a cusp at the throat, which is possible for the $\mathcal{TE}$ and $\mathcal{RSV}$ wormhole solutions, does not exist for the depicted parameter range. However it should be noted, that for the $\mathcal{TE}$ and $\mathcal{RSV}$ wormholes, disk configurations with a cusp at the throat only exist for $a > 0$. In their corresponding static solutions, where $a = 0$, also no cusp solutions exist. The existence of unstable orbits at the throat and thus also the existence of disk configurations with a cusp at the throat can therefore possible be linked to the rotating of the spacetime, and is for rotating wormholes more likely to appear.

We conclude that, there exists a plethora of possibilities for different disk configurations at and around wormholes, even solutions which represent connected structures of accretion tori and accumulations of matter at the throat. The formation of central accretion structures around perhaps natural occuring wormholes seems to be therefore within the bounds of possibility.

\section{Conclusion}

In this work we investigated possible matter accumulations and accretion tori in symmetric wormhole spacetimes. Methodically, we analyzed the presence of circular orbits at the throat, as well as the equatorial Keplerian orbit distribution and linked them to accretion structures with the use of the Polish Doughnut model. As exemplary wormholes we selected three traversable wormhole spacetimes, namely a constructed rotating wormhole from the Teo class, designated by $\mathcal{TE}$, the rotating Simpson-Visser wormhole, designated by $\mathcal{RSV}$, and a static traversable wormhole in beyond Horndeski theories, designated by $\mathcal{BH}$.

In case of symmetrical wormholes, where both sides are connected smoothly by a global coordinate, it can be shown that a spectrum of circular orbits may be present at the throat. We analyzed here the boundaries of this spectrum for the different wormhole solutions, where we varied the main wormhole parameters in view of a broad analysis across the parameter range. The varied parameters are the spin parameter $a$ for the $\mathcal{TE}$ and $\mathcal{RSV}$ solutions, the regularization parameter $\xi$ for the $\mathcal{RSV}$ solution, and the throat location $r_0$ as well as the wormhole parameter $\alpha$ for the $\mathcal{BH}$ solution.

For the $\mathcal{TE}$ wormholes we found a phase transition of the parameter space for circular orbits at the throat, caused by the appearance of an ergoregion for spin parameters above the critical spin parameter $a_E = 0.184$, which marks the onset of solutions with an ergoregion. The size of the parameter space grows for increasing $a$ and within the parameter space corresponding to an ergoregion, it extends from $-\infty$ to $\infty$ with the exclusion of a small interval close to zero. A small interval above zero also corresponds to a region of unstable orbits at the throat. Nevertheless, the vast majority of the parameter space corresponds to stable orbits. The presence of stable orbits can be extended through accretion disk models to the presence of matter accumulations at the throat, where the inflow of matter from both sides of the wormhole is in equilibrium at the throat. Due to the symmetry the acting forces cancel each other out, the pressure gradient vanishes and a deeply bounded potential minimum forms at the throat. The resulting accretion structures are closely surrounding the wormhole and appear more spherical for smaller values of the specific angular momentum of the matter particles. Due to their encapsulating nature with regard to the wormhole throat, possible observational signatures are dominated by a more spherical radiations profile of these structures. Thus, they could appear similar to star-like objects for a distant observer, due to the presence of a central bright region.

We found that similar structures may also form for the other analyzed wormhole spacetimes, where also a spectrum of bound circular orbits exists at the throat. In case of the $\mathcal{RSV}$ wormholes, we analyzed three different solutions, for $\xi \in \{1.8,2.1,2.5\}$, where the first one is representing a solution with an ergoregion. For the absence of an ergoregion we found again a growing parameter space for increasing $a$. However, as $\xi$ grows, the parameter space shrinks and is less affected by the wormhole rotation. Since the effective potential $\mathcal{W}$ of the disk is a quadratic function of the disk specific angular momentum $\ell$, the deepest minimum of $\mathcal{W}$ at the throat is exactly given by the midpoint of the parameter space interval, which we illustrated by three-dimensional plots across the parameter space.

In case of the static $\mathcal{BH}$ wormhole solutions we analyzed two different solutions with different throat radii, namely $r_0 \in \{2.1,2.5\}$ and varied the wormhole parameter $\alpha$. We found a decreasing parameter space with regard to an increase of $\alpha$ and the throat radius $r_0$. Moreover, for increasing $r_0$ the parameter space boundaries are less affected by $\alpha$ and follow thus a more constant distribution. 

The combination of the circular orbit spectrum at the throat and the equatorial Keplerian orbit distribution leads to a variety of different disk configurations. Here, we identified five different types. The two most simple ones are given by the usual accretion tori, which may form around the wormhole and by the matter accumulations which may form around the throat. Within small regions of the parameter space, there also exist disk configurations, that correspond to a combined structure of the matter accumulations at the throat and the accretion tori around it. This combined structure can be disconnected, representing a small torus at the throat and a large torus around it, or it can be connected. If it is a connected structure, the torus around the throat is joined through a cusp with the outer torus. However, in case of unstable geodesic motion at the throat, there exists a cusp exactly at the throat, connected to a density maximum of an outer torus. A matter flow through the cusp at the throat through the wormhole towards the other side of the wormhole is possible.

The phase diagrams show for each of the analyzed wormhole solutions, that vast regions of the parameter space can be associated with some type of accretion disk configuration at or around the wormhole. We conclude, that this may hint to central bright regions which may be associated with wormhole spacetimes, instead of central dark regions, and furthermore that the presence of circular orbits at the throat may hint to instabilities of the spacetime, as it can be linked to matter accumulations at the wormhole for a vast spectrum of the parameter space.


\begin{acknowledgments}
P.N. gratefully acknowledges support from the Bulgarian NSF Grant KP-06-H68/7.

\end{acknowledgments}    

\bibliography{literature}

\begin{thebibliography}{30}%
\makeatletter
\providecommand \@ifxundefined [1]{%
 \@ifx{#1\undefined}
}%
\providecommand \@ifnum [1]{%
 \ifnum #1\expandafter \@firstoftwo
 \else \expandafter \@secondoftwo
 \fi
}%
\providecommand \@ifx [1]{%
 \ifx #1\expandafter \@firstoftwo
 \else \expandafter \@secondoftwo
 \fi
}%
\providecommand \natexlab [1]{#1}%
\providecommand \enquote  [1]{``#1''}%
\providecommand \bibnamefont  [1]{#1}%
\providecommand \bibfnamefont [1]{#1}%
\providecommand \citenamefont [1]{#1}%
\providecommand \href@noop [0]{\@secondoftwo}%
\providecommand \href [0]{\begingroup \@sanitize@url \@href}%
\providecommand \@href[1]{\@@startlink{#1}\@@href}%
\providecommand \@@href[1]{\endgroup#1\@@endlink}%
\providecommand \@sanitize@url [0]{\catcode `\\12\catcode `\$12\catcode `\&12\catcode `\#12\catcode `\^12\catcode `\_12\catcode `\%12\relax}%
\providecommand \@@startlink[1]{}%
\providecommand \@@endlink[0]{}%
\providecommand \url  [0]{\begingroup\@sanitize@url \@url }%
\providecommand \@url [1]{\endgroup\@href {#1}{\urlprefix }}%
\providecommand \urlprefix  [0]{URL }%
\providecommand \Eprint [0]{\href }%
\providecommand \doibase [0]{https://doi.org/}%
\providecommand \selectlanguage [0]{\@gobble}%
\providecommand \bibinfo  [0]{\@secondoftwo}%
\providecommand \bibfield  [0]{\@secondoftwo}%
\providecommand \translation [1]{[#1]}%
\providecommand \BibitemOpen [0]{}%
\providecommand \bibitemStop [0]{}%
\providecommand \bibitemNoStop [0]{.\EOS\space}%
\providecommand \EOS [0]{\spacefactor3000\relax}%
\providecommand \BibitemShut  [1]{\csname bibitem#1\endcsname}%
\let\auto@bib@innerbib\@empty
\bibitem [{\citenamefont {Flamm}(2015)}]{Flamm}%
  \BibitemOpen
  \bibfield  {author} {\bibinfo {author} {\bibfnamefont {L.}~\bibnamefont {Flamm}},\ }\bibfield  {title} {\bibinfo {title} {Republication of: Contributions to einstein’s theory of gravitation},\ }\href {https://link.springer.com/article/10.1007/s10714-015-1908-2} {\bibfield  {journal} {\bibinfo  {journal} {General Relativity and Gravitation}\ }\textbf {\bibinfo {volume} {47}} (\bibinfo {year} {2015})}\BibitemShut {NoStop}%
\bibitem [{\citenamefont {Scholz}(2001)}]{Weyl}%
  \BibitemOpen
  \bibfield  {author} {\bibinfo {author} {\bibfnamefont {E.}~\bibnamefont {Scholz}},\ }\bibfield  {title} {\bibinfo {title} {Hermann weyl’s raum - zeit - materie and a general introduction to his scientific work},\ }\href {https://link.springer.com/book/10.1007/978-3-0348-8278-1} {\bibfield  {journal} {\bibinfo  {journal} {Springer DMV Seminar}\ }\textbf {\bibinfo {volume} {30}} (\bibinfo {year} {2001})}\BibitemShut {NoStop}%
\bibitem [{\citenamefont {Einstein}\ and\ \citenamefont {Rosen}(1935)}]{ERbridges}%
  \BibitemOpen
  \bibfield  {author} {\bibinfo {author} {\bibfnamefont {A.}~\bibnamefont {Einstein}}\ and\ \bibinfo {author} {\bibfnamefont {N.}~\bibnamefont {Rosen}},\ }\bibfield  {title} {\bibinfo {title} {The particle problem in the general theory of relativit},\ }\bibfield  {journal} {\bibinfo  {journal} {Physical Review}\ }\textbf {\bibinfo {volume} {48}},\ \href {https://doi.org/https://doi.org/10.1103/PhysRev.48.73} {https://doi.org/10.1103/PhysRev.48.73} (\bibinfo {year} {1935})\BibitemShut {NoStop}%
\bibitem [{\citenamefont {Fuller}\ and\ \citenamefont {Wheeler}(1962)}]{Fuller}%
  \BibitemOpen
  \bibfield  {author} {\bibinfo {author} {\bibfnamefont {R.~W.}\ \bibnamefont {Fuller}}\ and\ \bibinfo {author} {\bibfnamefont {J.~A.}\ \bibnamefont {Wheeler}},\ }\bibfield  {title} {\bibinfo {title} {Causality and multiply connected space-time},\ }\bibfield  {journal} {\bibinfo  {journal} {Physical Review}\ }\textbf {\bibinfo {volume} {128}},\ \href {https://doi.org/https://doi.org/10.1103/PhysRev.128.919} {https://doi.org/10.1103/PhysRev.128.919} (\bibinfo {year} {1962})\BibitemShut {NoStop}%
\bibitem [{\citenamefont {Ellis}(1973)}]{Ellis}%
  \BibitemOpen
  \bibfield  {author} {\bibinfo {author} {\bibfnamefont {H.~G.}\ \bibnamefont {Ellis}},\ }\bibfield  {title} {\bibinfo {title} {Ether flow through a drainhole: A particle model in general relativity},\ }\href {https://doi.org/https://doi.org/10.1063/1.1666161} {\bibfield  {journal} {\bibinfo  {journal} {Journal of Mathematical Physics}\ }\textbf {\bibinfo {volume} {14}},\ \bibinfo {pages} {104 } (\bibinfo {year} {1973})}\BibitemShut {NoStop}%
\bibitem [{\citenamefont {Bronnikov}(1973)}]{Bronnikov}%
  \BibitemOpen
  \bibfield  {author} {\bibinfo {author} {\bibfnamefont {K.~A.}\ \bibnamefont {Bronnikov}},\ }\bibfield  {title} {\bibinfo {title} {Scalar-tensor theory and scalar charge},\ }\href {https://www.actaphys.uj.edu.pl/fulltext?series=Reg&vol=4&page=251} {\bibfield  {journal} {\bibinfo  {journal} {Acta Physica Polonica B}\ }\textbf {\bibinfo {volume} {4}},\ \bibinfo {pages} {251 } (\bibinfo {year} {1973})}\BibitemShut {NoStop}%
\bibitem [{\citenamefont {Morris}\ and\ \citenamefont {Throne}(1988)}]{Morris-Throne}%
  \BibitemOpen
  \bibfield  {author} {\bibinfo {author} {\bibfnamefont {M.~S.}\ \bibnamefont {Morris}}\ and\ \bibinfo {author} {\bibfnamefont {K.~S.}\ \bibnamefont {Throne}},\ }\bibfield  {title} {\bibinfo {title} {Wormholes in spacetime and their use for interstellar travel: A tool for teaching general relativity},\ }\href {https://doi.org/https://doi.org/10.1119/1.15620} {\bibfield  {journal} {\bibinfo  {journal} {American Journal of Physics}\ }\textbf {\bibinfo {volume} {56}},\ \bibinfo {pages} {395 } (\bibinfo {year} {1988})}\BibitemShut {NoStop}%
\bibitem [{\citenamefont {Visser}(1989)}]{Visser1989}%
  \BibitemOpen
  \bibfield  {author} {\bibinfo {author} {\bibfnamefont {M.}~\bibnamefont {Visser}},\ }\bibfield  {title} {\bibinfo {title} {Traversable wormholes from surgically modified schwarzschild spacetimes},\ }\href {https://doi.org/https://doi.org/10.1016/0550-3213(89)90100-4} {\bibfield  {journal} {\bibinfo  {journal} {Nuclear Physics B}\ }\textbf {\bibinfo {volume} {328}},\ \bibinfo {pages} {203 } (\bibinfo {year} {1989})}\BibitemShut {NoStop}%
\bibitem [{\citenamefont {Teo}(1998)}]{Teo}%
  \BibitemOpen
  \bibfield  {author} {\bibinfo {author} {\bibfnamefont {E.}~\bibnamefont {Teo}},\ }\bibfield  {title} {\bibinfo {title} {Rotating traversable wormholes},\ }\bibfield  {journal} {\bibinfo  {journal} {Physical Review D}\ }\textbf {\bibinfo {volume} {58}},\ \href {https://doi.org/10.1103/physrevd.58.024014} {10.1103/physrevd.58.024014} (\bibinfo {year} {1998})\BibitemShut {NoStop}%
\bibitem [{\citenamefont {Cramer}\ \emph {et~al.}(1995)\citenamefont {Cramer} \emph {et~al.}}]{Cramer1995}%
  \BibitemOpen
  \bibfield  {author} {\bibinfo {author} {\bibfnamefont {J.~G.}\ \bibnamefont {Cramer}} \emph {et~al.},\ }\bibfield  {title} {\bibinfo {title} {Natural wormholes as gravitational lenses},\ }\bibfield  {journal} {\bibinfo  {journal} {Physical Review D}\ }\textbf {\bibinfo {volume} {51}},\ \href {https://doi.org/https://doi.org/10.1103/PhysRevD.51.3117} {https://doi.org/10.1103/PhysRevD.51.3117} (\bibinfo {year} {1995})\BibitemShut {NoStop}%
\bibitem [{\citenamefont {Gravanis}\ and\ \citenamefont {Willison}(2007)}]{Gravanis2007}%
  \BibitemOpen
  \bibfield  {author} {\bibinfo {author} {\bibfnamefont {E.}~\bibnamefont {Gravanis}}\ and\ \bibinfo {author} {\bibfnamefont {S.}~\bibnamefont {Willison}},\ }\bibfield  {title} {\bibinfo {title} {``mass without mass'' from thin shells in gauss-bonnet gravity},\ }\href {https://doi.org/10.1103/PhysRevD.75.084025} {\bibfield  {journal} {\bibinfo  {journal} {Phys. Rev. D}\ }\textbf {\bibinfo {volume} {75}},\ \bibinfo {pages} {084025} (\bibinfo {year} {2007})}\BibitemShut {NoStop}%
\bibitem [{\citenamefont {Bronnikov}\ and\ \citenamefont {Galiakhmetov}(2015)}]{Bronnikov_2015}%
  \BibitemOpen
  \bibfield  {author} {\bibinfo {author} {\bibfnamefont {K.~A.}\ \bibnamefont {Bronnikov}}\ and\ \bibinfo {author} {\bibfnamefont {A.~M.}\ \bibnamefont {Galiakhmetov}},\ }\bibfield  {title} {\bibinfo {title} {Wormholes without exotic matter in einstein–cartan theory},\ }\href {https://doi.org/10.1134/s0202289315040027} {\bibfield  {journal} {\bibinfo  {journal} {Gravitation and Cosmology}\ }\textbf {\bibinfo {volume} {21}},\ \bibinfo {pages} {283–288} (\bibinfo {year} {2015})}\BibitemShut {NoStop}%
\bibitem [{\citenamefont {Blázquez-Salcedo}\ \emph {et~al.}(2021)\citenamefont {Blázquez-Salcedo}, \citenamefont {Knoll},\ and\ \citenamefont {Radu}}]{Jose2021}%
  \BibitemOpen
  \bibfield  {author} {\bibinfo {author} {\bibfnamefont {J.~L.}\ \bibnamefont {Blázquez-Salcedo}}, \bibinfo {author} {\bibfnamefont {C.}~\bibnamefont {Knoll}},\ and\ \bibinfo {author} {\bibfnamefont {E.}~\bibnamefont {Radu}},\ }\bibfield  {title} {\bibinfo {title} {Traversable wormholes in einstein-dirac-maxwell theory},\ }\bibfield  {journal} {\bibinfo  {journal} {Physical Review Letters}\ }\textbf {\bibinfo {volume} {126}},\ \href {https://doi.org/10.1103/physrevlett.126.101102} {10.1103/physrevlett.126.101102} (\bibinfo {year} {2021})\BibitemShut {NoStop}%
\bibitem [{\citenamefont {Bakopoulos}\ \emph {et~al.}(2022)\citenamefont {Bakopoulos}, \citenamefont {Charmousis},\ and\ \citenamefont {Kanti}}]{Bakopoulos_2022}%
  \BibitemOpen
  \bibfield  {author} {\bibinfo {author} {\bibfnamefont {A.}~\bibnamefont {Bakopoulos}}, \bibinfo {author} {\bibfnamefont {C.}~\bibnamefont {Charmousis}},\ and\ \bibinfo {author} {\bibfnamefont {P.}~\bibnamefont {Kanti}},\ }\bibfield  {title} {\bibinfo {title} {Traversable wormholes in beyond horndeski theories},\ }\href {https://doi.org/10.1088/1475-7516/2022/05/022} {\bibfield  {journal} {\bibinfo  {journal} {Journal of Cosmology and Astroparticle Physics}\ }\textbf {\bibinfo {volume} {2022}}\bibinfo  {number} { (05)},\ \bibinfo {pages} {022}}\BibitemShut {NoStop}%
\bibitem [{\citenamefont {Mazza}\ \emph {et~al.}(2021)\citenamefont {Mazza}, \citenamefont {Franzin},\ and\ \citenamefont {Liberati}}]{Mazza_2021}%
  \BibitemOpen
\bibfield  {number} {  }\bibfield  {author} {\bibinfo {author} {\bibfnamefont {J.}~\bibnamefont {Mazza}}, \bibinfo {author} {\bibfnamefont {E.}~\bibnamefont {Franzin}},\ and\ \bibinfo {author} {\bibfnamefont {S.}~\bibnamefont {Liberati}},\ }\bibfield  {title} {\bibinfo {title} {A novel family of rotating black hole mimickers},\ }\href {https://doi.org/10.1088/1475-7516/2021/04/082} {\bibfield  {journal} {\bibinfo  {journal} {Journal of Cosmology and Astroparticle Physics}\ }\textbf {\bibinfo {volume} {2021}}\bibinfo  {number} { (04)},\ \bibinfo {pages} {082}}\BibitemShut {NoStop}%
\bibitem [{\citenamefont {Nedkova}\ \emph {et~al.}(2013)\citenamefont {Nedkova}, \citenamefont {Tinchev},\ and\ \citenamefont {Yazadjiev}}]{Petya2013}%
  \BibitemOpen
\bibfield  {number} {  }\bibfield  {author} {\bibinfo {author} {\bibfnamefont {P.~G.}\ \bibnamefont {Nedkova}}, \bibinfo {author} {\bibfnamefont {V.~K.}\ \bibnamefont {Tinchev}},\ and\ \bibinfo {author} {\bibfnamefont {S.~S.}\ \bibnamefont {Yazadjiev}},\ }\bibfield  {title} {\bibinfo {title} {Shadow of a rotating traversable wormhole},\ }\bibfield  {journal} {\bibinfo  {journal} {Physical Review D}\ }\textbf {\bibinfo {volume} {88}},\ \href {https://doi.org/10.1103/physrevd.88.124019} {10.1103/physrevd.88.124019} (\bibinfo {year} {2013})\BibitemShut {NoStop}%
\bibitem [{\citenamefont {Gyulchev}\ \emph {et~al.}(2018)\citenamefont {Gyulchev}, \citenamefont {Nedkova}, \citenamefont {Tinchev},\ and\ \citenamefont {Yazadjiev}}]{Gyulchev:2018}%
  \BibitemOpen
  \bibfield  {author} {\bibinfo {author} {\bibfnamefont {G.}~\bibnamefont {Gyulchev}}, \bibinfo {author} {\bibfnamefont {P.}~\bibnamefont {Nedkova}}, \bibinfo {author} {\bibfnamefont {V.}~\bibnamefont {Tinchev}},\ and\ \bibinfo {author} {\bibfnamefont {S.}~\bibnamefont {Yazadjiev}},\ }\bibfield  {title} {\bibinfo {title} {{On the shadow of rotating traversable wormholes}},\ }\href {https://doi.org/10.1140/epjc/s10052-018-6012-9} {\bibfield  {journal} {\bibinfo  {journal} {Eur. Phys. J. C}\ }\textbf {\bibinfo {volume} {78}},\ \bibinfo {pages} {544} (\bibinfo {year} {2018})},\ \Eprint {https://arxiv.org/abs/1805.11591} {arXiv:1805.11591 [gr-qc]} \BibitemShut {NoStop}%
\bibitem [{\citenamefont {Huang}\ \emph {et~al.}(2023)\citenamefont {Huang}, \citenamefont {Kunz}, \citenamefont {Yang},\ and\ \citenamefont {Zhang}}]{Hyat2023}%
  \BibitemOpen
  \bibfield  {author} {\bibinfo {author} {\bibfnamefont {H.}~\bibnamefont {Huang}}, \bibinfo {author} {\bibfnamefont {J.}~\bibnamefont {Kunz}}, \bibinfo {author} {\bibfnamefont {J.}~\bibnamefont {Yang}},\ and\ \bibinfo {author} {\bibfnamefont {C.}~\bibnamefont {Zhang}},\ }\bibfield  {title} {\bibinfo {title} {Light ring behind wormhole throat: Geodesics, images, and shadows},\ }\bibfield  {journal} {\bibinfo  {journal} {Physical Review D}\ }\textbf {\bibinfo {volume} {107}},\ \href {https://doi.org/10.1103/physrevd.107.104060} {10.1103/physrevd.107.104060} (\bibinfo {year} {2023})\BibitemShut {NoStop}%
\bibitem [{\citenamefont {Paul}\ \emph {et~al.}(2020)\citenamefont {Paul}, \citenamefont {Shaikh}, \citenamefont {Banerjee},\ and\ \citenamefont {Sarkar}}]{Paul_2020}%
  \BibitemOpen
  \bibfield  {author} {\bibinfo {author} {\bibfnamefont {S.}~\bibnamefont {Paul}}, \bibinfo {author} {\bibfnamefont {R.}~\bibnamefont {Shaikh}}, \bibinfo {author} {\bibfnamefont {P.}~\bibnamefont {Banerjee}},\ and\ \bibinfo {author} {\bibfnamefont {T.}~\bibnamefont {Sarkar}},\ }\bibfield  {title} {\bibinfo {title} {Observational signatures of wormholes with thin accretion disks},\ }\href {https://doi.org/10.1088/1475-7516/2020/03/055} {\bibfield  {journal} {\bibinfo  {journal} {Journal of Cosmology and Astroparticle Physics}\ }\textbf {\bibinfo {volume} {2020}}\bibinfo  {number} { (03)},\ \bibinfo {pages} {055–055}}\BibitemShut {NoStop}%
\bibitem [{\citenamefont {Deligianni}\ \emph {et~al.}(2021{\natexlab{a}})\citenamefont {Deligianni}, \citenamefont {Kunz}, \citenamefont {Nedkova}, \citenamefont {Yazadjiev},\ and\ \citenamefont {Zheleva}}]{Deligianni:2021}%
  \BibitemOpen
\bibfield  {number} {  }\bibfield  {author} {\bibinfo {author} {\bibfnamefont {E.}~\bibnamefont {Deligianni}}, \bibinfo {author} {\bibfnamefont {J.}~\bibnamefont {Kunz}}, \bibinfo {author} {\bibfnamefont {P.}~\bibnamefont {Nedkova}}, \bibinfo {author} {\bibfnamefont {S.}~\bibnamefont {Yazadjiev}},\ and\ \bibinfo {author} {\bibfnamefont {R.}~\bibnamefont {Zheleva}},\ }\bibfield  {title} {\bibinfo {title} {{Quasiperiodic oscillations around rotating traversable wormholes}},\ }\href {https://doi.org/10.1103/PhysRevD.104.024048} {\bibfield  {journal} {\bibinfo  {journal} {Phys. Rev. D}\ }\textbf {\bibinfo {volume} {104}},\ \bibinfo {pages} {024048} (\bibinfo {year} {2021}{\natexlab{a}})},\ \Eprint {https://arxiv.org/abs/2103.13504} {arXiv:2103.13504 [gr-qc]} \BibitemShut {NoStop}%
\bibitem [{\citenamefont {Deligianni}\ \emph {et~al.}(2021{\natexlab{b}})\citenamefont {Deligianni}, \citenamefont {Kleihaus}, \citenamefont {Kunz}, \citenamefont {Nedkova},\ and\ \citenamefont {Yazadjiev}}]{Deligianni:2021hwt}%
  \BibitemOpen
  \bibfield  {author} {\bibinfo {author} {\bibfnamefont {E.}~\bibnamefont {Deligianni}}, \bibinfo {author} {\bibfnamefont {B.}~\bibnamefont {Kleihaus}}, \bibinfo {author} {\bibfnamefont {J.}~\bibnamefont {Kunz}}, \bibinfo {author} {\bibfnamefont {P.}~\bibnamefont {Nedkova}},\ and\ \bibinfo {author} {\bibfnamefont {S.}~\bibnamefont {Yazadjiev}},\ }\bibfield  {title} {\bibinfo {title} {{Quasiperiodic oscillations in rotating Ellis wormhole spacetimes}},\ }\href {https://doi.org/10.1103/PhysRevD.104.064043} {\bibfield  {journal} {\bibinfo  {journal} {Phys. Rev. D}\ }\textbf {\bibinfo {volume} {104}},\ \bibinfo {pages} {064043} (\bibinfo {year} {2021}{\natexlab{b}})},\ \Eprint {https://arxiv.org/abs/2107.01421} {arXiv:2107.01421 [gr-qc]} \BibitemShut {NoStop}%
\bibitem [{\citenamefont {Delijski}\ \emph {et~al.}(2022)\citenamefont {Delijski}, \citenamefont {Gyulchev}, \citenamefont {Nedkova},\ and\ \citenamefont {Yazadjiev}}]{Delijski:2022}%
  \BibitemOpen
  \bibfield  {author} {\bibinfo {author} {\bibfnamefont {V.}~\bibnamefont {Delijski}}, \bibinfo {author} {\bibfnamefont {G.}~\bibnamefont {Gyulchev}}, \bibinfo {author} {\bibfnamefont {P.}~\bibnamefont {Nedkova}},\ and\ \bibinfo {author} {\bibfnamefont {S.}~\bibnamefont {Yazadjiev}},\ }\bibfield  {title} {\bibinfo {title} {{Polarized image of equatorial emission in horizonless spacetimes: Traversable wormholes}},\ }\href {https://doi.org/10.1103/PhysRevD.106.104024} {\bibfield  {journal} {\bibinfo  {journal} {Phys. Rev. D}\ }\textbf {\bibinfo {volume} {106}},\ \bibinfo {pages} {104024} (\bibinfo {year} {2022})},\ \Eprint {https://arxiv.org/abs/2206.09455} {arXiv:2206.09455 [gr-qc]} \BibitemShut {NoStop}%
\bibitem [{\citenamefont {{Fishbone}}\ and\ \citenamefont {{Moncrief}}(1976)}]{Fishbone1976}%
  \BibitemOpen
  \bibfield  {author} {\bibinfo {author} {\bibfnamefont {L.~G.}\ \bibnamefont {{Fishbone}}}\ and\ \bibinfo {author} {\bibfnamefont {V.}~\bibnamefont {{Moncrief}}},\ }\bibfield  {title} {\bibinfo {title} {{Relativistic fluid disks in orbit around Kerr black holes.}},\ }\href {https://doi.org/10.1086/154565} {\bibfield  {journal} {\bibinfo  {journal} {\apj}\ }\textbf {\bibinfo {volume} {207}},\ \bibinfo {pages} {962} (\bibinfo {year} {1976})}\BibitemShut {NoStop}%
\bibitem [{\citenamefont {{Abramowicz}}\ \emph {et~al.}(1978)\citenamefont {{Abramowicz}}, \citenamefont {{Jaroszynski}},\ and\ \citenamefont {{Sikora}}}]{Abramowicz1978}%
  \BibitemOpen
  \bibfield  {author} {\bibinfo {author} {\bibfnamefont {M.}~\bibnamefont {{Abramowicz}}}, \bibinfo {author} {\bibfnamefont {M.}~\bibnamefont {{Jaroszynski}}},\ and\ \bibinfo {author} {\bibfnamefont {M.}~\bibnamefont {{Sikora}}},\ }\bibfield  {title} {\bibinfo {title} {{Relativistic, accreting disks.}},\ }\href@noop {} {\bibfield  {journal} {\bibinfo  {journal} {Astronomy and Astrophysics}\ }\textbf {\bibinfo {volume} {63}},\ \bibinfo {pages} {221} (\bibinfo {year} {1978})}\BibitemShut {NoStop}%
\bibitem [{\citenamefont {{Kozlowski}}\ \emph {et~al.}(1978)\citenamefont {{Kozlowski}}, \citenamefont {{Jaroszynski}},\ and\ \citenamefont {{Abramowicz}}}]{Kozlowski1978}%
  \BibitemOpen
  \bibfield  {author} {\bibinfo {author} {\bibfnamefont {M.}~\bibnamefont {{Kozlowski}}}, \bibinfo {author} {\bibfnamefont {M.}~\bibnamefont {{Jaroszynski}}},\ and\ \bibinfo {author} {\bibfnamefont {M.~A.}\ \bibnamefont {{Abramowicz}}},\ }\bibfield  {title} {\bibinfo {title} {{The analytic theory of fluid disks orbiting the Kerr black hole.}},\ }\href@noop {} {\bibfield  {journal} {\bibinfo  {journal} {Astronomy and Astrophysics}\ }\textbf {\bibinfo {volume} {63}},\ \bibinfo {pages} {209} (\bibinfo {year} {1978})}\BibitemShut {NoStop}%
\bibitem [{\citenamefont {{Abramowicz}}\ \emph {et~al.}(1980)\citenamefont {{Abramowicz}}, \citenamefont {{Calvani}},\ and\ \citenamefont {{Nobili}}}]{Abramowicz1980}%
  \BibitemOpen
  \bibfield  {author} {\bibinfo {author} {\bibfnamefont {M.~A.}\ \bibnamefont {{Abramowicz}}}, \bibinfo {author} {\bibfnamefont {M.}~\bibnamefont {{Calvani}}},\ and\ \bibinfo {author} {\bibfnamefont {L.}~\bibnamefont {{Nobili}}},\ }\bibfield  {title} {\bibinfo {title} {{Thick accretion disks with super-Eddington luminosities}},\ }\href {https://doi.org/10.1086/158512} {\bibfield  {journal} {\bibinfo  {journal} {\apj}\ }\textbf {\bibinfo {volume} {242}},\ \bibinfo {pages} {772} (\bibinfo {year} {1980})}\BibitemShut {NoStop}%
\bibitem [{\citenamefont {{Paczy{\'n}sky}}\ and\ \citenamefont {{Wiita}}(1980)}]{Paczynsky1980}%
  \BibitemOpen
  \bibfield  {author} {\bibinfo {author} {\bibfnamefont {B.}~\bibnamefont {{Paczy{\'n}sky}}}\ and\ \bibinfo {author} {\bibfnamefont {P.~J.}\ \bibnamefont {{Wiita}}},\ }\bibfield  {title} {\bibinfo {title} {{Thick Accretion Disks and Supercritical Luminosities}},\ }\href@noop {} {\bibfield  {journal} {\bibinfo  {journal} {Astronomy and Astrophysics}\ }\textbf {\bibinfo {volume} {88}},\ \bibinfo {pages} {23} (\bibinfo {year} {1980})}\BibitemShut {NoStop}%
\bibitem [{\citenamefont {{Paczynski}}\ and\ \citenamefont {{Bisnovatyi-Kogan}}(1981)}]{Paczynski1981}%
  \BibitemOpen
  \bibfield  {author} {\bibinfo {author} {\bibfnamefont {B.}~\bibnamefont {{Paczynski}}}\ and\ \bibinfo {author} {\bibfnamefont {G.}~\bibnamefont {{Bisnovatyi-Kogan}}},\ }\bibfield  {title} {\bibinfo {title} {{A Model of a Thin Accretion Disk around a Black Hole}},\ }\href {https://ui.adsabs.harvard.edu/abs/1981AcA....31..283P} {\bibfield  {journal} {\bibinfo  {journal} {Acta Astronomica}\ }\textbf {\bibinfo {volume} {31}},\ \bibinfo {pages} {283} (\bibinfo {year} {1981})}\BibitemShut {NoStop}%
\bibitem [{\citenamefont {{Paczynski}}\ and\ \citenamefont {{Abramowicz}}(1982)}]{Paczysnki1982}%
  \BibitemOpen
  \bibfield  {author} {\bibinfo {author} {\bibfnamefont {B.}~\bibnamefont {{Paczynski}}}\ and\ \bibinfo {author} {\bibfnamefont {M.~A.}\ \bibnamefont {{Abramowicz}}},\ }\bibfield  {title} {\bibinfo {title} {{A model of a thick disk with equatorial accretion}},\ }\href {https://doi.org/10.1086/159689} {\bibfield  {journal} {\bibinfo  {journal} {\apj}\ }\textbf {\bibinfo {volume} {253}},\ \bibinfo {pages} {897} (\bibinfo {year} {1982})}\BibitemShut {NoStop}%
\bibitem [{\citenamefont {Gjorgjieski}\ \emph {et~al.}(2025)\citenamefont {Gjorgjieski}, \citenamefont {Kunz},\ and\ \citenamefont {Nedkova}}]{ThroatOrbits}%
  \BibitemOpen
  \bibfield  {author} {\bibinfo {author} {\bibfnamefont {K.}~\bibnamefont {Gjorgjieski}}, \bibinfo {author} {\bibfnamefont {J.}~\bibnamefont {Kunz}},\ and\ \bibinfo {author} {\bibfnamefont {P.}~\bibnamefont {Nedkova}},\ }\href {https://arxiv.org/abs/2505.07507} {\bibinfo {title} {Circular orbits and photon orbits at wormhole throats}} (\bibinfo {year} {2025}),\ \Eprint {https://arxiv.org/abs/2505.07507} {arXiv:2505.07507 [gr-qc]} \BibitemShut {NoStop}%
\end{thebibliography}%

\end{document}